\documentclass[preprints,review,accept,oneauthor,pdftex]{mdpi} 
\makeatletter

\let\jnl@style=\rm
\def\ref@jnl#1{{\jnl@style#1}}

\def\aj{\ref@jnl{\emph{Astron. J.}}}                   
\def\actaa{\ref@jnl{\emph{Acta Astron.}}}      
\def\araa{\ref@jnl{\emph{Annu. Rev. Astron Astrophys}}}             
\def\apj{\ref@jnl{\emph{Astrophys. J.}}}                 
\def\apjl{\ref@jnl{\emph{Astrophys. J. Lett.}}}                
\def\apjs{\ref@jnl{\emph{Astrophys. J. Suppl.}}}               
\def\ao{\ref@jnl{\emph{Appl. Opt.}}}           
\def\apss{\ref@jnl{\emph{Astrophys. Space Sci.}}}             
\def\aap{\ref@jnl{\emph{Astron. Astrophys.}}}                
\def\aapr{\ref@jnl{\emph{Astron. Astrophys. Rev.}}}          
\def\aaps{\ref@jnl{\emph{Astron. Astrophys. Suppl.}}}              
\def\azh{\ref@jnl{\emph{Astron. Zhurnal}}}                 
\def\baas{\ref@jnl{\emph{Bull. Am. Astron. Soc.}}}               
\def\bac{\ref@jnl{\emph{Bull. Astron. Inst. Czechoslov.}}}
\def\caa{\ref@jnl{\emph{Chin. Astron. Astrophys.}}}
\def\cjaa{\ref@jnl{\emph{Chin. J. Astron. Astrophys.}}}
\def\icarus{\ref@jnl{\emph{Icarus}}}           
\def\jcap{\ref@jnl{\emph{J. Cosmol. Astropart. Phys.}}}
\def\jrasc{\ref@jnl{\emph{J. RAS Can.}}}             
\def\memras{\ref@jnl{\emph{Mem. RAS}}}            
\def\mnras{\ref@jnl{\emph{Mon. Not. R. Astron. Soc.}}}             
\def\na{\ref@jnl{\emph{New Astron.}}}                
\def\nar{\ref@jnl{\emph{New Astron. Rev.}}}          
\def\pra{\ref@jnl{\emph{Phys. Rev. A Gen. Phys.}}}        
\def\prb{\ref@jnl{\emph{Phys. Rev. B Solid State}}}        
\def\prc{\ref@jnl{\emph{Phys. Rev. C}}}        
\def\prd{\ref@jnl{\emph{Phys. Rev. D}}}        
\def\pre{\ref@jnl{\emph{Phys. Rev. E}}}        
\def\prl{\ref@jnl{\emph{Phys. Rev. Lett.}}}    
\def\pasa{\ref@jnl{\emph{Publ. Astron. Soc. Aust.}}}               
\def\pasp{\ref@jnl{\emph{Publ. ASP}}}               
\def\pasj{\ref@jnl{\emph{Publ. ASJ}}}               
\def\rmxaa{\ref@jnl{\emph{Rev. Mex. Astron. Astrofis.}}}%
\def\qjras{\ref@jnl{\emph{Q. J. RAS}}}             
\def\skytel{\ref@jnl{\emph{Sky Telesc.}}}             
\def\solphys{\ref@jnl{\emph{Sol. Phys.}}}      
\def\sovast{\ref@jnl{\emph{Soviet Astron.}}}      
\def\ssr{\ref@jnl{\emph{Space Sci. Rev.}}}     
\def\zap{\ref@jnl{\emph{Z. Fuer Astrophys.}}}                 
\def\nat{\ref@jnl{\emph{Nature}}}              
\def\iaucirc{\ref@jnl{\emph{IAU Cirulars}}}       
\def\aplett{\ref@jnl{\emph{Astrophys. Lett.}}} 
\def\apspr{\ref@jnl{\emph{Astrophys. Space Phys. Res.}}}
\def\bain{\ref@jnl{\emph{Bull. Astron. Inst. Neth.}}} 
\def\fcp{\ref@jnl{\emph{Fundam. Cosm. Phys.}}}  
\def\gca{\ref@jnl{\emph{Geochim. Cosmochim. Acta}}}   
\def\grl{\ref@jnl{\emph{Geophys. Res. Lett.}}} 
\def\jcp{\ref@jnl{\emph{J. Chem. Phys.}}}      
\def\jgr{\ref@jnl{\emph{J. Geophys. Res.}}}    
\def\jqsrt{\ref@jnl{\emph{J. Quant. Spec. Radiat. Transf.}}}
\def\memsai{\ref@jnl{\emph{Mem. Soc. Astron. Ital.}}}
\def\nphysa{\ref@jnl{\emph{Nucl. Phys. A}}}   
\def\physrep{\ref@jnl{\emph{Phys. Rep.}}}   
\def\physscr{\ref@jnl{\emph{Phys. Scr.}}}   
\def\planss{\ref@jnl{\emph{Planet. Space Sci.}}}   
\def\procspie{\ref@jnl{\emph{Proc. SPIE}}}   

\makeatother

\setitemize{parsep=6pt,itemsep=0pt,leftmargin=*,labelsep=5.5mm}
\setenumerate{parsep=6pt,itemsep=0pt,leftmargin=*,labelsep=5.5mm}
\setlist[description]{itemsep=0mm}
\usepackage{environ}
\NewEnviron{myequation}{%
\begin{equation}
\scalebox{0.9}{$\BODY$}
\end{equation}}

\firstpage{1} 
\pubvolume{05}
\issuenum{7}
\articlenumber{0}
\pubyear{2019}
\copyrightyear{2019}
\history{Received: 31 May 2019; Accepted: 9 July 2019; Published: {date}}

\pdfoutput=1

\Title{A Model-Independent Characterisation of Strong Gravitational Lensing by Observables}

\Author{Jenny Wagner 
 \orcidA{}}

\AuthorNames{Jenny Wagner}

\address[1]{%
Astronomisches Rechen-Institut, Universit\"at Heidelberg, Zentrum f\"ur Astronomie, M\"onchhofstr. 12--14, 69120~Heidelberg, Germany; j.wagner@uni-heidelberg.de}

\abstract{When light from a distant source object, like a galaxy or a supernova, travels towards us, it~is deflected by massive objects that lie in its path. 
When the mass density of the deflecting object exceeds a certain threshold, multiple, highly distorted images of the source are observed. 
This~strong gravitational lensing effect has so far been treated as a model-fitting problem. 
Using~the observed multiple images as constraints yields a self-consistent model of the deflecting mass density and the source object. 
As several models meet the constraints equally well, we develop a lens characterisation that separates data-based information from model assumptions.
The observed multiple images allow us to determine local properties of the deflecting mass distribution on any mass scale from one simple set of equations. 
Their solution is unique and free of model-dependent degeneracies. 
The reconstruction of source objects can be performed completely model-independently, enabling us to study galaxy evolution without a lens-model bias. 
Our approach reduces the lens and source description to its data-based evidence that all models agree upon, simplifies an automated treatment of large datasets, and allows for an extrapolation to a global description resembling model-based~descriptions.}


\begin{document}

\section{Introduction}\label{sec:introduction}

Forty years ago, in~1979, two images of the quasar QSO 0957+561 were observed, \cite{bib:Walsh}, which~marked the discovery of strong gravitational lenses. {The first description of the image configuration in form of a lens model followed in~\cite{bib:Young}. In~1986, luminous arcs, highly magnified images of background galaxies, were discovered behind the galaxy clusters Abell 370 and Cl2244-02,~\cite{bib:Lynds,bib:Soucail}. The~first lens models for Abell 370 were set up in~\cite{bib:Hammer,bib:Kovner}.}
Since then, observations of multiple images have been routinely used to probe the deflecting mass-density distribution, in~particular in order to infer its dark matter content and dark matter properties---see, e.g.,\@ \cite{bib:Bolton,bib:Gavazzi} for galaxy-scale lens surveys, or~\cite{bib:Postman,bib:Lotz} for galaxy-cluster-scale lens surveys.  
Recent benchmarks with simulations of cluster-scale lenses show that a lot of different approaches accurately and precisely reconstruct the mass-density distribution in the vicinity of extended multiple images up to a few percent,~\cite{bib:Meneghetti}. 
For galaxy-scale lenses, an~analogous project is currently being pursued,~\cite{bib:Ding}. 
Furthermore, strong gravitational lensing of time varying objects has been used as a cosmological probe to determine the Hubble-Lemaître constant, $H_0$, \cite{bib:Refsdal,bib:Suyu,bib:Grillo}. 
Attempts to infer the cosmic spatial curvature, the~cosmic matter density parameter, and~dark energy properties of the current cosmological concordance model and its potential extensions are also being pursued with galaxy-scale and cluster-scale lenses,~\cite{bib:Collett,bib:Rasanen,bib:Magana}.
Even after forty years of strong gravitational lensing studies, it still is a subject of high research interest because unprecedented observations, like supernova (SN) Refsdal,~\cite{bib:Kelly}, or~the recently discovered fast radio bursts (FRBs), \cite{bib:Lorimer}, have continuously contributed to develop lens reconstruction methods further and thereby broaden the range of strong-lensing~applications.

In our research, as~being developed in~\cite{bib:Wagner0,bib:Wagner1,bib:Wagner2,bib:Wagner3,bib:Wagner4,bib:Wagner5,bib:Wagner6}, we pursue the question which properties of a strong gravitational lens can be directly inferred from the observables that describe a multiple-image configuration. 
Contrary to most approaches, we do not aim for a \emph{global} mass-density reconstruction of the entire lens that usually employs the observables as constraints in a model-fitting problem. 
For~galaxy-scale lenses, such global lens reconstructions using parametric models or free-form methods can be found in~\cite{bib:Keeton,bib:Saha,bib:Suyu_glee,bib:Vegetti,bib:Birrer3, bib:Tessore2}, for~instance.
Galaxy-cluster mass densities can be reconstructed with global lens reconstructions as, for~instance, detailed in~\cite{bib:Liesenborgs_grale,bib:Merten_sawlens,bib:Jullo_lenstool,bib:Zitrin_ltm}. References to further approaches and a comparison between the approaches can be found in~\cite{bib:Meneghetti}. 
Some of the approaches originally developed for galaxy-scale lenses can also be employed and extended to reconstruct cluster-scale lenses and vice~versa, see~\cite{bib:Grillo_glee} for an example.
Due to the high symmetry of the lenses on galaxy scale, fully automatic characterisations are easier to implement than for cluster-scale lenses.
The~latter usually require at least a visual inspection of the results, or~manual fine-tuning at intermediate stages. 
In~addition, aiming at a global characterisation, the~sparsely distributed multiple images are not sufficient to fully constrain the mass density within the cluster,~\cite{bib:Liesenborgs1}.
As a consequence, different approaches employ different additional information and assumptions to reconstruct the mass-density distribution, which complicates the comparison of the~results.

Instead of finding the global deflecting mass density that can cause the multiple images observed, we separate data-based information from model-based assumptions to find those lens properties that all approaches agree upon. 
We employ the most general equations of the standard gravitational lensing formalism to directly determine \emph{local} lens properties from the observables.
These local properties are unique and calculated in the same way for all lenses from symmetric galaxy-scale lenses to amorphous galaxy-cluster-scale lenses.
Thus, the~approach is highly efficient and robust to handle large datasets with a minimum amount of manual~intervention. 

The cosmic distances between us, the~lens, and~the source are usually determined according to the cosmological standard model and, as~such, are not model-independent, see e.g.,\@ \cite{bib:Bonvin}. 
To make them independent of any assumption about the nature and the overall abundance of dark matter and dark energy in the universe, we set up data-based cosmic distances from the most recent Pantheon sample of SNe type Ia,~\cite{bib:Scolnic}. 
This covers SNe up to redshift 2.3 such that data-based distances are available for a lot of observed lensing configurations.
In this way, our approach is independent of a specific lens model and of a specific parametrisation of the cosmological background model, which is only assumed to be spatially homogeneous and isotropic and should fulfil the Einstein field equations. 
To set up the data-based cosmic distances, we employ the analytic orthonormal basis introduced in~\cite{bib:Mignone} for the sake of run-time efficiency and numerical stability. 
Other approaches to develop distance measures based on the standardisable candles are, for~instance,~\cite{bib:Wang,bib:Ishida,bib:Porqueres,bib:Gomez}.

In this work, we summarise the status quo of our approach, show a comparison to different global lens reconstruction ansatzes, based on~\cite{bib:Wagner_cluster,bib:Wagner_quasar}, and~detail its contribution to determine the invariance transformations of the most general equations of the gravitational lensing formalism, especially the time-delay equation which is used to infer $H_0$. 
Based on these results, we outline future developments of lens reconstructions that could be enabled and tested by the recently discovered FRBs, as~further described in~\cite{bib:Wagner_frb}.

The next sections are organised as follows: 
In Section~\ref{sec:materials}, we introduce the theory of the model-independent lens characterisation for the leading-order configurations of multiple images. 
We list the local lens characteristics that can be determined by the different observables in each multiple-image configuration.
Section~\ref{sec:results} shows examples for these different cases on galaxy and on galaxy-cluster scale and the results obtained by our approach.
We also estimate to which precision $H_0$ can be determined from the time-delay equation employing the Pantheon-data-based angular diameter distances. 
After these proof-of-concept examples, we summarise the assets and applications of our approach in Section~\ref{sec:discussion} and conclude the review with a diagrammatic summary of our~method.

\section{Materials and~Methods}\label{sec:materials}

\begin{figure}[H]
\centering
\includegraphics[width=0.77\textwidth]{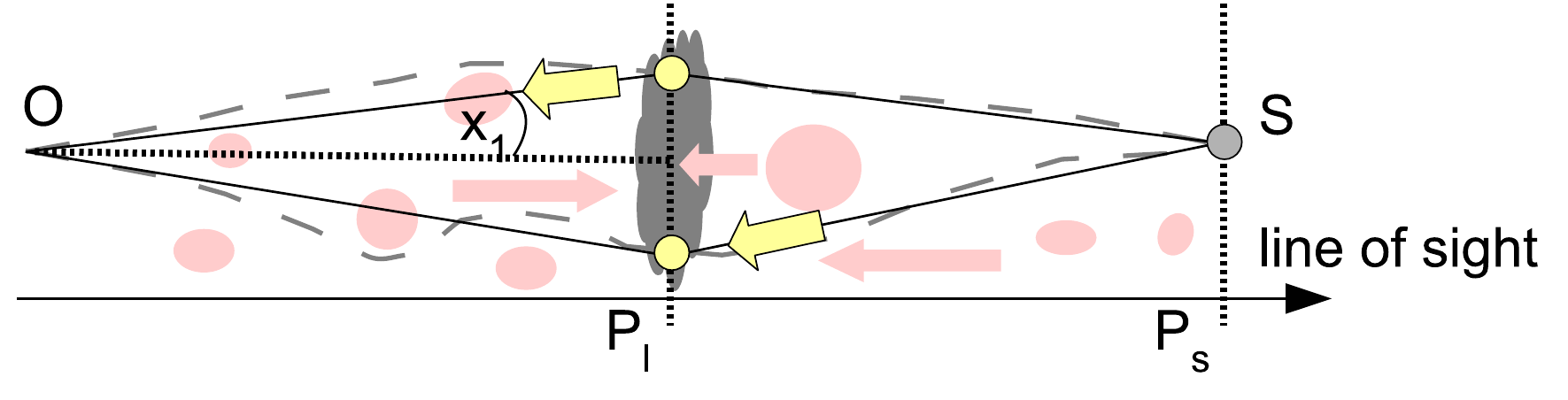}
\caption{Strong gravitational lenses can cause a deflection of light rays from a source $S$ (grey circle) in the source plane $P_\mathrm{s}$ into two images (yellow circles) in the lens plane $P_\mathrm{l}$. The~upper image is observed at the angular position $x_1$ on the sky. The~deflecting mass density (grey cloud) is assumed to consist of all masses along the line of sight (red circles) projected into $P_\mathrm{l}$, such that the light deflection is assumed to happen instantaneously at $P_\mathrm{l}$ and not at the individual masses (as indicated by the grey dashed paths). Differences in the arrival times of the light rays from the images at the observer $O$ (indicated by the yellow arrows) occur due to the different light~paths.}
\label{fig:lensing_sketch}
\end{figure}

In this section, we introduce the theoretical concepts to determine local properties of a strong gravitational lens without assuming a specific lens model and without assuming a specific parametrisation of a spatially homogeneous and isotropic Friedmann background cosmology. 
Most~cases of strong gravitational lensing are well described by an effective theory of light deflection by a mass distribution in the static, thin-screen, weak-field limit in a spatially homogeneous and isotropic background cosmology, as~summarised in~\cite{bib:SEF,bib:Narayan}. 
This means that the deflecting mass distribution is assumed to be located in one (or several) plane(s) orthogonal to the line of sight, the~lens plane(s). 
The~mass density in these lens planes can be described by a Newtonian gravitational potential on top of the cosmic background metric which is usually based on a Friedmann--Lemaître--Robertson--Walker metric (FLRW metric). 
Depending on the mass-density distribution, as~investigated in~\cite{bib:Subramanian}, this~Newtonian potential can deflect light rays emitted from a source, lying in a source plane, into~multiple images.
It is assumed that the distances on the sky between the angular positions of the source and the multiple images are small compared to the distances that the light rays travel along the line of sight.
For the light deflection to be static, all motions of the observer, the~source, and~the lens with respect to each other are considered negligible during the observing time.
If the source object has an intrinsic time-varying intensity, the~deflected light rays arrive at different times at the observer because they cross environments of different mass densities and travel distances of varying length from the source to the observer. 
Figure~\ref{fig:lensing_sketch} depicts an example of the strong gravitational lensing effect in which one background source is mapped to two images by a deflecting mass density in one lens~plane. 

In principle, there are other ways to describe the deflection of light rays in our universe which are detailed in~\cite{bib:Wagner6}. 
For instance, the~approach in which the background metric accounts for the entire deflecting mass-density distribution is an equivalent ansatz. 
However,  determining the geodesics in such an inhomogeneous universe is more difficult than treating gravitational lensing in an effective theory based on lens planes in a homogeneous and isotropic background cosmology. 
Another option partitions the mass density along the line of sight into a main deflecting object and small-scale perturbations scattered along the light path in an FLRW background metric without projecting the masses to lens~planes.

In order to avoid degeneracies within the description, we should choose the background metric, small-scale weakly deflecting masses, and~deflecting masses that cause multiple images in such a way that the observables can directly constrain all parts of the chosen partition.
We detail our choices of an FLRW background cosmology and a single-lens-plane Newtonian deflection potential in \mbox{Sections~\ref{sec:observables} and \ref{sec:formalism}}.

Given these prerequisites, singularity theory has been established as the mathematical foundation to characterise the generic, nonlinear mappings between multiple-image configurations and their sources,~\cite{bib:Petters}. 
Originally developed by Vladimir Arnold to describe the set of singular points arising on manifolds that undergo (sudden) changes, e.g.,~projections to lower dimensions, singularity theory is the ideal tool to describe gravitational lensing in the lens-plane formalism introduced above.
It~thus provides a framework to solve the under-constrained inverse problem of reconstructing the gravitational lens and the source from an observed multiple-image~configuration.

\subsection{Observables in Multiple~Images}
\label{sec:observables}

As the (time-varying) brightness profile of multiple images caused by the strong gravitational lensing effect depends on the (time-varying) brightness profile of the source object and the deflecting mass distribution, one may expect a large variance in the appearances of multiple images and an equally high variance in the arrival time patterns of time-varying multiple images. 
However,  the~current measurement precision and resolution allow us to categorise the observations into three types of multiple images sharing the same observable features as summarised in Table~\ref{tab:image_types}.
The number of multiple-image configurations is also limited and most of the occurring cases can be categorised into the three configurations shown in Table~\ref{tab:image_configurations}. 
Here, we focus on the observational properties of these configurations.
In Section~\ref{sec:formalism}, we give an explanation for the limited amount of possible configurations based on singularity theory.
Furthermore, general rules about the order in which light from these multiple-image configurations reaches the observer can also be derived from singularity theory,~\cite{bib:Petters}. 

\begin{table}[H]
\caption{Types of multiple images and quantitative observables that can be extracted from each image. Each type can occur in different multiple-image configurations shown in Table~\ref{tab:image_configurations}. \textit{Image credits: NASA/ESA/HST}.}
\centering
\begin{tabular}{ccc}
\toprule
 \textbf{Point Image}	& \textbf{Unresolved Image} & \textbf{Resolved Image} \\
\midrule
\includegraphics[width=0.24\textwidth]{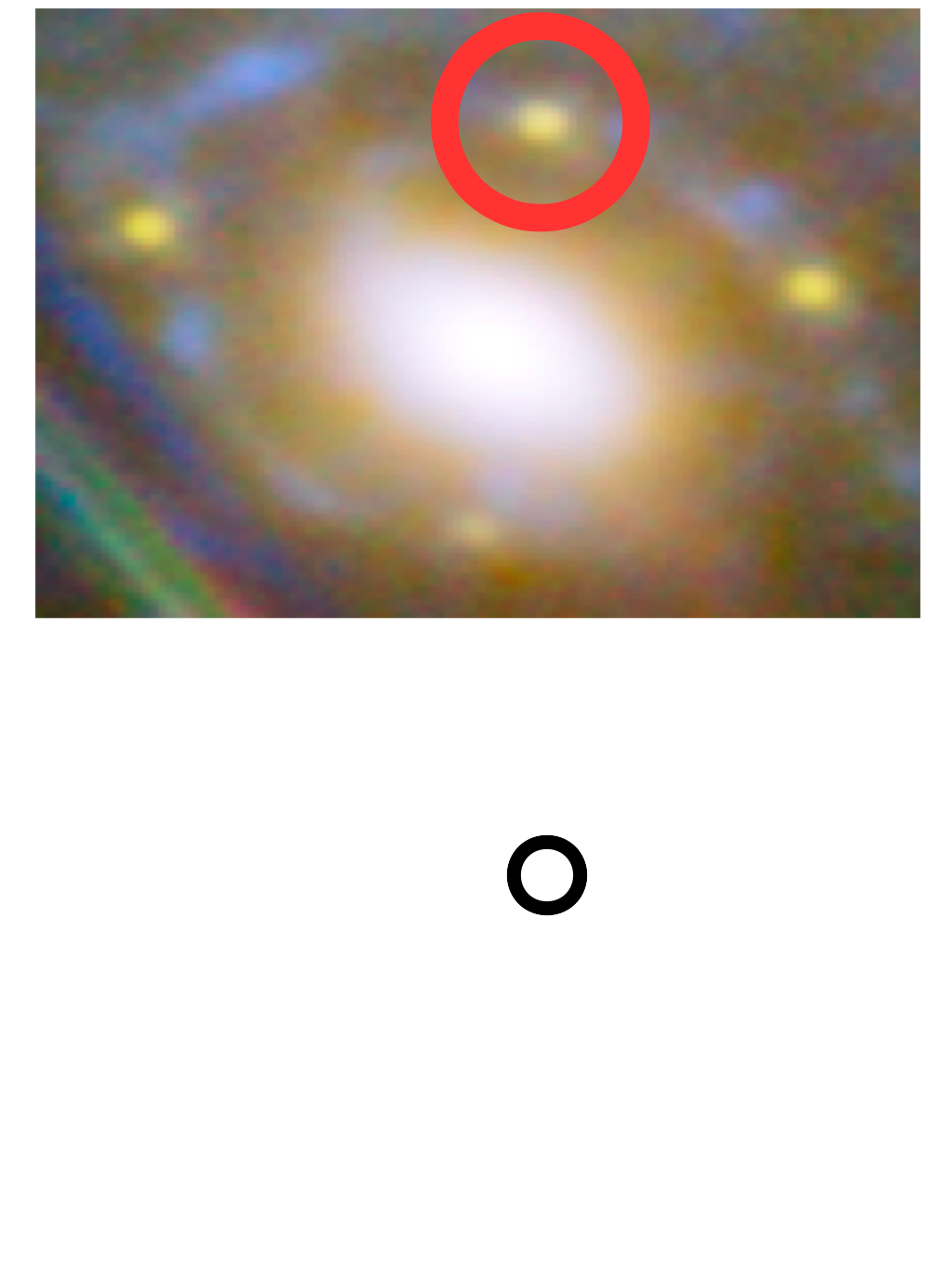} & \includegraphics[width=0.24\textwidth]{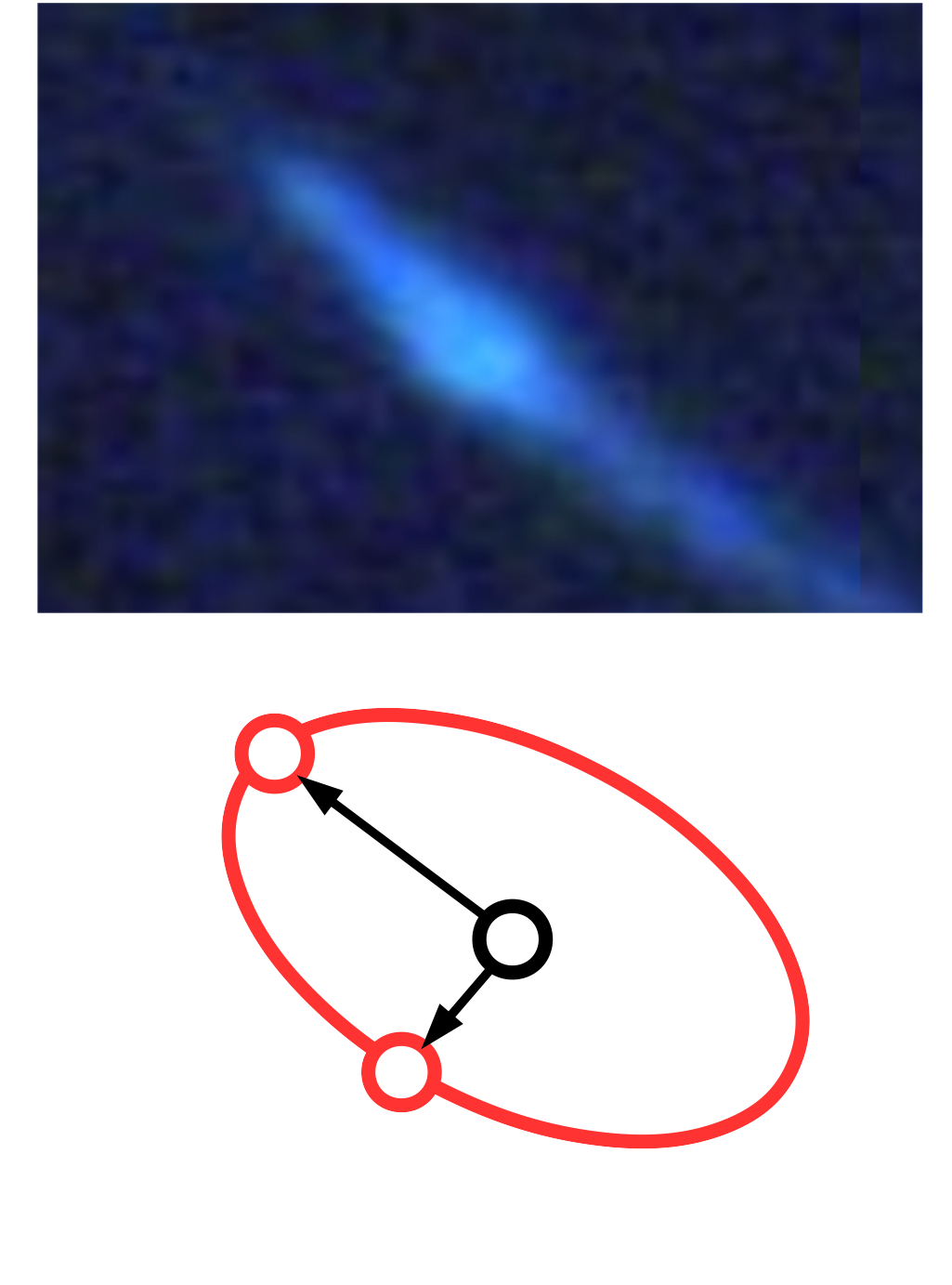}  & \includegraphics[width=0.24\textwidth]{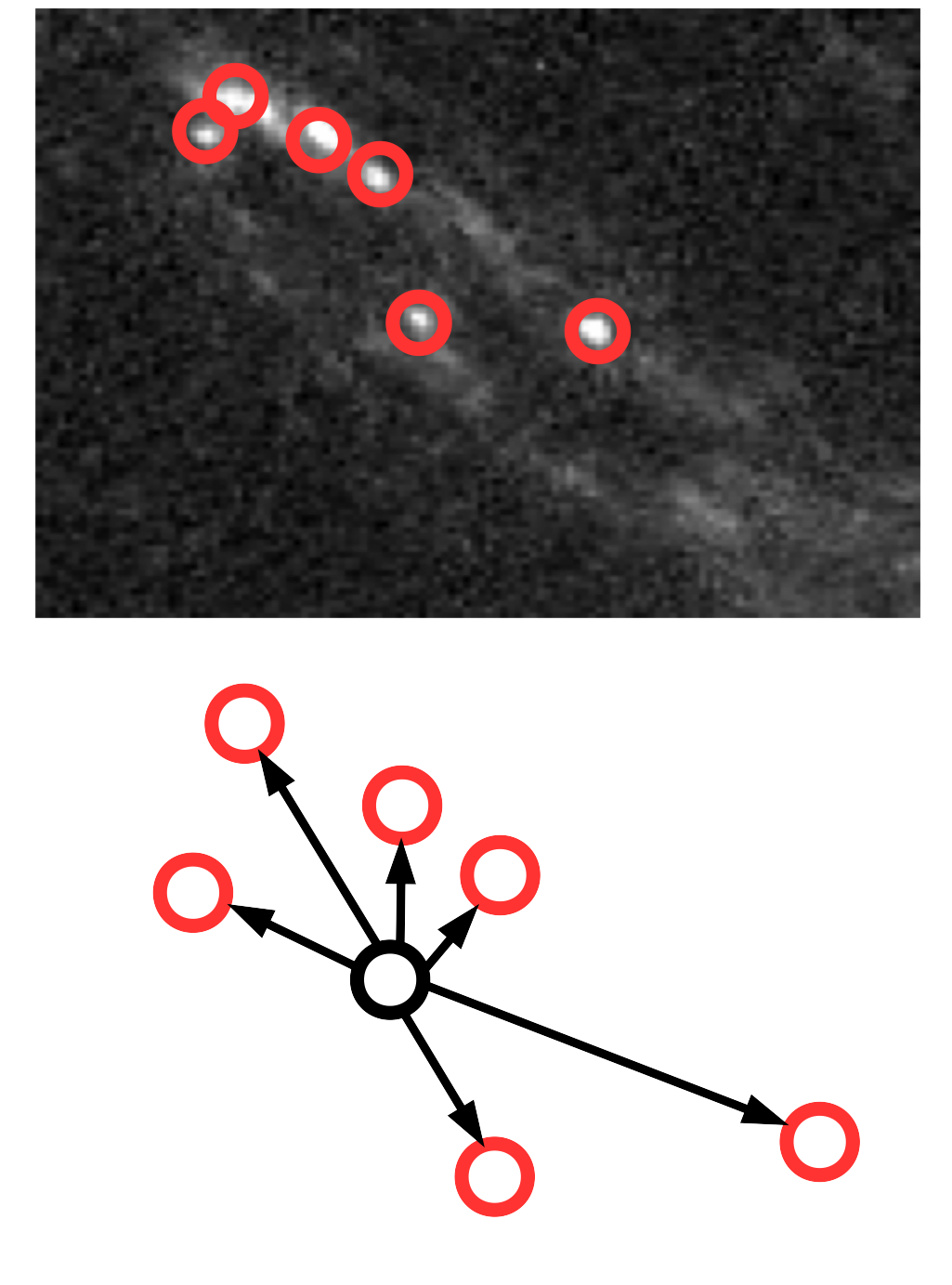}  \\
\midrule
angular position & centre of light, & vectors of distances \\
on the sky & quadrupole & between features \\
\bottomrule
\end{tabular}
\label{tab:image_types}
\end{table}

Any point source or object far below the resolution limit of the telescope will be mapped to multiple image points by the gravitational lensing effect with the angular position on the sky as the quantitative observable. 
Examples for such cases are multiple images of SNe, FRBs, or~most quasars (QSOs), i.e.,\@ objects of short duration or with a time-varying intensity, see Table~\ref{tab:image_types} (left).

We denote images above the resolution limit of the data acquisition as unresolved images if these extended images do not show intrinsic features in the brightness profile above noise level, as~shown in Table~\ref{tab:image_types} (centre).
The brightness profile of these images can be decomposed into multipoles. 
We~obtain the centre of light of the brightness profiles of the images and the quadrupole around the centre of brightness as observables. 
Higher-order moments may be extractable, as~we are currently investigating.
From the quadrupole, we obtain the vectors of the semi-minor and semi-major axes with respect to the centre of light. 
The ellipticity of an unresolved image is calculated from the axis ratio between the semi-minor and the semi-major axis.
The orientation of the image is defined as the direction of the semi-major axis in the coordinate system of the telescope observation. 
Decomposing the brightness profiles into other sets of basis functions that rely on specific profile models may lead to biases when inferring lens properties, as~found e.g.,\@ in~\cite{bib:Andrae,bib:Melchior} for weakly-distorted galaxy brightness profiles. 
Therefore, any set of basis functions that is based on non-parametric, statistical properties of the brightness profiles is preferred, see e.g.,\@ \cite{bib:Melchior2} for a moment-based decomposition of weakly-distorted galaxy brightness profiles, or~\cite{bib:Rahman,bib:Schroeder} for another suitable basis~set.

In the optimal case, the~source object has a brightness profile with intrinsic features and the lensing effect is so strong that these features can be identified in all images like in Table~\ref{tab:image_types} (right).
As quantitative observables in each image, we obtain the vectors of the angular distances between the brightness features.
Our approach requires at least two such non-parallel vectors per multiple image. 
This implies that any unresolved image can also be treated as a resolved image by means of the semi-major and semi-minor axes of the quadrupole.  
If the source is a galaxy, the~brightness features can be star-forming regions. 
Thus far, three cases of resolved multiple images of background galaxies caused by a galaxy cluster as lens have been observed: five multiple images in the galaxy cluster CL0024+17 as found by~\cite{bib:Tyson}, a~giant arc and its counter image in the galaxy cluster RCS2 032727-132623 as~discovered by~\cite{bib:Wuyts}, and~three images of a spiral galaxy in the galaxy cluster MACS J1149.5+2223 as first detailed in~\cite{bib:Smith}.
In the radio bands, some multiple images of QSOs show resolved features like core-jet structures, see e.g.,\@ \cite{bib:Gorenstein} or~\cite{bib:Biggs}.

Table~\ref{tab:image_configurations} shows the most common multiple-image configurations, i.e.,\@ the relative positions and orientations of multiple images with respect to each other. 
The configurations are shown for unresolved images.
However,  they are equally observed for point-like and resolved images. 
As we detail in Section~\ref{sec:formalism}, these configurations are derived from a leading order approximation to the general nonlinear lens mapping. 
In addition to these configurations, central images close to the lens centre can occur.
They remain unobserved in most cases because they are demagnified by the lens and are usually outshone or occluded by a bright galaxy-scale lens or by the brightest cluster galaxy in the centre of a galaxy-cluster-scale~lens.

\begin{table}[H]
\caption{Special multiple-image configurations of a common source. The~figure shows an observed example, the~schematic underneath the observables for unresolved images. Point-like and resolved images can also form these configurations. The~bottom rows list the quantitative observables for each case. For~arcs, we assume the centre of brightness of the lens is known. \textit{Image credits: NASA/ESA/HST.}}
\centering
\begin{tabular}{ccc}
\toprule
\textbf{Fold} & \textbf{Cusp} & \textbf{Giant Arcs}\\
\midrule
\includegraphics[width=0.24\textwidth]{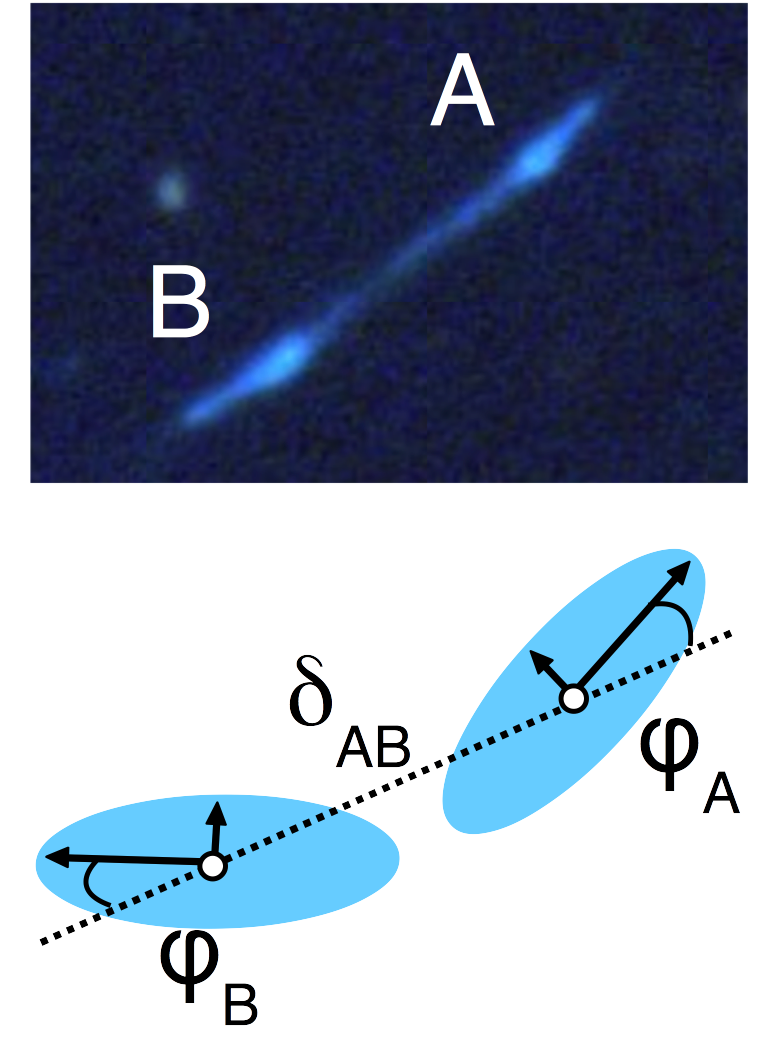}  & \includegraphics[width=0.24\textwidth]{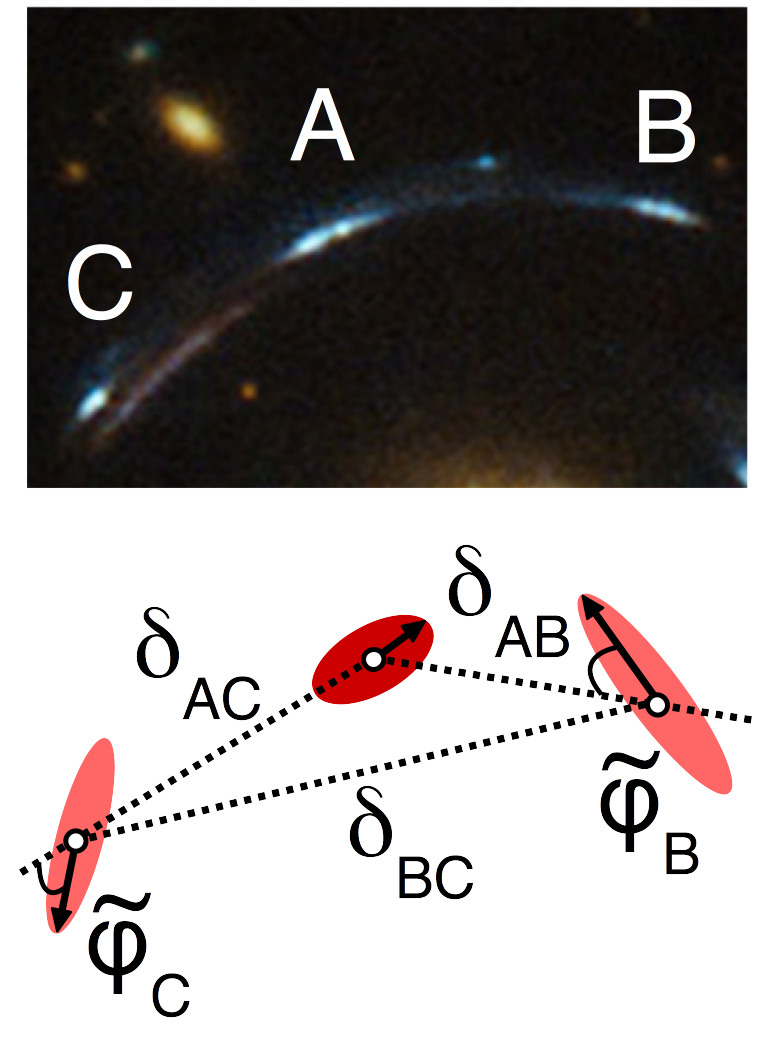}  &  \includegraphics[width=0.24\textwidth]{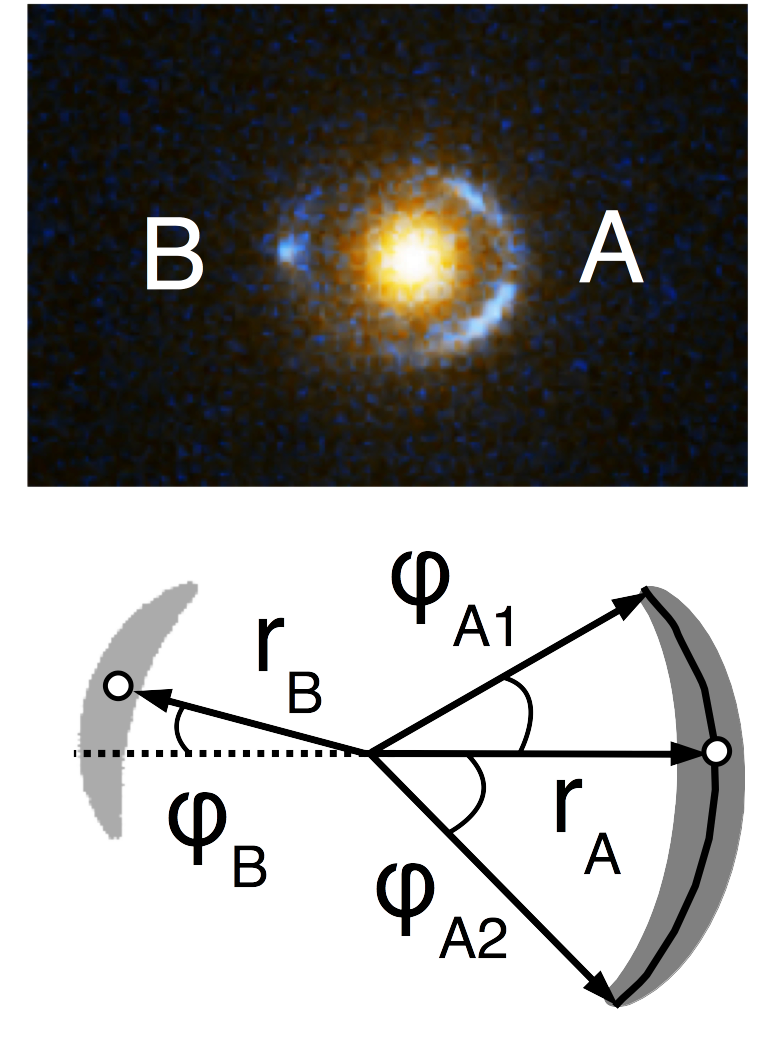}  \\
\midrule
rel.\@ distances $\delta_{ij}$  & rel.\@ distances $\delta_{ij}$ & radial distances $r_i$,  \\
rel.\@ orientations $\varphi_i$ & rel.\@ orientations $\tilde{\varphi}_i$ & arc lengths $r_i \varphi_{i}$ \\
(time delays) & (time delays) & (time delays) \\
\bottomrule
\end{tabular}
\label{tab:image_configurations}
\end{table}

The first case in Table~\ref{tab:image_configurations} (left) is a fold configuration of two images that are mirror images of each other.
This is often expressed as images $A$ and $B$ having opposite parity (handedness).
The~second case in Table~\ref{tab:image_configurations} (centre) is a cusp configuration of three multiple images arranged in a parabolic shape. 
To~leading order, this configuration can be decomposed into the superposition of two fold configurations formed by images $A$ and $B$ and $A$ and $C$, as~further detailed in Section~\ref{sec:cc}.
The last case in Table~\ref{tab:image_configurations} (right) occurs for (almost) axisymmetric lenses if the position of the background source falls close to the point in the source plane to which the symmetry centre of the lens is back-projected. 
For a perfect alignment of the source position and the symmetry centre of the lens along the observer's line of sight, the~source is mapped into an Einstein ring.
Small deviations from this alignment or from a perfect axisymmetric lens lead to two giant arcs, a~larger one and a smaller one.
The latter is called the counter-image.
These arcs surround the lens as shown in Table~\ref{tab:image_configurations} (right).
Sometimes, the~larger arc can be split further into a cusp configuration, if~the deviations from an axisymmetric mass-density distribution are so strong that the ellipticity of the deflecting mass profile becomes~relevant.

Measuring the redshift of the images and the lens, the~distances between the observer, the~source, and~the lens can be determined, employing the distance measures of the background metric. For~a spatially flat FLRW metric, the~angular diameter distances between two redshifts $z_1$ and $z_2$ with $z_2 > z_1$ are given as
\begin{equation}
D_\mathrm{A}(z_1, z_2) = \dfrac{c}{H_0} \dfrac{z_1+1}{z_2+1} \int \limits_{z_1}^{z_2} \dfrac{\mathrm{d}z}{E(z)} \;,
\label{eq:D_A}
\end{equation}
in which $E(z)$ is the expansion function of the universe. For~the standard cold-dark-matter model with a cosmological constant $\Lambda$ ($\Lambda$CDM) based on a spatially flat FLRW metric, it reads
\begin{equation}
E(z) = \sqrt{\Omega_r (1+z)^4 + \Omega_m (1+z)^3 + \Omega_\Lambda } \;.
\label{eq:E}
\end{equation}

The $\Omega_i$, $i=r,m,\Lambda$, denote the dimensionless contributions of radiation, matter, and~the cosmological constant to the total energy density of the universe, normalised such that $E(0) = 1$. The~term for the spatial curvature with $\Omega_K$ is discarded, as~\cite{bib:Planck} found that the curvature is compatible with zero.
Alternatively, any data-based expansion function can be inserted, as, for~instance set up for our approach in~\cite{bib:Wagner5}. 
Section~\ref{sec:distances} details these data-based distance measures.
The distances between the observer, the~source, and~the lens are only relevant in the time-delay equation, i.e.,\@ for cases involving images of time-varying intensity, as~detailed in Section~\ref{sec:cc}.

\subsection{The Standard Gravitational Lensing~Formalism}
\label{sec:formalism}

Based on the physical prerequisites of Section~\ref{sec:materials} and the detailed derivation in~\cite{bib:Wagner6}, we set up the standard gravitational lensing formalism for one lens plane as follows. 
First, the~light propagation along the line of sight is divided into the light propagation through the flat FLRW metric and the light deflection at the lens plane. 
Relative to the unperturbed light ray that propagates along a null-geodesic of the background metric, let the light be subject to a time delay $t_\mathrm{s}$ due to the deflection in the lens plane.
Due to this deflection, the~light also travels the additional time $t_\mathrm{g}$ in the flat FLRW metric relative to the unperturbed ray. 
Minimising the total travel time with respect to an unperturbed light ray, $t = t_\mathrm{g} + t_\mathrm{s}$, we obtain
\begin{equation}
t = \dfrac{(1+z_\mathrm{l})}{c} \dfrac{D_\mathrm{A}(0,z_\mathrm{s})D_\mathrm{A}(0,z_\mathrm{l})}{D_\mathrm{A}(z_\mathrm{l},z_\mathrm{s})} \left( \dfrac12 \boldsymbol{\alpha}(\boldsymbol{x})^2 - \psi(\boldsymbol{x}) \right) \;,
\label{eq:travel_time}
\end{equation}
for a light ray starting at the source redshift $z_\mathrm{s}$ and being deflected by a two-dimensional gravitational potential $\psi(\boldsymbol{x})$ at redshift $z_\mathrm{l}$, as~shown in Figure~\ref{fig:lensing_sketch}.
$\boldsymbol{\alpha}(\boldsymbol{x}) \in \mathbb{R}^2$ denotes the deflection angle caused by the gravitational potential in the lens plane.
It is given by the difference of the angular position of a multiple image in the lens plane, $\boldsymbol{x} = (x_1,x_2) \in \mathbb{R}^2$, and~the angular position of the source in the source plane, $\boldsymbol{y} = (y_1,y_2) \in \mathbb{R}^2$,
\begin{equation}
\boldsymbol{\alpha}(\boldsymbol{x}) = \boldsymbol{x} - \boldsymbol{y} \;.
\label{eq:alpha}
\end{equation}

In addition, the~deflection angle $\boldsymbol{\alpha}(\boldsymbol{x})$ is given as the gradient of the projected deflection potential $\psi(\boldsymbol{x})$ with respect to $\boldsymbol{x}$, $\boldsymbol{\alpha}(\boldsymbol{x}) = \nabla \psi(\boldsymbol{x})$ such that Equation~\eqref{eq:alpha} is often stated as
\begin{equation}
\boldsymbol{y} = \boldsymbol{x} - \nabla \psi(\boldsymbol{x}) \;.
\label{eq:lens_equation}
\end{equation}

For the sake of convenience, we abbreviate
\begin{equation}
\Gamma  \equiv \dfrac{(1+z_\mathrm{l})}{c} \dfrac{D_\mathrm{A}(0,z_\mathrm{s})D_\mathrm{A}(0,z_\mathrm{l})}{D_\mathrm{A}(z_\mathrm{l},z_\mathrm{s})} \;, \quad \phi(\boldsymbol{x},\psi) \equiv \dfrac12 \left( \boldsymbol{x}- \boldsymbol{y} \right)^2 - \psi(\boldsymbol{x}) = \dfrac12  \left( \boldsymbol{\alpha}(\boldsymbol{x}) \right)^2 - \psi(\boldsymbol{x})\;.
\label{eq:abbreviations}
\end{equation}
$\phi(\boldsymbol{x},\psi)$ is called the Fermat potential.
From the viewpoint of singularity theory, $\phi(\boldsymbol{x},\psi)$ is a function of $\boldsymbol{x}$ parametrised by $\psi(\boldsymbol{x})$.
Its critical points, $\nabla \phi(\boldsymbol{x},\psi)=0$, are the positions of the multiple images given by Equation~\eqref{eq:lens_equation}. 
The second derivative, the~Hessian matrix of $\phi(\boldsymbol{x}, \psi)$, also called distortion matrix,
\begin{equation}
A(\boldsymbol{x}) = \left( \begin{matrix} \phi_{11}(\boldsymbol{x}) & \phi_{12}(\boldsymbol{x}) \\ \phi_{12}(\boldsymbol{x}) & \phi_{22}(\boldsymbol{x}) \end{matrix} \right) \;, \quad \phi_{\alpha \beta}(\boldsymbol{x}) \equiv \dfrac{\partial^2 \phi(\boldsymbol{x},\psi)}{\partial x_\alpha \partial x_\beta} \;, \quad \alpha, \beta=1,2 \;,
\label{eq:A}
\end{equation}
is a two-dimensional mapping from vectors in the lens plane around $\boldsymbol{x}$ to vectors in the source plane around $\boldsymbol{y}$.
Hence, it describes the distortion of light bundles in the vicinity of $\boldsymbol{x}$ caused by the gravitational lensing effect.
$A(\boldsymbol{x})$ is often parametrised by the scaled, projected mass density called convergence $\kappa(\boldsymbol{x})$ and the shear $\boldsymbol{\gamma}(\boldsymbol{x}) = (\gamma_1(\boldsymbol{x}), \gamma_2(\boldsymbol{x}))$, such that $A(\boldsymbol{x})$ is decomposed into a magnifying trace part and a distorting symmetric part
\begin{equation}
A(\boldsymbol{x}) = \left(  \begin{matrix} 1-\kappa(\boldsymbol{x}) & 0 \\ 0 & 1 - \kappa(\boldsymbol{x}) \end{matrix} \right) - \left(  \begin{matrix} \gamma_1(\boldsymbol{x}) & \gamma_2(\boldsymbol{x}) \\ \gamma_2(\boldsymbol{x}) & -\gamma_1(\boldsymbol{x}) \end{matrix} \right) = (1-\kappa(\boldsymbol{x}) )\left(  \begin{matrix} 1-g_1(\boldsymbol{x}) & -g_2(\boldsymbol{x}) \\ -g_2(\boldsymbol{x}) & 1 + g_1(\boldsymbol{x}) \end{matrix} \right) \;.
\label{eq:A_kappa_gamma}
\end{equation}
$\kappa(\boldsymbol{x})$ is related to the deflection potential by the Poisson equation $\Delta \psi(\boldsymbol{x}) = 2 \kappa(\boldsymbol{x})$.
$\boldsymbol{\gamma}(\boldsymbol{x})$ can be physically interpreted as the deflection caused by parts of the mass density surrounding $\boldsymbol{x}$. 
The shear thus represents the non-local part of the deflection potential that is not included in the Poisson equation.
It enters $\psi(\boldsymbol{x})$ via the boundary conditions, as~detailed in~\cite{bib:Wagner6}. $\boldsymbol{g}(\boldsymbol{x})=(g_1(\boldsymbol{x}),g_2(\boldsymbol{x}))$ is the reduced shear, which is the quantity constrained by observables, as~detailed in Section~\ref{sec:images}. 
Given $A(\boldsymbol{x})$ either as parametrised by Equations~\eqref{eq:A} or \eqref{eq:A_kappa_gamma}, it is possible to reconstruct vectors in the source plane in the vicinity of $\boldsymbol{x}$ up to an overall scale factor, as~also detailed in Section~\ref{sec:images}.

The set of degenerate critical points $\boldsymbol{x}_0$ for which $\nabla \phi(\boldsymbol{x}_0,\psi) = 0$ and $\det(A(\boldsymbol{x}_0)) = 0$ are called the critical curves. Using Equations~\eqref{eq:A} and \eqref{eq:A_kappa_gamma}, they are given by all $\boldsymbol{x}_0$ fulfilling
\begin{equation}
\det (A(\boldsymbol{x}_0)) = \phi_{11}(\boldsymbol{x}_0) \, \phi_{22}(\boldsymbol{x}_0) - \phi_{12}(\boldsymbol{x}_0)^2 = \left( 1 - \kappa(\boldsymbol{x}_0) \right)^2 - \boldsymbol{\gamma}(\boldsymbol{x}_0)^2 = 0 \;.
\label{eq:cc}
\end{equation}

As shown in~\cite{bib:Petters}, the~critical curves described by Equation~\eqref{eq:cc} consist of only two kinds of critical points, so-called fold points, $\boldsymbol{x}_\mathrm{f}$, and~cusp points, $\boldsymbol{x}_\mathrm{c}$. 
Irrespective of the specific form of $\psi(\boldsymbol{x})$, $\boldsymbol{x}_\mathrm{f}$ and $\boldsymbol{x}_\mathrm{c}$ are determined by a Taylor expansion of the Fermat potential of order four around each critical~point. 

Thus, we have two possibilities to determine local lens properties without assuming a specific form of $\psi(\boldsymbol{x})$: 
first, we assume that $A(\boldsymbol{x})$ is constant in the vicinity of the centre of light of a multiple image.
This approximation is valid if the extensions of a multiple image are small compared to the scale on which the deflecting mass density changes. 
A comprehensive synopsis of the validity of these approximations for extended sources and finite beam sizes in gravitational lensing can be found in~\cite{bib:Fleury}. 
In \cite{bib:Tessore} the most general equations to constrain local lens properties from Equation~\eqref{eq:lens_equation} were derived. 
In~\cite{bib:Wagner2}, we~developed this idea further by assuming a constant $A(\boldsymbol{x})$ in the area spanned by the quadrupole of an unresolved multiple image or in the area spanned by the convex hull of all identifiable reference points of a resolved multiple image. 
Section~\ref{sec:images} details the ansatz and shows the local lens properties that are determined in the vicinity of the multiple images. 
They are calculated using the quadrupoles of at least two unresolved images in a fold configuration. 
Alternatively, the~ansatz leads to the same local lens properties for at least three identifiable reference points in at least two images in a fold~configuration.

Second, we perform the Taylor expansion to fourth order around a critical point and employ the observables of the multiple images to infer local properties of the critical curves, as~detailed in Section~\ref{sec:cc}.  
Consequently, local lens properties in the vicinity of the multiple images and approximations to the critical curves can be determined independently of the specific global form of $\psi(\boldsymbol{x})$, i.e.,\@ without any prior assumptions about the decomposition of the deflecting gravitational~potential.


\subsection{Local Lens Properties from a Taylor Expansion around the Centre of Light of the Multiple~Images}
\label{sec:images}
The work in Ref.~\cite{bib:Gorenstein} already described the possibility to extract the local lens properties contained in $A(\boldsymbol{x})$ by transforming multiple images onto each other. 
In~\cite{bib:Wagner2,bib:Wagner_cluster}, we realised this idea for unresolved and resolved multiple images that contain at least two non-parallel vectors, as~indicated by the black arrows in the figures of Table~\ref{tab:image_types}.
We assume to have $m$ reference points within each multiple image $i$ which are located at positions $\boldsymbol{x}_{i\alpha}$, $\alpha = 1,...,m$.
Using the quadrupole of an unresolved image as observables, the~three reference points are the centre of light and the end points of the semi-major and semi-minor axis of the quadrupole. 
Furthermore, we assume that the multiple images have a very small extent compared to the scale on which the mass density changes.
Sections~\ref{sec:app_cluster} and \ref{sec:app_quasar} show the validity of this assumption in two practical applications on cluster and on galaxy scale. 
This implies that the entries in $A(\boldsymbol{x})$ are approximately constant over the area spanned by the $\boldsymbol{x}_{i\alpha}$ in each multiple image.
Since the vectors in all multiple images $\boldsymbol{x}_{i\alpha} - \boldsymbol{x}_{i\beta}$, $\alpha, \beta  = 1,...,m$, originate from the same, corresponding vectors $\boldsymbol{y}_\alpha - \boldsymbol{y}_\beta$ in the source plane, we note that for each of these vectors
\begin{equation}
\boldsymbol{y}_\alpha - \boldsymbol{y}_\beta = A(\boldsymbol{x}_j) \left( \boldsymbol{x}_{j\alpha} - \boldsymbol{x}_{j\beta} \right) = A(\boldsymbol{x}_i) \left( \boldsymbol{x}_{i\alpha} - \boldsymbol{x}_{i\beta} \right) \;, \forall i,j=A,B,C,... \, \; \forall \alpha,\beta = 1,2,...,m
\label{eq:transformations}
\end{equation}
holds.
As the distortion matrix is assumed to be constant over the area spanned by the reference points, $\boldsymbol{x}_i$ and $\boldsymbol{x}_j$ represent an arbitrary point within this area. 
Reformulating the part on the right-hand side of Equation~\eqref{eq:transformations} yields a linear transformation $T_{ij}$ between vectors in image $j$ to vectors in image $i$ as
\begin{equation}
 \boldsymbol{x}_{j\alpha} - \boldsymbol{x}_{j\beta} = A^{-1}(\boldsymbol{x}_j) A(\boldsymbol{x}_i) \left( \boldsymbol{x}_{i\alpha} - \boldsymbol{x}_{i\beta} \right) \equiv T_{ij} \left( \boldsymbol{x}_{i\alpha} - \boldsymbol{x}_{i\beta} \right) \;, \forall i,j=A,B,C,... \, \; \forall \alpha,\beta = 1,2,...,m\;.
 \label{eq:transformations2}
\end{equation}

Using Equation~\eqref{eq:transformations2} to solve for the entries in $A(\boldsymbol{x})$ given the reference points $\boldsymbol{x}_{i\alpha}$ is a total-least-squares parameter estimation problem. 
It differs from the standard least-squares parameter estimation because the vectors between the reference points on both sides of the equation are subject to measurement uncertainties. 
To account for this, we introduce one latent variable $\boldsymbol{x}_i$ per multiple image, the~so-called anchor points in~\cite{bib:Wagner2,bib:Wagner_cluster}.

Solving the total-least-squares parameter estimation yields ratios of convergences, $f_{ij}$, between~pairs of multiple images $i$ and $j$ and the reduced shear $g(\boldsymbol{x})$ at the positions of the multiple~images
\begin{equation}
f^{(\kappa)}_{ij} \equiv \dfrac{1-\kappa(\boldsymbol{x}_i)}{1-\kappa(\boldsymbol{x}_j)}\;, \quad g_i \equiv g(\boldsymbol{x}_i) = \dfrac{\boldsymbol{\gamma}(\boldsymbol{x}_i)}{1-\kappa(\boldsymbol{x}_i)} \quad \forall i,j=A,B,C,... \;.
\label{eq:kappa_fgs}
\end{equation}

It also yields the positions of the anchor points $\boldsymbol{x}_{i}$.
Alternatively, expressed in derivatives of the Fermat potential, we can infer the analogous quantities
\begin{equation}
f^{(\phi)}_{ij} = \dfrac{\phi_{11}(\boldsymbol{x}_i)}{\phi_{11}(\boldsymbol{x}_j)}  \;, \quad \tilde{\phi}_{12}(\boldsymbol{x}_i) = \dfrac{\phi_{12}(\boldsymbol{x}_i)}{\phi_{11}(\boldsymbol{x}_i)} \;, \quad \tilde{\phi}_{22}(\boldsymbol{x}_i) = \dfrac{\phi_{22}(\boldsymbol{x}_i)}{\phi_{11}(\boldsymbol{x}_i)} \quad \forall i,j=A,B,C,...\;.
\label{eq:phi_fgs}
\end{equation}

\begin{figure}[H]
\centering
\includegraphics[width=0.6\textwidth]{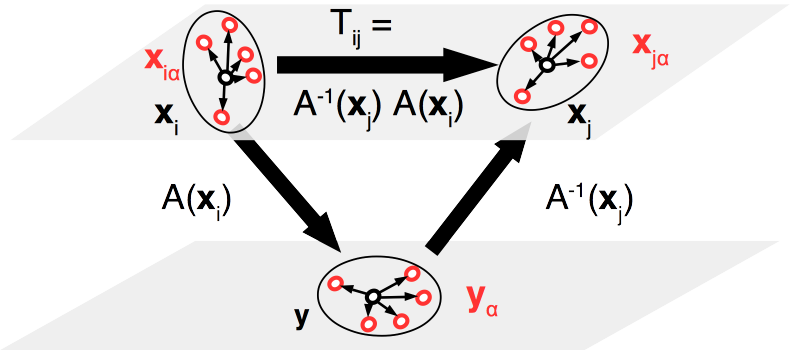}
\caption{A linear transformation $T_{ij}$ between resolved or unresolved multiple images located around $\boldsymbol{x}_i$ and $\boldsymbol{x}_j$ (black circles) in the image plane of the same source located around $\boldsymbol{y}$ (black circle) in the source plane is equal to a product of distortion matrices, $A(\boldsymbol{x})$. Thus, without~employing any source properties, the~local lens properties represented by the matrix entries of $A(\boldsymbol{x})$ are inferred from the $T_{ij}$.}
\label{fig:images_transformation}
\end{figure}

In~\cite{bib:Wagner2}, we performed a comparison between the different parametrisations of $A(\boldsymbol{x})$ by Equations~\eqref{eq:kappa_fgs} and \eqref{eq:phi_fgs} on simulated data. We found that both parametrisations have their advantages and drawbacks, so that it depends on the image configuration, which one yields more accurate, precise, and~robust results.
Furthermore, we showed that the larger the area of each multiple image from which the observables are extracted, the~smaller the confidence bounds on the inferred local lens properties. On~the other hand, the~area should be so small that $A(\boldsymbol{x})$ is still approximately constant for all $\boldsymbol{x}$ within the area. 
In any case, the~approach needs at least three non-collinear reference points $\boldsymbol{x}_{i\alpha}$, $\alpha \ge 3$, in~each multiple image $i$ and at least three images to return a unique solution for the local lens properties as given by Equations~\eqref{eq:kappa_fgs} or \eqref{eq:phi_fgs}.
The system of equations is under-determined, if~$T_{ij}$ is diagonal, so~that we cannot determine local lens properties from aligned multiple-image configurations created by an axisymmetric lens (see~\cite{bib:Wagner2} for a detailed explanation).   
In addition, $\kappa(\boldsymbol{x}_j) = 1$ should be avoided in the denominator of $f_{ij}$ in Equation~\eqref{eq:kappa_fgs}, such that $f_{ij}$ does not become singular. 
Apart from these restrictions, it handles any configuration of at least three images including central maxima or counter images of cusp configurations. 
Figure~\ref{fig:images_transformation} visualises the principle to obtain Equations~\eqref{eq:kappa_fgs} and \eqref{eq:phi_fgs}.
An implementation to solve for Equation~\eqref{eq:kappa_fgs} is publicly available at \url{https://github.com/ntessore/imagemap} and details
about it can be found in~\cite{bib:Wagner_cluster}. 

As mentioned in Section~\ref{sec:formalism}, Equation~\eqref{eq:transformations} shows that vectors in the source plane, $\boldsymbol{y}_\alpha - \boldsymbol{y}_\beta$ can be determined from the back-projection of the respective vectors in the image plane, $\boldsymbol{x}_{i\alpha} - \boldsymbol{x}_{i\beta}$, using $A(\boldsymbol{x}_i)$ as constrained by the local lens properties in Equations~\eqref{eq:kappa_fgs} or \eqref{eq:phi_fgs}. Comparing \mbox{Equations~\eqref{eq:kappa_fgs} and \eqref{eq:phi_fgs}} with Equations~\eqref{eq:A} and \eqref{eq:A_kappa_gamma}, we find that $\boldsymbol{y}_\alpha - \boldsymbol{y}_\beta$ are only determined up to an overall scale factor because the observables do not constrain $\kappa(\boldsymbol{x})$ but only ratios of convergences, $f_{ij}$.
An example source reconstruction using this method is performed in Section~\ref{sec:app_cluster} and clearly shows that the reconstructed morphology of the source can give useful information about the relative shapes and sizes of the star-forming regions of a multiply-imaged galaxy.
A more detailed analysis and comparison to lens-model-based source reconstructions is currently in~preparation.

From the local lens properties given by Equations~\eqref{eq:kappa_fgs} and \eqref{eq:phi_fgs}, we calculate the relative image parity information between the multiple images $i$ and $j$,
\begin{equation}
\mathcal{J}_{ij} \equiv \det{T_{ij}} = \det(A(\boldsymbol{x}_j))^{-1} \det(A(\boldsymbol{x}_i)) \quad \forall i,j=A,B,C,...\;.
\end{equation}

For resolved multiple images, the~relative parity can be read off the telescope observation by visual inspection as a consistency check. 
In addition, $\mathcal{J}_{ij}$ is also the magnification ratio between the two images $i$ and $j$.
Hence, it can be compared to the measured flux density ratios, if~available, to~indicate extinction due to dust, micro-lensing, or~higher order lensing effects beyond convergence and~shear. 

\unskip   

\subsubsection{The Fold~Case}
\label{sec:images_fold}

In order to uniquely solve Equation~\eqref{eq:transformations2} for a fold configuration, we have to reduce the amount of variables in Equations~\eqref{eq:kappa_fgs} and \eqref{eq:phi_fgs}.
Therefore, we approximate $f_{ij} \approx 1$ in Equations~\eqref{eq:kappa_fgs} and \eqref{eq:phi_fgs}, assuming that the convergence or the $\phi_{11}(\boldsymbol{x})$ are equal for the areas covered by the two fold images. 
Thus, only the reduced shear can be determined for the two images at a~fold.

The validity of the approximation $f_{ij} \approx 1$ was investigated in~\cite{bib:Wagner2} for two analytic mass density models (the ones discussed in Appendix~\ref{appendix:lens_model_examples}, for~choices of parameters, see~\cite{bib:Wagner2}).
We found that the decrease in accuracy of the approximation $f_{ij} \approx 1$ depends on the underlying mass density. Even~for our simulated extreme case of a highly elliptical lens, at~least one of the parameterisations in Equations~\eqref{eq:kappa_fgs} and \eqref{eq:phi_fgs} yielded local lens properties that were within the 2$-\sigma$ confidence intervals around the true~values.

\subsection{Local Lens Properties from a Taylor Expansion around a Critical~Point}
\label{sec:cc}

Next, we Taylor-expand the Fermat potential around a critical point $\boldsymbol{x}_0$, whose source is located at the so-called caustic point $\boldsymbol{y}_0$. 
Without loss of generality, we choose the coordinate systems in the lens and the source plane such that the critical and the caustic point are at the origin, $\boldsymbol{x}_0 = \boldsymbol{y}_0 = \boldsymbol{0}$. 
Inserting Equation~\eqref{eq:lens_equation} into the Fermat potential, the~Taylor series to order four reads
\begin{align}
\phi_\mathrm{T}(\boldsymbol{x},\psi) \equiv& \; \phi(\boldsymbol{x}_0,\psi) + \tfrac12 \boldsymbol{y}^2 - \boldsymbol{x}\boldsymbol{y} + \tfrac12 \phi_{11}(\boldsymbol{x}_0) x_1^2 + \phi_{12}(\boldsymbol{x}_0) x_1 x_2 + \tfrac12 \phi_{22}(\boldsymbol{x}_0) x_2^2 \label{eq:taylor_series} \\
  &+ \tfrac16 \phi_{111}(\boldsymbol{x}_0) x_1^3 + \tfrac12 \phi_{112}(\boldsymbol{x}_0) x_1^2x_2 +  \tfrac12 \phi_{122}(\boldsymbol{x}_0) x_1x_2^2 +  \tfrac16 \phi_{222}(\boldsymbol{x}_0) x_2^3  \nonumber \\
  &+ \tfrac{1}{24} \phi_{1111}(\boldsymbol{x}_0) x_1^4  + \tfrac16 \phi_{1112}(\boldsymbol{x}_0) x_1^3x_2 +  \tfrac14 \phi_{1122}(\boldsymbol{x}_0) x_1^2x_2^2  +  \tfrac16 \phi_{1222}(\boldsymbol{x}_0) x_1x_2^3 + \tfrac{1}{24} \phi_{2222}(\boldsymbol{x}_0) x_2^4 \nonumber \;.
\end{align} 

As $\boldsymbol{x}_0$ is a critical point, Equation~\eqref{eq:cc} must be fulfilled for the second-order derivatives in Equation~\eqref{eq:taylor_series}.
Using Equation~\eqref{eq:taylor_series} as the local Fermat potential, the~time-delay difference between light rays arriving from two multiple images $i$ and $j$ that are located close to $\boldsymbol{x}_0$ is given by
\begin{align}
\tau_{ij} = t_i - t_j = \Gamma \phi_\mathrm{T}(\boldsymbol{x}_i,\psi) - \Gamma \phi_\mathrm{T}(\boldsymbol{x}_j\, \psi) \;.
\label{eq:tdd}
\end{align}
\begin{equation}
\nabla \phi_\mathrm{T}(\boldsymbol{x},\psi)=0
\label{eq:taylor_le}
\end{equation}
yields the lensing equation to determine the positions of multiple images in the vicinity of $\boldsymbol{x}_0$. 
From Equations~\eqref{eq:tdd} and \eqref{eq:taylor_le}, a~system of equations is set up that contains $\boldsymbol{y}$ and derivatives of the Fermat potential at $\boldsymbol{x}_0$ as unknowns. 
The observables, introduced in Section~\ref{sec:observables}, are employed to determine the angular positions $\boldsymbol{x}$ of the multiple images with respect to $\boldsymbol{x}_0$, as~detailed in the following~sections.

Depending on the type of critical point, $\boldsymbol{x}_\mathrm{f}$ or $\boldsymbol{x}_\mathrm{c}$, different further approximations to Equation~\eqref{eq:taylor_series} are necessary to reduce the number of unknowns, such that the system can be uniquely solved. 
Table~\ref{tab:taylor_configs} shows these approximations to Equation~\eqref{eq:taylor_le} to obtain the lensing equations for \mbox{Sections~\ref{sec:cc_fold} and \ref{sec:cc_cusp}}.
The approximations are inspired by axisymmetric and elliptical lens models (see Appendix~\ref{appendix:lens_model_examples} for two examples) and can be motivated by the fact that Newtonian gravity (as also employed for the description of the gravitational lens), is a central force. The~goodness of the approximating assumptions will increase with decreasing distance of the multiple images to the centre of the potential of the deflecting galaxy or galaxy cluster and with increasing depth of this potential with respect to neighbouring gravitationally interacting~objects.

\begin{table}[H]
\caption{Synopsis of approximations to Equation~\eqref{eq:taylor_series} and embedding of the observables for fold and cusp critical points, $\boldsymbol{x}_\mathrm{f}$ and $\boldsymbol{x}_\mathrm{c}$, in~the coordinate system of Equation~\eqref{eq:coordinates}. The~critical point is marked by a black dot in the figures. Degeneracies arise in the sign of the slope of the critical curve at $\boldsymbol{x}_\mathrm{f}$. Depending on the parity of image $A$, there are two possible configurations for a~cusp.}
\centering
\begin{tabular}{cc}
\toprule
\textbf{Fold} & \textbf{Cusp} \\
\midrule
 & $\qquad \qquad$ \textbf{(Positive)} \hfill \textbf{(Negative)} $\qquad \qquad$ \\
\midrule
\includegraphics[width=0.29\textwidth]{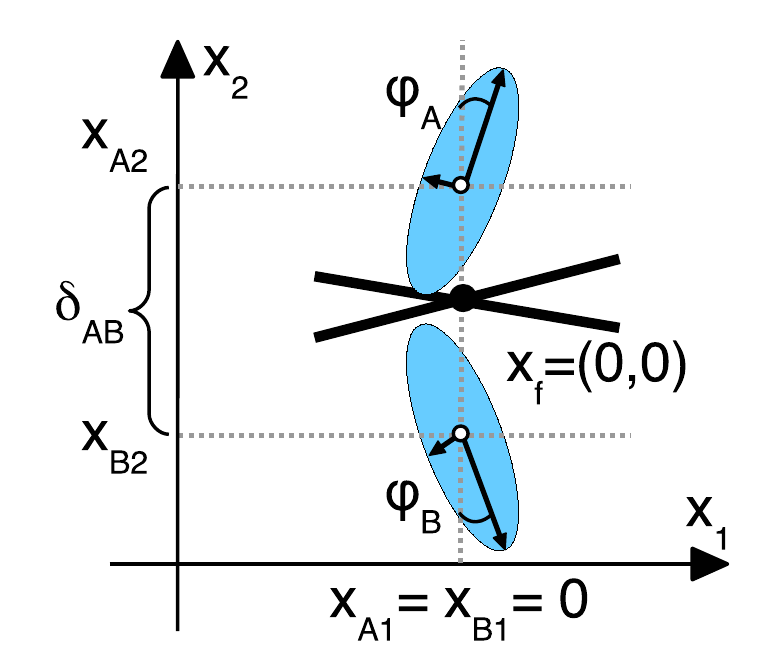} &  \includegraphics[width=0.29\textwidth]{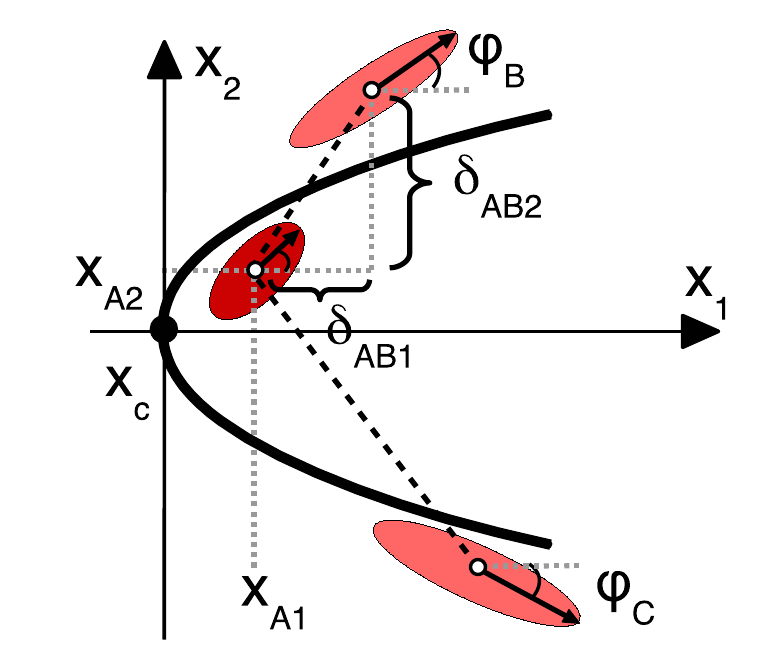}   \includegraphics[width=0.29\textwidth]{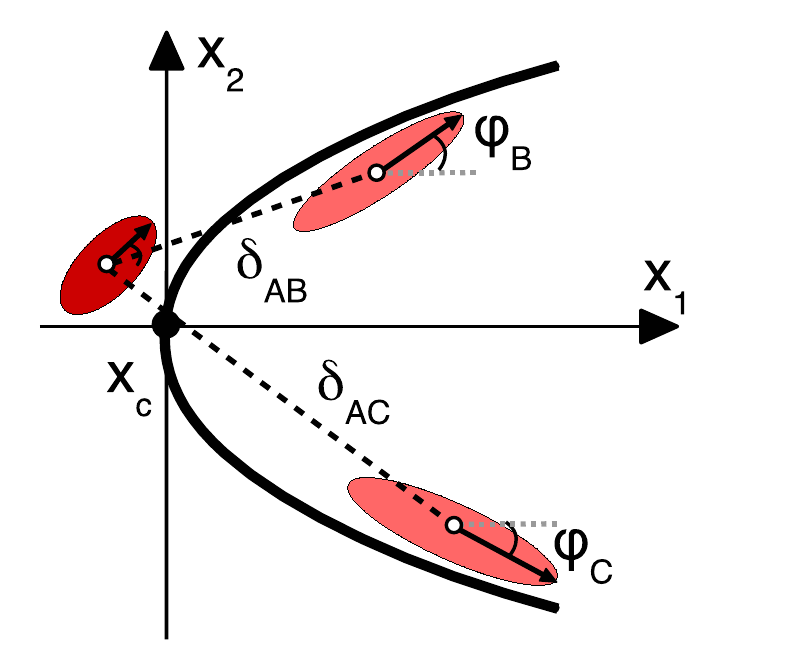} \\
\midrule
coordinate system specifications & coordinate system specifications \\[0.1cm]
$\phi_{222}(\boldsymbol{x}_0) > 0$, & $\phi_{122}(\boldsymbol{x}_0) < 0$, $\phi_{2222}(\boldsymbol{x}_0) > 0$, \\[0.1cm]
 & $\phi_{122}(\boldsymbol{x}_0)^2 - \tfrac13 \phi_{11}(\boldsymbol{x}_0) \phi_{2222}(\boldsymbol{x}_0) \ne 0$ \\ 
\midrule
leading order $\phi_\mathrm{T}(\boldsymbol{x},\psi) =$ & leading order $\phi_\mathrm{T}(\boldsymbol{x},\psi) =$\\[0.1cm]
$- \boldsymbol{x}\boldsymbol{y} + \tfrac12 \phi_{11}(\boldsymbol{x}_0) x_1^2 + \tfrac12 \phi_{112}(\boldsymbol{x}_0) x_1^2x_2$ 
&  
$- \boldsymbol{x}\boldsymbol{y} + \tfrac12 \phi_{11}(\boldsymbol{x}_0) x_1^2$ \\[0.1cm]
$ +  \tfrac12 \phi_{122}(\boldsymbol{x}_0) x_1x_2^2 +  \tfrac16 \phi_{222}(\boldsymbol{x}_0) x_2^3$ 
& 
$ +  \tfrac12 \phi_{122}(\boldsymbol{x}_0) x_1x_2^2 +  \tfrac{1}{24} \phi_{2222}(\boldsymbol{x}_0) x_2^4$
\\
\bottomrule
\end{tabular}
\label{tab:taylor_configs}
\end{table}

Diagonalising $A(\boldsymbol{x}_0)$ introduces a special coordinate system, which we choose, as~in~\cite{bib:SEF},
\begin{equation}
\phi_1(\boldsymbol{x}_0)  = \phi_2(\boldsymbol{x}_0) = \phi_{12}(\boldsymbol{x}_0) = \phi_{22}(\boldsymbol{x}_0) = 0\;, \quad \phi_{11}(\boldsymbol{x}_0) \ne 0 \;
\label{eq:coordinates}
\end{equation} 
without loss of generality.
The transformation into this coordinate frame simplifies the equations, but~also complicates the combination of several multiple-image configurations in a mutual telescope observation because each multiple-image system has its own frame of~reference. 

Table~\ref{tab:taylor_configs} summarises the specifications that we employ to embed the observables of Section~\ref{sec:observables} into the coordinate system given by Equation~\eqref{eq:coordinates}. 

Similarly to Section~\ref{sec:images}, the~observables allow for determining the distortion matrices at the image positions, $A(\boldsymbol{x}_i)$, $i=A, B, C$, as~described in Sections~\ref{sec:cc_fold} and \ref{sec:cc_cusp}. Here, $\boldsymbol{x}_i$ is the centre of light of a multiple image. Again, these distortion matrices can be employed to reconstruct vectors in the source plane from corresponding vectors in the image plane close to $\boldsymbol{x}_i$ up to an overall scale~factor.

\subsubsection{The Fold~Case}
\label{sec:cc_fold}

The work in Ref.~\cite{bib:SEF}
derived that, to~leading order, the~fold configuration consists of two images that are symmetrically located around $x_\mathrm{f}$ as shown in Table~\ref{tab:taylor_configs} (left figure). 
Employing $\phi_\mathrm{T}(\boldsymbol{x},\psi)$ in the coordinate system as introduced in Table~\ref{tab:taylor_configs}, we derive the time delay difference between light rays arriving from the two images by Equation~\eqref{eq:tdd} and the lensing equation by Equation~\eqref{eq:taylor_le}. The~distortion matrix as the second derivative of $\phi_\mathrm{T}(\boldsymbol{x},\psi)$ with respect to $\boldsymbol{x}$ reads
\begin{equation}
A(\boldsymbol{x}_i)=  \phi_{11}(\boldsymbol{x}_\mathrm{f}) \left( \begin{matrix} 1 &  \tfrac{\phi_{122}(\boldsymbol{x}_\mathrm{f})}{\phi_{11}(\boldsymbol{x}_\mathrm{f})} x_{i2} \\  \tfrac{\phi_{122}(\boldsymbol{x}_\mathrm{f})}{ \phi_{11}(\boldsymbol{x}_\mathrm{f})} x_{i2}  & \tfrac{\phi_{222}(\boldsymbol{x}_\mathrm{f})}{ \phi_{11}(\boldsymbol{x}_\mathrm{f})} x_{i2} \end{matrix}\right) \;, \quad i=A,B \;.
\label{eq:fold_A}
\end{equation}

Hence, we make the approximation that $\phi_{11}(\boldsymbol{x}_i) \approx \phi_{11}(\boldsymbol{x}_\mathrm{f})$, $i=A,B$ and interpolate $A(\boldsymbol{x})$ linearly between $\boldsymbol{x}_\mathrm{f}$ and $\boldsymbol{x}_i$.
We factor $A(\boldsymbol{x}_i)$ into an overall scale, $\phi_{11}(\boldsymbol{x}_\mathrm{f})$ and a reduced distortion matrix, analogous to Equation~\eqref{eq:A_kappa_gamma}.
Comparing Equation~\eqref{eq:fold_A} to Equation~\eqref{eq:phi_fgs}, we arrive again at the approximation made in Section~\ref{sec:images_fold}. 
In addition, the~mirror symmetry implies that the images are of different parity, which can easily be verified by inserting $x_{i2}$ into Equation~\eqref{eq:fold_A}.

Due to the symmetry, $x_{A2} = \delta_{AB}/2$ and $x_{B2} = - \delta_{AB}/2$, in~which the distance $\delta_{AB}$ between $\boldsymbol{x}_A$ and $\boldsymbol{x}_B$ is an observable. 
Further observables are the quadrupole of each image around the centre of light, $\boldsymbol{x}_i$, parametrised by the axis ratio $r_{i}$ of the semi-minor to the semi-major axis and the orientation angle of the semi-major axis to $\delta_{AB}$, $\varphi_i$, $i=A, B$. 
The validity of the symmetry assumption can thus be checked by comparing the observables of the two quadrupoles with each other. 
A measured flux ratio close to one between images $A$ and $B$ additionally corroborates the symmetry assumption, if~dust extinction and micro-lensing are negligible effects (see also Section~\ref{sec:images}).
If the symmetry assumption is applicable, the~images are close enough to the critical curve to assume that the source properties are negligible. 
In the appendix of~\cite{bib:Wagner1}, we generally showed that the influence of the intrinsic quadrupole of the source decreases for decreasing distance to the critical point. 
Consequently, we assume that the observed quadrupole of the multiple images is caused by the gravitational lensing effect.
We set it equal to the reduced distortion matrix in Equation~\eqref{eq:fold_A}. 

Altogether, we can determine the following leading-order local lens properties from a fold configuration of resolved or unresolved multiple images with a measured time delay difference of $\tau_{AB}$
\begin{align}
\phi_{222}(\boldsymbol{x}_\mathrm{f}) &= \dfrac{12  \, \tau_{AB}}{\Gamma \delta_{AB}^3} \;, \label{eq:fold1}\\
\left| \tilde{\phi}_{122}(\boldsymbol{x}_\mathrm{f}) \right| &\equiv \left| \dfrac{\phi_{122}(\boldsymbol{x}_\mathrm{f})}{\phi_{11}(\boldsymbol{x}_\mathrm{f})} \right| = \dfrac{2 \left| \tilde{\phi}_{12}(\boldsymbol{x}_i) \right|}{\delta_{AB}} = \dfrac{2}{\delta_{AB}} \cdot \left| \dfrac{\tan(\varphi_i)(r_i -1)}{r_i \tan^2(\varphi_i) + 1}\right| \;, \quad i=A, B \;, \quad \varphi_i \in \left] -\tfrac{\pi}{2}, \tfrac{\pi}{2}\right[ \label{eq:fold2} \;,\\
\tilde{\phi}_{222}(\boldsymbol{x}_\mathrm{f}) &\equiv \dfrac{\phi_{222}(\boldsymbol{x}_\mathrm{f})}{\phi_{11}(\boldsymbol{x}_\mathrm{f})} = \dfrac{2 \, \tilde{\phi}_{22}(\boldsymbol{x}_i)}{\delta_{AB}}  = \dfrac{2}{\delta_{AB}} \cdot \dfrac{r_i + \tan^2(\varphi_i)}{r_i \tan^2 (\varphi_i) + 1} \approx  \dfrac{2 \, r_i}{\delta_{AB}}\;, \label{eq:fold3} \\
m_\mathrm{f} &\equiv \dfrac{\left| \phi_{122}(\boldsymbol{x}_\mathrm{f})  \right|}{\phi_{222}(\boldsymbol{x}_\mathrm{f})} \;, \label{eq:fold4}
\end{align}
in which the slope of the critical curve at $\boldsymbol{x}_\mathrm{f}$, $m_\mathrm{f}$, is determined by the ratio of Equations~\eqref{eq:fold2} and \eqref{eq:fold3}. 
The sign of the slope cannot be determined.
This requires a third image, as~discussed in Section~\ref{sec:cc_cusp}. 
In~\cite{bib:Wagner1}, we found that the approximation assuming $\varphi_i = 0$ on the right-hand side of Equation~\eqref{eq:fold3} yields the most accurate results for this ratio of~derivatives.

We note that Equations~\eqref{eq:fold2}--\eqref{eq:fold4} do not contain any information about the angular diameter distances to the lens or the source plane. 
As derived in~\cite{bib:Wagner4} and explained in~\cite{bib:Wagner6}, the~former fact arises because the Fermat potential is an angular, projected potential, and~the multiple-image shapes are the corresponding angular observables on the celestial sphere. 
The entire geometric information is contained in $\Gamma$ (see Equation~\eqref{eq:abbreviations}) and therefore only relevant in Equation~\eqref{eq:fold1}.
Hence, the~gravitational lensing geometry, and~properties of the underlying cosmology, can only be constrained with measured time delay differences by this model-independent~approach.

Furthermore, Equations~\eqref{eq:fold2}--\eqref{eq:fold4} only constrain ratios of derivatives. 
This is a consequence of the fact that the deflecting mass density is an inhomogeneity defined on top of a background mass density. 
Rescaling the background mass density and scaling the deflecting mass density accordingly is a choice of parametrisation and does not change the observables.
As detailed in~\cite{bib:Wagner6}, Equation~\eqref{eq:fold1} breaks this degeneracy for fixed $\Gamma$, i.e.,~fixing the background cosmology and thereby the background mass density at $\boldsymbol{x}_A$ and $\boldsymbol{x}_B$. 

Linearising Equation~\eqref{eq:lens_equation} and applying it to fields of single weakly distorted galaxies,~\cite{bib:Schneider3} already found that weak-gravitational-lensing observations only constrain the reduced shear $\boldsymbol{g}(\boldsymbol{x})$ and not $\gamma(\boldsymbol{x})$ because the weak-lensing equations are also subject to the mass-sheet degeneracy (MSD) that is intrinsic in the lensing formalism,~\cite{bib:Falco}. This finding and the MSD can now be understood with the results as stated above. Further details can be found in Section~\ref{sec:app_H_0} and in~\cite{bib:Wagner6}.


\subsubsection{The Cusp~Case}
\label{sec:cc_cusp}

Analogously to the derivations performed for Section~\ref{sec:cc_fold}, the~leading-order local lens properties for the cusp configuration as embedded into the coordinate system shown in Table~\ref{tab:taylor_configs} (centre and right figure) can be determined.
The distortion matrix which is set equal to the quadrupoles of the multiple images is given by
\begin{equation}
A(\boldsymbol{x}_i)=  \phi_{11}(\boldsymbol{x}_\mathrm{c}) \left( \begin{matrix} 1 &  \tfrac{\phi_{122}(\boldsymbol{x}_\mathrm{c})}{\phi_{11}(\boldsymbol{x}_\mathrm{c})} x_{i2} \\  \tfrac{\phi_{122}(\boldsymbol{x}_\mathrm{c})}{ \phi_{11}(\boldsymbol{x}_\mathrm{c})} x_{i2}  & \tfrac{\phi_{122}(\boldsymbol{x}_\mathrm{c})}{\phi_{11}(\boldsymbol{x}_\mathrm{c})} x_{i1} + \tfrac12 \tfrac{\phi_{2222}(\boldsymbol{x}_\mathrm{c})}{\phi_{11}(\boldsymbol{x}_\mathrm{c})}x_{i2}^2\end{matrix}\right) \;, \quad i=A,B, C \;.
\label{eq:cusp_A}
\end{equation}

The $x_2$-axis of the special coordinate system of the fold is aligned with the relative distance between the two images, such that the observables directly correspond to the variables in the equations. Contrary to that, the~coordinate system for the cusp configuration is constructed such that its $x_1$-axis is the symmetry axis of the parabolic approximation of the critical curve around $\boldsymbol{x}_\mathrm{c}$. 
Therefore, the~quadrupoles of the multiple images have to be rotated into this coordinate system first, i.e.,~calculating the $\varphi_i$ with respect to the positive $x_1$-axis from the relative angles $\tilde{\varphi}_i$ as shown in Table~\ref{tab:image_types}. 
Appendix~\ref{sec:appendix_cusp_rotation} shows the equation to obtain the rotation angle which transforms a configuration as shown in Table~\ref{tab:image_types} into the coordinate system shown in Table~\ref{tab:taylor_configs}.
In this appendix, we also give the coordinates of the image closest to the cusp, $\boldsymbol{x}_A$. 
This determines the position of $\boldsymbol{x}_\mathrm{c}$ with respect to $\boldsymbol{x}_A$ from the observables, analogous to $\boldsymbol{x}_\mathrm{f}$ being half the distance between image $A$ and $B$ in Section~\ref{sec:cc_fold}. 

Without loss of generality, we denote the two images that are closest to $\boldsymbol{x}_\mathrm{c}$ by $A$ and $B$, as~shown in Table~\ref{tab:taylor_configs}.
Any combination of two images from $A, B, C$ can be used to determine the following leading-order local lens properties. 
The most accurate results are obtained for images $A$ and $B$ closest to the expansion point of the Taylor approximation, as~we showed in~\cite{bib:Wagner1}. 
Summarising the calculations of~\cite{bib:Wagner0,bib:Wagner1,bib:Wagner2}, we obtain as leading-order lens properties from the relative distances between the multiple images, the~quadrupoles, and~a time delay difference of $\tau_{AB}$ between images $A$ and $B$
\begin{align}
\phi_{2222}(\boldsymbol{x}_\mathrm{c}) &= \dfrac{8 \, \tau_{AB}}{\Gamma \delta_{AB2}^ 4} \;, \label{eq:cusp1}  \\
\tilde{\phi}_{122}(\boldsymbol{x}_\mathrm{c}) &\equiv \dfrac{\phi_{122}(\boldsymbol{x}_\mathrm{c})}{\phi_{11}(\boldsymbol{x}_\mathrm{c})} = \dfrac{\tilde{\phi}_{12}(\boldsymbol{x}_A) - \tilde{\phi}_{12}(\boldsymbol{x}_B) }{\delta_{AB2}} = \dfrac{1}{\delta_{AB2}}\left(  \dfrac{\cot(\varphi_A) (r_A - 1) }{r_A \cot^2(\varphi_A) + 1} - \dfrac{\cot(\varphi_B) (r_B - 1) }{r_B \cot^2(\varphi_B) + 1} \right) \;, \label{eq:cusp2} \\
\tilde{\phi}_{2222}(\boldsymbol{x}_\mathrm{c}) &\equiv \dfrac{\phi_{2222}(\boldsymbol{x}_\mathrm{c})}{\phi_{11}(\boldsymbol{x}_\mathrm{c})} =  \dfrac{-6}{\delta_{AB2}^{2}}\left( \dfrac{\delta_{AB1}}{\delta_{AB2}} \tilde{\phi}_{12}(\boldsymbol{x}_B) + \tilde{\phi}_{22}(\boldsymbol{x}_A) \right) \label{eq:cusp3} \\ 
&=\dfrac{-6}{\delta_{AB2}^{2}}\left( \dfrac{\delta_{AB1}}{\delta_{AB2}} \dfrac{\cot(\varphi_B) (r_B - 1) }{r_B \cot^2(\varphi_B) + 1} \stackrel{-}{+} \dfrac{r_A + \cot^2(\varphi_A) }{r_A \cot^2(\varphi_A) + 1} \right) \;, \quad \varphi_A, \varphi_B \in \left] 0, \pi \right[ \;, \nonumber \\
m_\mathrm{c} &\equiv -\dfrac{\tfrac12 \tilde{\phi}_{2222}(\boldsymbol{x}_\mathrm{c}) - \tilde{\phi}_{122}(\boldsymbol{x}_\mathrm{c})^2}{\tilde{\phi}_{122}(\boldsymbol{x}_\mathrm{c})}. \label{eq:cusp4}
\end{align}

As in Section~\ref{sec:cc_fold} and for the same reasons, we only obtain ratios of derivatives from the quadrupole observables and, as~before, the~time delay difference breaks this degeneracy. 
To leading order, the~cusp configuration is a linear superposition of two fold configurations, as~investigated in detail in~\cite{bib:Wagner1,bib:SEF}.
This also becomes apparent when we compare Equations~\eqref{eq:fold2} and \eqref{eq:cusp2}\footnote{The expansion point is a different one in the two cases, yet, the~underlying principle to determine $\tilde{\phi}_{122}(\boldsymbol{x}_0)$ from the off-diagonal entries of the quadrupole moment of images $A$ and $B$ is the same.}.
We see that Equation~\eqref{eq:cusp2} is the sum of the $\left| \tilde{\phi}_{12}(\boldsymbol{x})\right|$ of images $A$ and $B$, while, for~Equation~\eqref{eq:fold2}, images $A$ and $B$ are described as mirror images of each other, so that $\left| \tilde{\phi}_{12}(\boldsymbol{x})\right|$ of one image, $A$ or $B$, is sufficient to determine $\tilde{\phi}_{122}(\boldsymbol{x}_0)$.
Treating images $A$ and $B$ as a fold configuration, the~sign of the slope in Equation~\eqref{eq:fold2} can be determined from the position of image $C$, as~the sign of the slope at $\boldsymbol{x}_\mathrm{f}$ between images $A$ and $B$ should be consistent with the slope of the parabola given by $m_\mathrm{c}$.
The leading-order Taylor expansion of the cusp configuration requires a rotation into a coordinate system that is not directly aligned with an observable and it needs a fourth order derivative, i.e.,\@ an expansion to one order more than the fold configuration.
Consequently, the~local lens properties determined at $\boldsymbol{x}_\mathrm{c}$ are less accurate than the ones at $\boldsymbol{x}_\mathrm{f}$, see~\cite{bib:Wagner1,bib:Wagner2} for~details. 

The remaining degeneracy is shown in Equation~\eqref{eq:cusp3}. 
Consistently with Sections~\ref{sec:images} and \ref{sec:cc_fold}, only~the relative parities between the three images are known. 
Hence, there are two possible pathways for the critical curve around the three images, as~further detailed in~\cite{bib:Petters}.
The first with the minus-sign in Equation~\eqref{eq:cusp3} encloses image $A$ in the parabolic approximation to the critical curve (called a positive cusp, as~shown in Table~\ref{tab:taylor_configs} (centre)).
The second with the plus-sign in Equation~\eqref{eq:cusp3} encloses images $B$ and $C$, such that image $A$ lies on the negative side of the $x_1$-axis (called a negative cusp, as~shown in Table~\ref{tab:taylor_configs} (right)).


\subsubsection{The Giant-Arc~Case}
\label{sec:cc_giant_arc}

``Golden lenses'' are highly (axi-)symmetric galaxies that show multiple images in form of giant arcs (see Table~\ref{tab:image_configurations} (right)) close to the (almost) circular critical curve, or~even complete Einstein rings, see e.g.,\@ \cite{bib:Williams} or~\cite{bib:Kochanek}. 
Giant arcs and Einstein rings as highly distorted and extended images provide non-local information in the sense that they trace the critical curves in a larger region compared to less distorted and therefore locally more constrained multiple images that are located at larger distances to the critical curves.
As noted in Section~\ref{sec:formalism}, potential biases in gravitational lens modelling decrease for decreasing distance to the model-invariant critical curve, such that Einstein rings are model-independent probes of the total enclosed mass.
For static sources, it is advantageous that the intrinsic source properties become negligible with decreasing distance to the critical curve as well. 
Multiple images of a time-varying source close to the critical curve embedded in a host galaxy that forms giant arcs or an Einstein ring are ideal configurations to constrain $H_0$, see e.g.,\@ \cite{bib:Suyu}, or~to calculate the total enclosed mass with Equation~\eqref{eq:tdd}. 
Example applications are given in Section~\ref{sec:app_H_0}.

Symmetric multiple-image configurations with a small perturbation also constrain the properties of the perturbing mass. Using lens models, upper bounds on the temperature and the sub-galactic distribution of dark matter can be inferred, see e.g.,\@ \cite{bib:Vegetti2,bib:Bose,bib:Xu,bib:Minor}.

In~\cite{bib:Wagner3}, we investigated which properties of a perturber to an axisymmetric mass-density distribution could be determined from the observables shown in Table~\ref{tab:image_configurations} (right) without a lens model.
Due to the underlying axisymmetry, we used polar coordinates $(r, \varphi)$ and perturbed the symmetric deflection potential, $\psi_\mathrm{a}(r)$, around the Einstein radius $r_\mathrm{E}$ with a small deflection potential $\psi_\mathrm{p}(r,\varphi)$, $\left| \psi_\mathrm{p}(r,\varphi) \right| \ll \left| \psi_\mathrm{a}(r) \right|$, such that the total deflection potential $\psi(r,\varphi)$ reads
\begin{equation}
\psi(r, \varphi) = \sum \limits_{n=0}^{\infty} \left( a_n + p_n (\varphi) \right) (r - r_\mathrm{E})^n
\label{eq:giant_arc_potential}
\end{equation}
with the coefficients $a_n$ and $p_n(\varphi)$ determined as derivatives of the two potentials at $r = r_\mathrm{E}$
\begin{align}
a_n = \left. \dfrac{1}{n!} \dfrac{\mathrm{d}^n \psi_\mathrm{a}(r)}{\mathrm{d}r^n}  \right|_{r = r_\mathrm{E}}\;, \quad p_n (\varphi) =  \left. \dfrac{1}{n!}  \dfrac{\partial^n \psi_\mathrm{p}(r,\varphi) }{\partial r^n} \right|_{r = r_\mathrm{E}} \;.
\label{eq:giant_arc_coeff}
\end{align}

To leading order, the~lens equation yields the model-independent local lens properties
\begin{align}
r_i - r_j &= \dfrac{\alpha_{\mathrm{p}, r}(r_\mathrm{E}, \varphi_i) -  \alpha_{\mathrm{p}, r}(r_\mathrm{E}, \varphi_j)}{2(1-\kappa_\mathrm{a}(r_\mathrm{E}))} \;,\label{eq:ce1} \\
\varphi_i - \varphi_j &= \dfrac{\alpha_{\mathrm{p}, \varphi}(r_\mathrm{E}, \varphi_i)}{r_i^2} -  \dfrac{\alpha_{\mathrm{p}, \varphi}(r_\mathrm{E}, \varphi_j)}{r_j^2} \;.\label{eq:ce2}
\end{align}
$\kappa_a(r_\mathrm{E})$ denotes the convergence of the axisymmetric part of the potential at $r_\mathrm{E}$.
$\boldsymbol{\alpha}_\mathrm{p}=(\alpha_{\mathrm{p},r},\alpha_{\mathrm{p},\varphi})$ is the deflection angle caused by the perturber.
$r_i - r_j$ can be the difference between the observed radial positions of arcs $A$ and $B$. 
Alternatively, the~difference between the radial positions of the centre-close and centre-far isophote delineating one arc can also be inserted as $r_i - r_j$.
Dividing Equation~\eqref{eq:ce1} for the radial differences of the delineating isophotes of arc $A$ by the same equation for arc $B$, we arrive again at Equation~\eqref{eq:kappa_fgs}. Since this procedure maps the two arcs onto each other, it is not surprising that the ratio of Equation~\eqref{eq:ce1} of arc $A$ to $B$ is a special case of Equation~\eqref{eq:kappa_fgs} for lenses with almost axisymmetric mass-density~distributions.

Equation~\eqref{eq:ce1} implies a degeneracy between the deflection angle of the perturber and the convergence of the axisymmetric lens. 
As further detailed in~\cite{bib:Wagner3}, this degeneracy can be interpreted as an exact source-position transformation, introduced in~\cite{bib:Schneider}. 
As explained in~\cite{bib:Wagner6}, this degeneracy originates from an a priori arbitrary split of the total deflection potential into a main lens and a perturber.
Without additional assumptions or information on the mass-density profile of the axisymmetric lens at $r_\mathrm{E}$ or on the deflection angle of the perturber, Equations~\eqref{eq:ce1} and \eqref{eq:ce2} are under-constrained. 
By~means of simulations, this was noted in~\cite{bib:Schneider2}. 
Applying Equations~\eqref{eq:ce1} and \eqref{eq:ce2} to their data, as~detailed in~\cite{bib:Wagner3}, the~observed degeneracies can be~explained.

At best, different models based on a single smooth deflecting mass density or including perturbing inhomogeneities can be ranked with respect to each other in a Bayesian framework, see e.g.,\@ \cite{bib:Vegetti} or~\cite{bib:Suyu2}. 
In this way, the~most likely explanation for the perturbed, symmetric multiple-image configuration is determined from a certain class of models.
In Section~\ref{sec:degeneracies}, we apply Equations~\eqref{eq:ce1} and \eqref{eq:ce2} to an example, which is further detailed in~\cite{bib:Wagner3}. 
Assuming a specific model for the perturber and the main lens and inserting them into the model-independent equations, we can compare the total mass of the perturber with existing estimates obtained by the Bayesian approach of~\cite{bib:Vegetti}.

\subsection{Lensing Distance Ratios in a General Friedmann~Universe}
\label{sec:distances}

SNe are standardisable candles which allow for a data-based distance measure that is independent of any specific parametrisation of the FLRW metric. 
Replacing the distances based on the FLRW metric (see Equation~\eqref{eq:D_A}, using Equation~\eqref{eq:E} as $E(z)$) by data-based distance measures makes the angular diameter distances in Equation~\eqref{eq:travel_time} independent of any parametrisation of the FLRW metric.
This means that the data-based distances are agnostic about any partition of the total cosmic energy content into radiation, matter, or~additional contributions (as done for Equation~\eqref{eq:E}).
Thus, using these SNe-based distance measures, we free our lens-model-independent approach from a parametrisation of the FLRW metric as well.
In particular, we avoid setting up particular assumptions about the cosmological constant or potential dark energy~models.

In the standardisation of the SNe, an~overall distance scale for an ensemble of observed SNe has to be fixed by complementary data. 
It can either be given by the absolute magnitude of a standard SNe (e.g.,\@ from measurements in our local cosmic neighbourhood, see~\cite{bib:Richardson}), or~by $H_0$ (as determined in our local cosmic neighbourhood,~\cite{bib:Riess}, or~from the cosmic microwave background,~\cite{bib:Planck}). 
This implies that data sets of SNe determine $E(z)$ and leave $H_0$ as a free parameter.
The Pantheon sample, as~detailed in~\cite{bib:Scolnic}, is the most recent compilation of 1048 unscaled SNe with redshifts $z < 2.3$.
This redshift range covers a large fraction of the universe, in~which gravitational lensing effects occur.
Therefore, we~determined angular diameter distances based on the Pantheon sample in a flat FLRW~metric.

First, we expanded the scale-free, observable luminosity distances $\tilde{D}_\mathrm{L}(0,z)$ into a complete, orthonormal set of analytic basis functions $b_j$, $j=0,...,n_\mathrm{b}$, the~so-called Einstein--de-Sitter polynomial basis as introduced in~\cite{bib:Mignone}
\begin{equation}
\tilde{D}_\mathrm{L}(0,z) \equiv \dfrac{H_0}{c} D_\mathrm{L}(0,z) = (1+z) \int \limits_0^z \dfrac{\mathrm{d}x}{E(x)} \equiv \sum \limits_{j=0}^{n_\mathrm{b}} c_j \, b_j(z) \;.
\label{eq:D_L}
\end{equation}

From the Pantheon sample of SNe, four coefficients for the first four basis functions can be significantly determined given the measurement precision and the covariances in the sample.
Inserting these scale-free luminosity distances represented by the Einstein--de-Sitter basis into the equation
\begin{equation}
\dfrac{\mathrm{d} \tilde{D}_\mathrm{L}(0,z)}{\mathrm{d} z} = \dfrac{\tilde{D}_\mathrm{L}(0,z)}{1+z} + \dfrac{1+z}{E(z)}
\label{eq:D_L_E}
\end{equation}
and requiring that $E(0)=1$, we obtained a data-based cosmic expansion function, $E(z)$, as~first noted by~\cite{bib:Starobinski}.
Subsequently, we inserted this $E(z)$ into Equation~\eqref{eq:D_A} together with a value for $H_0$, e.g.,\@ as constrained by~\cite{bib:Planck} or~\cite{bib:Riess} to obtain a distance~measure.

Replacing the model-based angular diameter distances in Equation~\eqref{eq:travel_time} by the data-based ones, we~obtain the data-based lensing distance ratio, as~detailed in~\cite{bib:Wagner5}. 
Leaving $H_0$ as a free parameter, we can use time delay differences to constrain its value, if~the difference of the Fermat potentials between the image positions is known.
Alternatively, with~a fixed value for $H_0$, the~time delay differences between two images constrain the difference between their Fermat potentials (see Equation~\eqref{eq:tdd}).

In Section~\ref{sec:app_distances}, we compare the relative precision of the lensing distance ratio of angular diameter distances as obtained from the Pantheon sample with the lensing distance ratio based on the angular diameter distances of the Friedmann model as established by~\cite{bib:Planck}.
Alternatively, given the Fermat potential difference between two images, a~measured time delay difference between these images infers a value for $H_0$. 
Constraints on its precision are estimated in Section~\ref{sec:app_H_0}.

\section{Results}
\label{sec:results}

In this section, we discuss example applications for our approach, theoretically outlined in Section~\ref{sec:materials}, to~observational data. 
In Section~\ref{sec:app_cluster}, we show the application of the approach of Section~\ref{sec:images} for five resolved images of a background galaxy behind the galaxy cluster CL0024 observed with the Hubble Space Telescope (HST). 
Further details about this example can be found in~\cite{bib:Wagner_cluster}. 
Section~\ref{sec:app_quasar} subsequently treats the application of our approach of Section~\ref{sec:images} to four unresolved, extended images of a quasar background source behind the spiral galaxy B0128, as~further detailed in~\cite{bib:Wagner_quasar}. 
For this application, very-long-baseline interferometry (VLBI) observations in the radio bands by~\cite{bib:Biggs}, were employed. 
In Section~\ref{sec:degeneracies}, we discuss the degeneracies detailed in Section~\ref{sec:cc} that arise for the two giant arcs of a background galaxy caused by the early-type galaxy B1938+666.
We review the precision up to which the lensing distance ratio in Equation~\eqref{eq:abbreviations} can be determined from the Pantheon sample of SNe type Ia, \cite{bib:Scolnic}, in~Section~\ref{sec:app_distances}. 
The details to set up the angular diameter distance measure from the sample are found in~\cite{bib:Wagner5}. 
Section~\ref{sec:app_H_0} sets up theoretical limits on the precision to which $H_0$ can be determined by Equation~\eqref{eq:tdd}. This section is based on the findings in~\cite{bib:Wagner6} and supported by data from quasar time delay monitoring, see e.g.,\@ the paper series of~\cite{bib:Eigenbrod}, and~FRB observations, discussed in~\cite{bib:Wagner_frb}.

In~\cite{bib:Wagner1,bib:Wagner2}, we used simulation data based on simple lens models to test the accuracy of the approximations detailed in Sections~\ref{sec:images} and \ref{sec:cc}. 
The relative image distances and the parameters employed for the lens models considered in these simulations were extreme cases compared to observations and lens model parameters as inferred by observations. 
Despite this, we obtained inaccuracies of about 10\% and mostly less for the ratios of third order derivatives at folds and cusps.
Most of the $f_{ij}$ and $g(\boldsymbol{x}_i)$, $i,j=A, B, C$, lay within the 1$-\sigma$ confidence bounds around the true value. 
As a consequence, we expect our approach to yield more accurate results in average-case scenarios considered in this section, so that the local lens properties and pathways of the critical curves in the vicinity of multiple images can be constrained more~tightly.

\subsection{Data~Preprocessing}
\label{sec:app_preprocessing}

Before any observables of Tables~\ref{tab:image_types} and \ref{tab:image_configurations} can be extracted, the~observational data have to be preprocessed. 
Effects of the data acquisition, such as shot noise or the impact of the point spread function (PSF), have to be taken into account. 
In addition, astrophysical effects like stray light from or partial occlusion by neighbouring objects must be handled, see e.g.,\@ \cite{bib:Melchior3} or~\cite{bib:Plazas} for a detailed explanation of all~effects. 

In order to keep our approach model-independent, the~data preprocessing should not rely on any model. Data processing methods that employ a model-based separation of the foreground deflecting galaxy from the multiple images for galaxy-scale lensing can be found in~\cite{bib:Bolton2,bib:Auger, bib:Lagattuta}. 
Furthermore, we separate the extraction of the observables from the application of our approach to these data. 
In~this way, the~lensing analysis pipeline is kept as flexible as possible to use our approach for data of various ground-based and space-based instruments and the impact of different preprocessing methods on the local lens properties can be easily~compared. 

Since the PSF of the HST is of the order of one pixel and thus much smaller than the extensions of the multiple images, effects of the convolution with the PSF are negligible.  
For Section~\ref{sec:app_cluster}, the~reference points for in the multiple images were identified manually directly from the FITS-file provided in the HST archive.
The same applies to the observables in Section~\ref{sec:degeneracies}. 
The observables in the radio band, including the quadrupoles of the multiple images as required for the application in Section~\ref{sec:app_quasar}, were already measured by~\cite{bib:Biggs} using the National Radio Astronomy Observatory Astronomical Image Processing Software package \textsc{AIPS}.

\subsection{Comparison to Lens Modelling Approaches for the Cluster-Scale Lens~CL0024}
\label{sec:app_cluster}

\begin{figure}[H]
\centering
\includegraphics[width=0.73\textwidth]{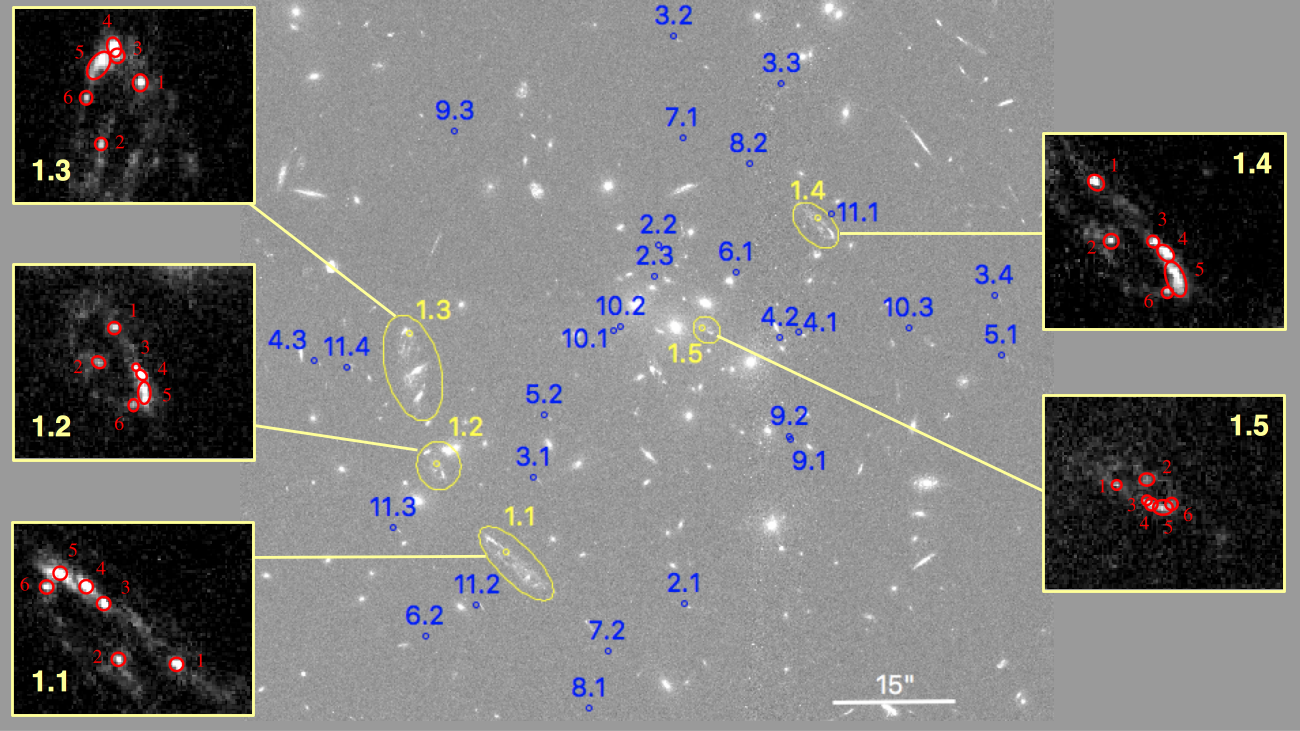}
\caption{Galaxy cluster CL0024 and its multiple-image sets (MISs): MIS~1 (yellow ellipses) with five images with six reference points each (red circles in image details) to apply our approach as described in Section~\ref{sec:images} and with MISs~2 to 11 (blue points) as predicted by~\cite{bib:Zitrin}, but~not spectroscopically confirmed. MIS~3, 4, 5, 8, and~10 were used to set up lens models that were compared to the model-independent approach. \textit{Image credits: NASA/ESA/HST.}}
\label{fig:CL0024}
\end{figure}

CL0024+1654, abbreviated as CL0024, is a massive galaxy cluster at redshift $z_\mathrm{l}=0.39$.
It consists of two components merging along the line of sight (as detailed e.g.,\@ in~\cite{bib:Czoske,bib:Zhang,bib:Umetsu}) and it magnifies a blue galaxy at redshift $z_\mathrm{s}$ = 1.675 into five highly-distorted, well-resolved images, as~first described by~\cite{bib:Colley} (see images 1.1 to 1.5 in the yellow ellipses in Figure~\ref{fig:CL0024}). 
In Figure~\ref{fig:CL0024}, the~yellow points in the ellipses mark the coordinates of the multiple images when they are treated as point-like constraints of global lens reconstructions. 
In the details at the sides, the~six reference points that can be identified in all multiple images are marked by red circles.
We extracted the reference points in all multiple images as observables from an HST ACS/WFC picture in the F475W band\footnote{Based on observations made with the NASA/ESA Hubble Space Telescope, and~obtained from the Hubble Legacy Archive, which is a collaboration between the Space Telescope Science Institute (STScI/NASA), the~Space Telescope European Coordinating Facility (ST-ECF/ESA) and the Canadian Astronomy Data Centre (CADC/NRC/CSA).}.
Using these reference points, we applied our approach of Section~\ref{sec:images} to these multiple images to constrain the local lens properties in Equation~\eqref{eq:kappa_fgs}.
World-coordinate-system coordinates of all points can be found in~\cite{bib:Wagner_cluster}.
The work in Ref.~\cite{bib:Zitrin}
predicted ten additional multiple-image sets (MISs) from sources with photometrically determined mean redshifts $z_\mathrm{s} \in \left[ 0.25, 4.16 \right]$ (see blue points 2.1 to 11.4 in Figure~\ref{fig:CL0024}, coordinates of these points can be found in~\cite{bib:Zitrin}). 
From these additional ten MISs, we employed MIS~3, 4, 5, 8, and~10 as point-like multiple-image constraints to set up global mass-density reconstructions. 
We selected these MISs because they are scattered broadly in the region covered by the cluster on the sky so that we expected well-constrained global mass-density reconstructions. 
In addition, these MISs are located close to the resolved multiple images of set~1, so that a potential influence of constraints from neighbouring MISs on the local lens properties at MIS~1 should be~observable.

As global lens reconstruction methods, we chose the parametric reconstruction method \textsc{Lenstool}, described in detail in~\cite{bib:Jullo_lenstool} and references therein, and~the free-form reconstruction approach \textsc{Grale}, as~detailed in~\cite{bib:Liesenborgs_grale}. Appendix~\ref{appendix:lens_reconstructions} contains a brief characterisation of both algorithms in terms of their peculiarities that are relevant for our comparison. 
Both approaches used the same data and yielded maps of the convergence, $\kappa(\boldsymbol{x})$, and~the shear, $\boldsymbol{\gamma}(\boldsymbol{x})$, in~the region covered by the multiple images.
We calculated the local lens properties of Equation~\eqref{eq:kappa_fgs} from the $\kappa(\boldsymbol{x})$- and $\boldsymbol{\gamma}(\boldsymbol{x})$-maps of the global reconstruction approaches at the positions of the multiple images of MIS~1 marked by the yellow points in Figure~\ref{fig:CL0024}.
Then, we compared these local lens properties obtained by the global methods with the local lens properties at the same position obtained by our model-independent approach detailed in Section~\ref{sec:images}.

Implementing different constraints in the \textsc{Lenstool} and \textsc{Grale} reconstructions and comparing the results among each other and with the results of our approach, we~investigated 
\begin{enumerate}
\item whether neighbouring MISs influence the reconstruction of local lens properties at the image positions of MIS~1 in \textsc{Lenstool} and \textsc{Grale} and thus introduce correlations between MISs that are not accounted for in our approach,
\item whether the local lens properties at MIS~1 as obtained by our local approach and by the two global reconstructions coincide, or~if image distortions beyond leading order play a significant role (apart from potential correlations between MISs mentioned under 1.),
\item whether the light-traces-mass (LTM) assumption used in \textsc{Lenstool} is corroborated and, if~so, which mass-to-light ratio (MLR) is favoured for the cluster member galaxies, and~\item whether the small-scale fine-tuning  of the mass density implemented in \textsc{Grale} overfits the free-form lens reconstruction to the multiple-image constraints.
\end{enumerate}

\begin{figure}[H]
\centering
\includegraphics[width=0.63\textwidth]{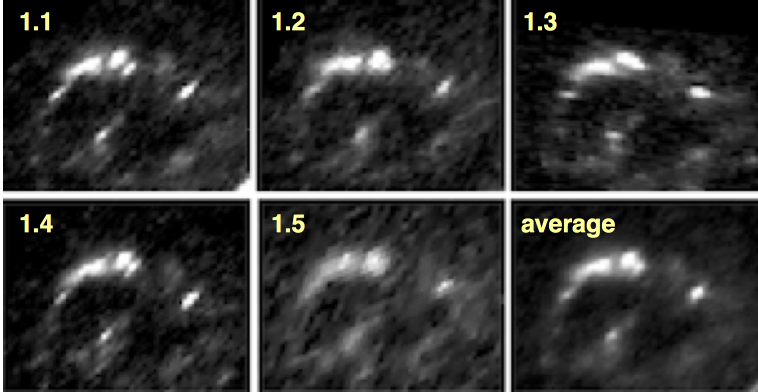}
\caption{Back-projected multiple images of MIS~1 to the source plane using the distortion matrix of local lens properties of Equation~\eqref{eq:A_kappa_gamma} for each image, normalised such that the transformation between 1.1. and 1.2 has unit determinant. Comparing the back-projections for images 1.1 and 1.5, we see that highly-resolved, large images with high signal-to-noise ratio (SNR) yield higher resolutions in the source reconstruction than smaller images with a low~SNR.}
\label{fig:CL0024_source}
\end{figure}

The local lens properties that we obtained showed broad confidence bounds for the multiple images close to the $\kappa(\boldsymbol{x})=1$ isocontour, which was to be expected from Equation~\eqref{eq:kappa_fgs}. The~tightest confidence bounds were obtained using all six reference points which delineate the largest area in each multiple image in accordance with the statements made in Section~\ref{sec:images}. Relative sizes of the confidence bounds with respect to the quantities in Equation~\eqref{eq:kappa_fgs} ranged from 8\% to 150\% with an average relative size of 9.4\%.
The implementation of our approach took less than half a second to  determine the quantities in Equation~\eqref{eq:kappa_fgs}.
Assuring that all reconstructions are of similar quality, we~found a high degree of coincidence in the local lens properties obtained by all three approaches. 
From this result, we concluded~that
\begin{enumerate}
\item non-local influences of neighbouring MISs are negligible at the current precision of the quantities in Equation~\eqref{eq:kappa_fgs}. 
Thus, the~local lens properties are mainly derived from the local constraints of the respective multiple images. 
Hence, we can reconstruct the morphology of the source up to an overall scale as detailed in Section~\ref{sec:images}. The~back-projections of all images to the source plane by their model-independently reconstructed $A(\boldsymbol{x}_i)$ and their pixel-wise averaged source are shown in Figure~\ref{fig:CL0024_source} (for comparison with model-based source reconstructions, see~\cite{bib:Colley,bib:Liesenborgs2}; a~physical analysis of the \textsc{Lenstool}-reconstructed source can be found in~\cite{bib:Jones,bib:Richard}).
\item the leading-order local lens properties of our model-independent approach mostly agree to local lens properties as obtained by \textsc{Lenstool} and \textsc{Grale} within their 1$-\sigma$ confidence bounds. 
\item the LTM assumption is corroborated by the high degree of agreement of the local lens properties as obtained by \textsc{Lenstool} and the other two approaches that do not make any assumptions about the relation between dark and luminous matter.
A comparison of the local lens properties of all approaches favours a constant MLR for the brightest cluster member galaxies over a non-constant MLR. 
This result was also found in~\cite{bib:Umetsu,bib:Kneib}.
A non-constant MLR that reproduces the fundamental plane (see~\cite{bib:Caminha} and references therein) yields a higher degree of agreement between the redshifts as estimated by \textsc{Lenstool} and the photometric redshifts of the additional MISs. 
This indicates that spectroscopic redshift observations are necessary to further constrain the redshifts and the~MLR.
\item the small-scale fine-tuning of the mass density in \textsc{Grale} does not significantly change the local lens properties, nor does it tighten the confidence bounds. As~it does not significantly alter the pathways of the critical curves, either, it can be omitted for run-time optimisation. 
\end{enumerate}
\unskip

Consequently, assuming that our results for CL0024 are representative, we demonstrated that a comparison of the local lens properties between the three conceptually different reconstruction ansatzes allows for corroborating or refuting assumptions of global reconstruction methods.
Furthermore, we showed that the source object can also be reconstructed by a back-projection of the multiple images with the local distortion matrix.  
Hence, a~full reconstruction of the deflecting mass density is not necessary to study the morphology of individual background objects. 
The back-projections shown in Figure~\ref{fig:CL0024_source} are the most general leading-order source reconstructions directly based on the standard lensing formalism and the observables in the multiple images. 
In addition, as~our source reconstruction does not assume a lens model, studying star formation and stellar populations in faint and distant, otherwise unobservable galaxies in the early universe (even at redshifts as high as 9--10) without a potential lens-model bias becomes possible.
The source reconstruction is still subject to the local MSD of this MIS (see~\cite{bib:Wagner4} for this local generalisation of the global MSD as defined by~\cite{bib:Falco}).
However, the latter can be easily broken by rescaling the distortion matrices, as~soon as the actual scaling is known, e.g.,\@ from time delay difference~observations.

\subsection{Transfer of the Method to Unresolved Multiple Images for the Galaxy-Scale Lens~B0128}
\label{sec:app_quasar}

\begin{figure}[H]
\centering
\includegraphics[width=0.64\textwidth]{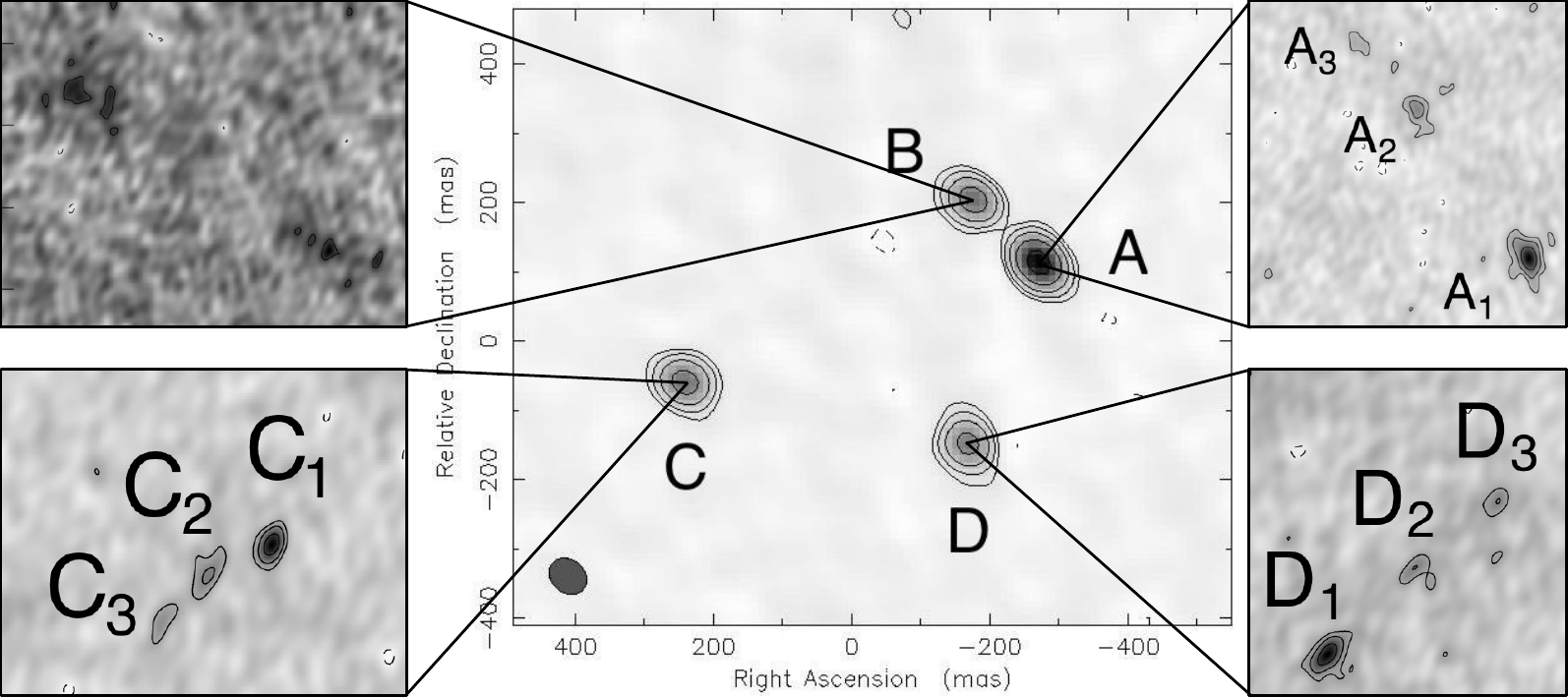} \hfill
\includegraphics[width=0.3\textwidth]{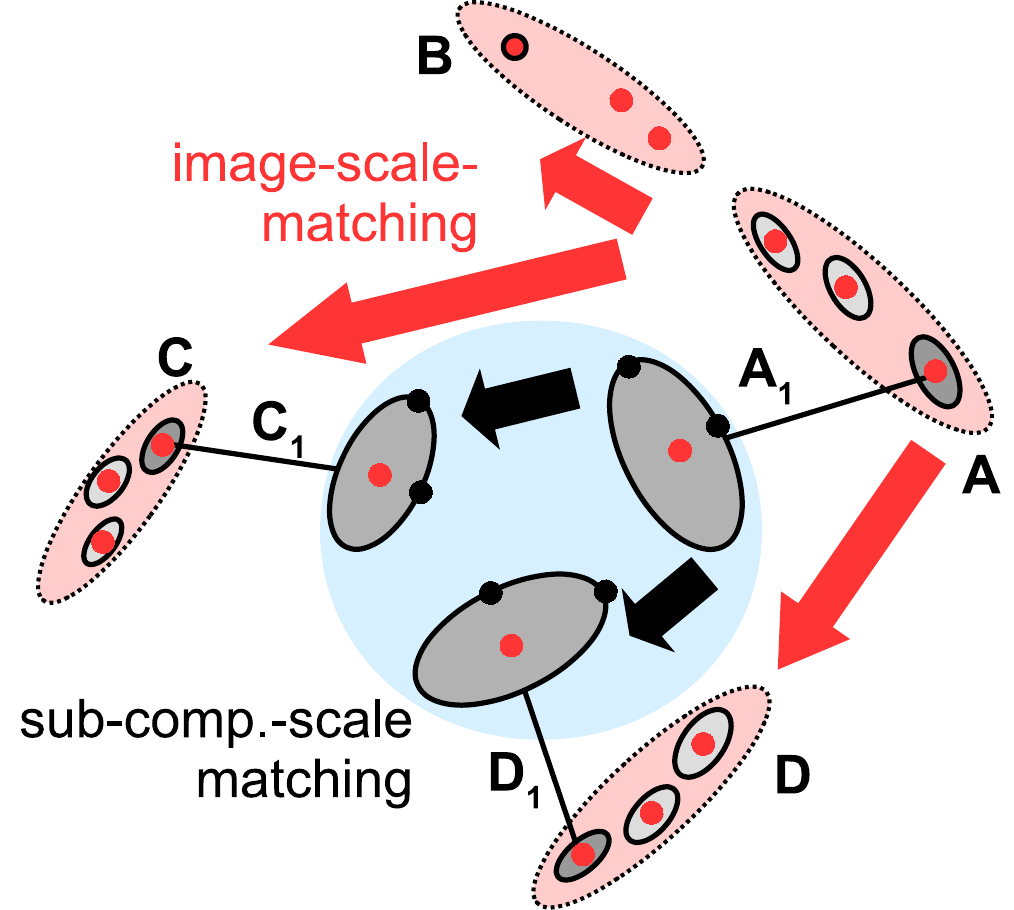}
\caption{Observational data of B0128 (\textbf{left}): MERLIN 5 GHz band showing four images of a quasar at $z_\mathrm{s}=3.124$ (\textbf{centre}) and VLBA 8.4 GHz details of all multiple images, revealing that images $A$, $C$, and~$D$ are resolved into three sub-components on milli-arcsecond scale, while $B$ is most likely scatter-broadened. North is up and East is to the left. Schematic of applications of our model-independent approach (\textbf{right}): on image scale, the~positions of the sub-components are used as reference points to apply our approach detailed in Section~\ref{sec:images}, on~sub-component scale, the~quadrupole moments of images $A$, $C$, and~$D$ are employed to determine local lens properties for the sub-components 1 and 3 using our approach detailed in Section~\ref{sec:images}. \textit{Image credits:~\cite{bib:Biggs,bib:Phillips}}.}
\label{fig:B0128}
\end{figure}

In the second example which is discussed in detail in~\cite{bib:Wagner_quasar}, we demonstrate that our approach can also be applied to galaxy-scale gravitational lenses with multiple-image configurations as observed in the radio bands. 
Figure~\ref{fig:B0128} (left) shows the configuration of four images of a quasar at $z_\mathrm{s}=3.124$ created by the lenticular or late-type galaxy B0128+437, B0128 for short.
Most probably, its redshift is $z_\mathrm{l}=1.145$, \cite{bib:Lagattuta2}.
The observation was made in the Multi-Element Radio Linked Interferometer Network (MERLIN) 5 GHz band, as~detailed in~\cite{bib:Phillips}. 
The multiple images denoted by $A$, $C$, and~$D$ in Figure~\ref{fig:B0128} are resolved into three clearly identifiable sub-components in VLBI observations in the 5 GHz and 8.4~GHz bands (see details on the left and right of the MERLIN observation), see~\cite{bib:Biggs}. 
Image $B$ is most likely scatter-broadened. 
The coordinates of all observational data are summarised in~\cite{bib:Wagner_quasar}. 

Analogously to the example in Section~\ref{sec:app_cluster}, we used the example to compare the local lens properties obtained by our approach according to Section~\ref{sec:images} with the local lens properties obtained by the global free-form lens reconstruction approach \textsc{PixeLens}, \cite{bib:Saha}.
We investigated whether the lens reconstruction by previous global parametric lens reconstruction methods was oversimplified because~no global parametric lens-modelling approach could explain the sub-components in images $A$, $C$, and~$D$ to their sub-milli-arcsecond precision, while the configuration of the four images shown in Figure~\ref{fig:B0128} (left, large figure) could be described by a singular isothermal elliptical (SIE) mass-density profile with external shear, as~detailed in~\cite{bib:Biggs}.
According to the simulations performed in~\cite{bib:Xu}, it~was found that the anomalous flux ratio between images $A$ and $B$ is not likely to be caused by small-scale dark-matter clumps.
Therefore,~\cite{bib:Xu} concluded that the scatter-broadening in image $B$ and an oversimplified lens reconstruction could cause the discrepancies between the fits of smooth symmetric models and the~observations.

Using the sub-components of images $A$, $C$, and~$D$ as reference points, we calculated the local lens properties according to Equation~\eqref{eq:kappa_fgs}.
In addition, as~the sub-components 1 and 3 in these three images were characterised by their quadrupole moment, we also calculated the local lens properties for both sub-components using the centre of light and the end points of the semi-minor and the semi-major axis as reference points (see Section~\ref{sec:observables}).
Figure~\ref{fig:B0128} (right) sketches the transformations on image and sub-component scale to obtain the respective local lens properties.
Due to the almost parallel alignment of the sub-components, the~local lens properties determined from the sub-components as reference points were subject to broad confidence bounds. Their relative sizes compared to the $f$- and $g$-values range from 8\% to over 1000\%.
Comparing these confidence bounds to the ones obtained from the quadrupole moments, which mostly have relative sizes on the order of 20\%, we found that the latter are smaller because the vectors spanned by the reference points are orthogonal to each other.
Half of the local lens properties at the positions of the two sub-components agree within their confidence bounds, so that it remains undecided if the distortion matrix can be assumed constant over the area of all sub-components or not. 
Only a weak upper limit to the scale of potential higher order perturbations, like gradients, in~the surface mass density at the position of the images A, C, and~D may be inferred from this result.
However,  the~result confirms the hypothesis of~\cite{bib:Xu} that the global parametric lens reconstructions were indeed oversimplified.
A comparison with our \textsc{PixeLens} lens reconstruction and an additional symmetry analysis according to the method based on relative opening angles between the multiple images as detailed in~\cite{bib:Gomer} both further corroborate the hypothesis.
All~results concordantly hint at a mass-density distribution that deviates from the simple elliptical models usually~assumed.

\subsection{Degeneracies Arising in the Detection of Perturbers for the Galaxy-Scale Lens B1938+666}
\label{sec:degeneracies}

\begin{figure}[H]
\centering
\includegraphics[width=0.25\textwidth]{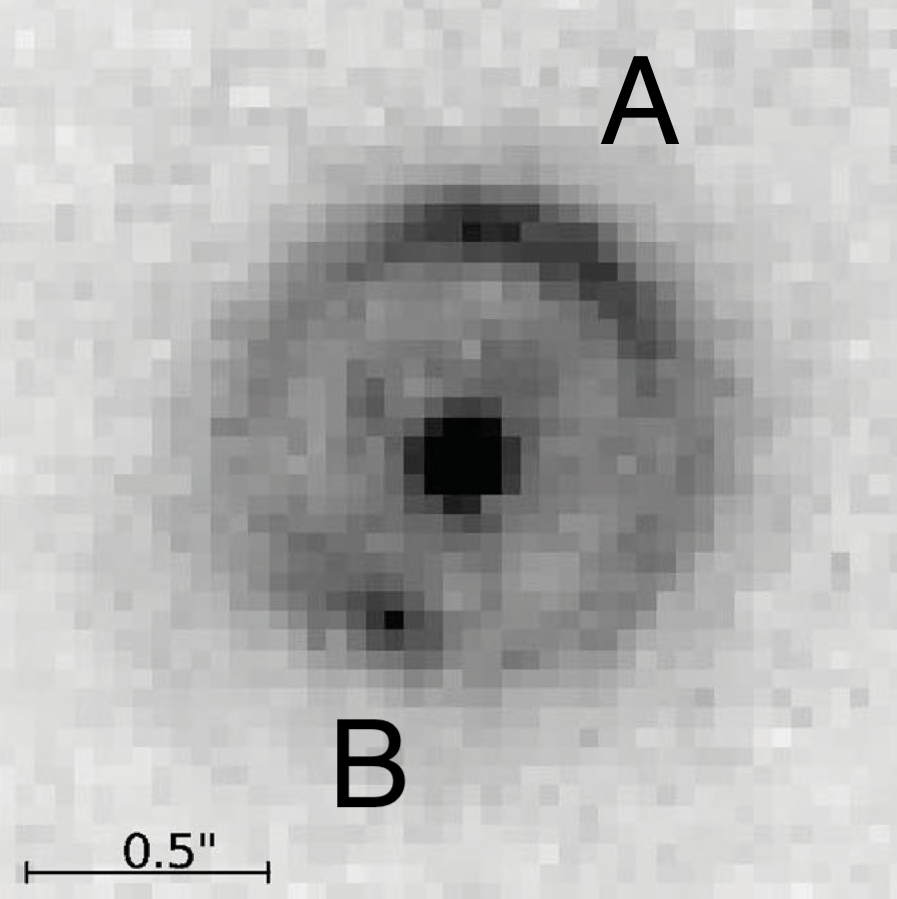} \hspace{2ex}
\includegraphics[width=0.25\textwidth]{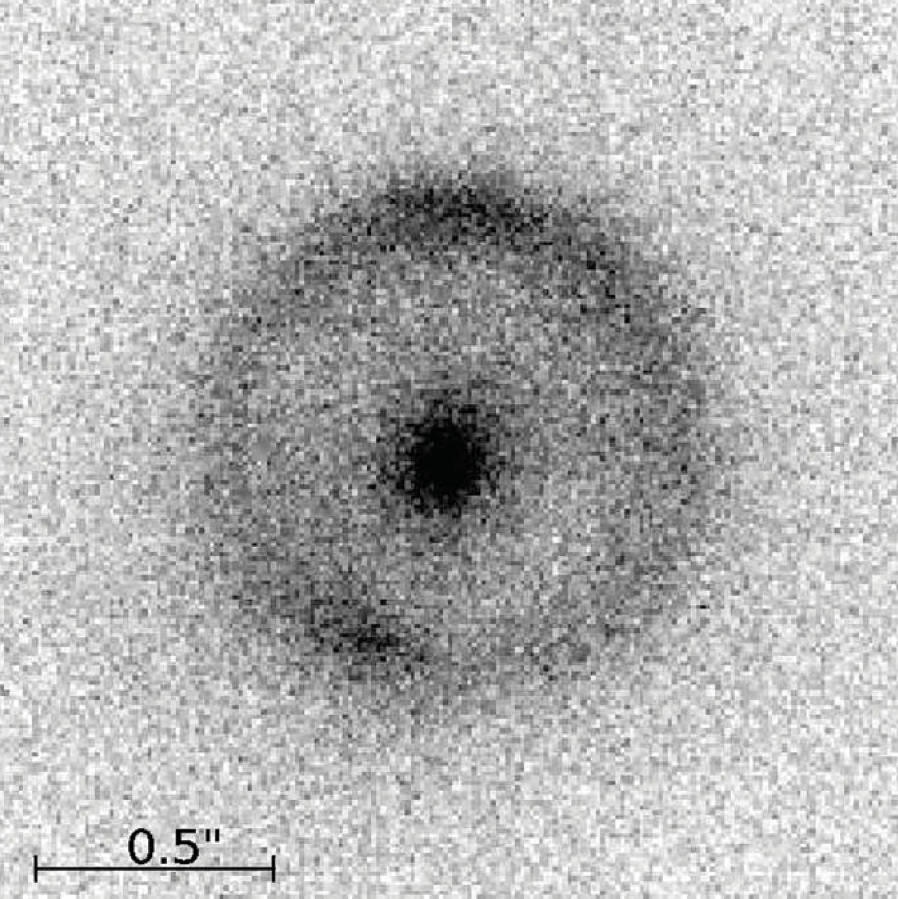} \hspace{2ex}
\includegraphics[width=0.25\textwidth]{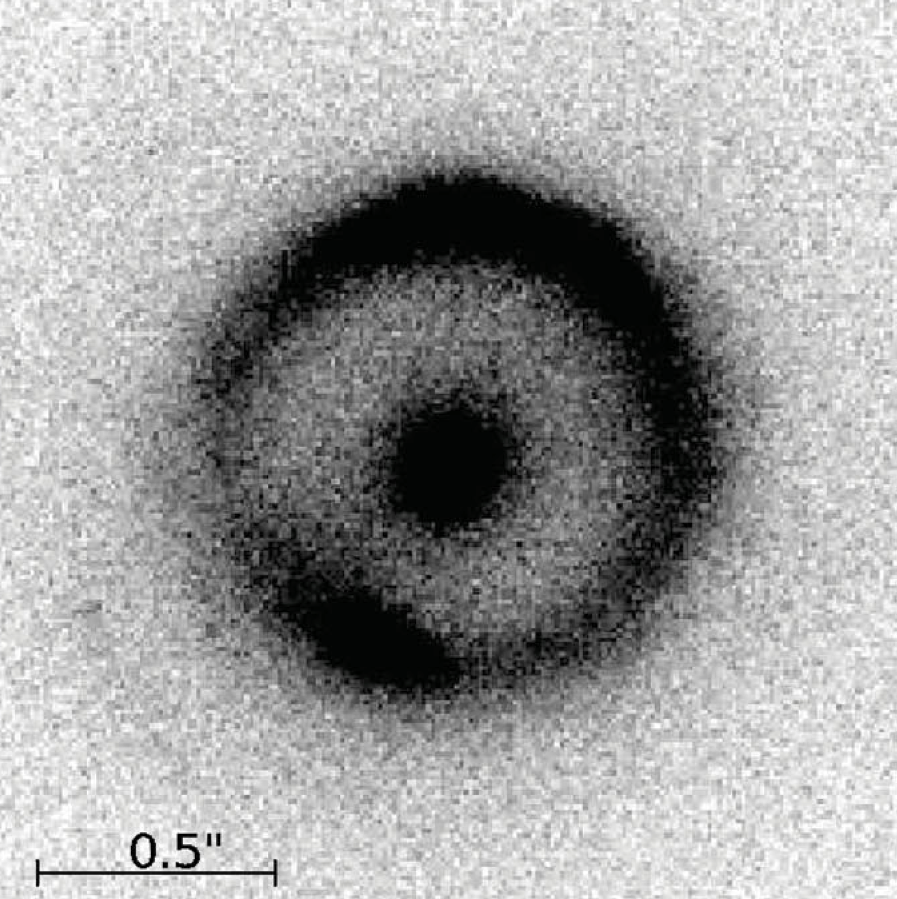}
\caption{Observational data of B1938+666: F160W band filter of HST NIC1 (\textbf{left}); H-band of the Keck NIRC2 camera with Laser Guide Star Adaptive Optics System (\textbf{centre}); K'-band of the Keck NIRC2 camera with Laser Guide Star Adaptive Optics System (\textbf{right}). North is up and East is to the left. \textit{Image credits:~\cite{bib:Lagattuta}}.}
\label{fig:B1938}
\end{figure}

B1938+666, detected in the Jodrell Bank Very Large Array Astrometric Survey,~\cite{bib:Patnaik}, is an early-type gravitational lens at redshift $z_\mathrm{l}=0.88$, \cite{bib:Tonry}. The~background source, a~quasar embedded in its host galaxy at redshift $z_\mathrm{s}=2.06$, \cite{bib:Riechers}, is assumed to consist of two components,~\cite{bib:Narasimha}.
Multifrequency Very Large Array (VLA), MERLIN and VLBI radio observations revealed that the first source component is mapped into four images and the second into two images,~\cite{bib:King}.
As observed in~\cite{bib:King2} by HST observations and later on confirmed by observations with the NIRC2 camera on the Keck II Telescope, using the Laser Guide Star Adaptive Optics System in~\cite{bib:Lagattuta}, the~source galaxy forms an Einstein ring visible in the near infrared band.
Figure~\ref{fig:B1938} shows the observational data in the optical and near~infra-red.

Perturbations in the symmetric brightness distribution of the Einstein ring were used to infer properties of a potential perturber, like its location and its mass in~\cite{bib:Lagattuta,bib:Vegetti3}. 
As there is no evidence for luminous perturbers equal and above the mass estimate determined in~\cite{bib:Vegetti3} (see below), \cite{bib:Lagattuta}, the~existence and the properties of a potential perturber were solely inferred from the observable brightness profiles of the giant arcs.
In \cite{bib:Lagattuta,bib:Vegetti3}, the Bayesian lens-reconstruction algorithm of~\cite{bib:Vegetti} was employed to infer the most likely properties of a perturber.
Using the HST data, \cite{bib:Vegetti3} arrived at
\begin{equation} 
m_\mathrm{p} = \left( 1.9 \pm 0.1 \right) \cdot 10^8 \, \mathrm{M}_\odot
\label{eq:mass_Vegetti}
\end{equation}
assuming that the perturber is located close to the critical curve in the lens plane and the projected distance between the symmetry centre of the main lens and the perturber in the lens plane is the total separation. 
Hence, Equation~\eqref{eq:mass_Vegetti} is a lower limit on the perturbing mass. 
As  also confirmed in~\cite{bib:Lagattuta}, the~mass estimates derived from the H- and K'-band Keck data are consistent with the value in Equation~\eqref{eq:mass_Vegetti}. 
Furthermore,~\cite{bib:Lagattuta} followed up on these results with Keck observations and compared the precision to which the parameters of the most likely lens reconstruction can be determined to the precision obtainable when using the HST observations. 
For this highly symmetric case, the~Keck observations were found to yield more precise parameter estimates. 
Nevertheless, due to the high symmetry of the main lens and the featureless, i.e.,\@ unresolved, giant arcs, degeneracies between the lens model parameters still remained large. 
This implies that significantly different models can fit the data equally well. 
To arrive at this result and test the robustness of the obtained properties of the perturber,~\cite{bib:Lagattuta,bib:Vegetti3} had to set up different models and probe the multi-dimensional parameter space of each model within given prior~ranges.  

Employing the approach outlined in Section~\ref{sec:cc_giant_arc} to estimate the mass of a potential perturber, we~first fitted a circle to the Einstein ring of the HST data and obtained $r_\mathrm{E} = 0.45''$.
This agrees well with $r_\mathrm{E} = 0.465''$, as~determined in~\cite{bib:King2}.  
In~\cite{bib:Vegetti3}, a lower value of $3.39~\mathrm{kpc}$ was used, which corresponds to $0.444''$~at $z_\mathrm{l}=0.88$, assuming a flat $\Lambda$CDM with $\Omega_m = 0.25$, $\Omega_\Lambda=0.75$, $H_0 = 73$~km/s/Mpc.
Fitting~two circles to the upper westward and lower eastward arc, denoted as image $A$ and $B$, respectively (see~\mbox{Figure~\ref{fig:B1938} (left)}), $r_\mathrm{A} = (0.50 \pm 0.1)''$ and $r_\mathrm{B} = (0.41 \pm 0.1)''$ were obtained.
Since the signal-to-noise ratio is to low to resolve brightness features in the arcs like the ones observable in the multiple images in CL0024, Equation~\eqref{eq:ce2} cannot be applied to this case.
Assuming that the perturber is a satellite galaxy with a singular isothermal sphere (SIS) as mass-density profile close to a SIS main lens, we inserted the $\boldsymbol{\alpha}_\mathrm{p}(\boldsymbol{x})$ and $\kappa_\mathrm{a}(r_\mathrm{E})$ for a SIS (see e.g.,\@ \cite{bib:SEF}) into Equation~\eqref{eq:ce1}.
Solving for the Einstein radius of the perturber, we obtained
\begin{equation}
m_\mathrm{p} = \left( 5.89^{+ 2.91}_{-2.30} \right) \cdot 10^8 \, \mathrm{M}_\odot
\label{eq:mass_Wagner}
\end{equation}
as a mass estimate for the perturber, if~it is optimally aligned, i.e.,\@ located on the line connecting the centre of light of the giant arc and the symmetry centre of the lens in the lens plane.
Comparing Equations~\eqref{eq:mass_Vegetti} and \eqref{eq:mass_Wagner}, we find that both are of the same order of~magnitude. 

Summarising the results for this example, as~further detailed in~\cite{bib:Wagner3}, we showed that our simple Equations~\eqref{eq:ce1} and \eqref{eq:ce2} to relate lens properties with observables allow us to infer lens properties that are in agreement with sophisticated lens reconstruction algorithms like~\cite{bib:Vegetti}, which implement more realistic classes of main lens and perturbing mass-density profiles. 
In addition, Equations~\eqref{eq:ce1} and \eqref{eq:ce2} clearly show the most general transformations between the perturber and the main, axisymmetric lens that leave the observables of the giant-arc brightness profiles invariant.
Similar equations could be analogously derived for main-lens mass profiles with other symmetries (e.g.,\@ ellipses).
However,  as~also noted in~\cite{bib:Vegetti2}, without~additional non-lensing information on the properties of the perturber, like its location (on the sky and also along the line of sight) and its brightness profile, the~properties of the potential perturbers remain highly~degenerate.

\subsection{Lensing Distance Ratios from the Pantheon Sample of~Supernovae}
\label{sec:app_distances}

As outlined in Section~\ref{sec:distances} and detailed in~\cite{bib:Wagner5}, we used the Pantheon sample of SNe type Ia of~\cite{bib:Scolnic} and $H_0$ from~\cite{bib:Planck2015} to set up data-based angular diameter distances for the lensing distance ratio in Equation~\eqref{eq:abbreviations}. 
We abbreviate the lensing distance ratio by
\begin{equation}
D(z_\mathrm{l},z_\mathrm{s}) \equiv \dfrac{D_\mathrm{A}(0,z_\mathrm{s}) D_\mathrm{A}(0,z_\mathrm{l})}{D_\mathrm{A}(z_\mathrm{l},z_\mathrm{s})} \;.
\label{eq:lensing_dr}
\end{equation}

To investigate the precision of the data-based $D(z_\mathrm{l},z_\mathrm{s})$ as established with the Pantheon sample, we implemented a Monte Carlo simulation of 1000 Pantheon-like data sets, each containing 1048~simulated, scale-free luminosity distances at the same redshifts as the Pantheon sample. 
Each of the 1048 data points was generated by drawing a random number from a Gauss distribution around the measured scale-free luminosity distance from the Pantheon sample with a standard deviation given by its measurement uncertainty. 
With these simulated data sets, the~confidence bounds on $D(z_\mathrm{l},z_\mathrm{s})$ can be determined for any lens and source redshift in the range of the Pantheon sample ($z \le 2.3$), given a value for $H_0$.
As confidence bounds, we calculated the standard deviation around the mean $D(z_\mathrm{l},z_\mathrm{s})$ and the confidence bounds based on percentiles, given by the $68\%$, $95\%$, and~$99\%$ confidence levels. 
In principle, the~precision of $D(z_\mathrm{l},z_\mathrm{s})$ also depends on $H_0$ and its uncertainty.
However,  due to the still ongoing debate over the tension in the value for $H_0$, \cite{bib:Planck,bib:Riess}, we only considered the relative precision of $D(z_\mathrm{l},z_\mathrm{s})$ and did not include the imprecision of $H_0$ in its~uncertainty. 

To compare the precision of the data-based $D(z_\mathrm{l},z_\mathrm{s})$ with the one obtained from the $\Lambda$CDM standard cosmological model as determined by~\cite{bib:Planck}, we developed a second Monte Carlo simulation. 
We drew 1000 different cosmological models from a Gaussian distribution around the cosmological parameters as determined by~\cite{bib:Planck2015}, $\Omega_m = (0.3089 \pm 0.0062)$, $\Omega_\Lambda = (0.6911 \pm 0.0062)$, with~their measurement uncertainty as standard deviation\footnote{As we consider the late-time cosmology with $z \in \left[ 0, 2.3\right]$, $\Omega_r = 0$ in Equation~\eqref{eq:E} to good approximation.}.
For each of the 1000 cosmological models, we determined model-based luminosity distances at the redshifts of the Pantheon sample, given $H_0 = 67.74$~km/s/Mpc. 
With this catalogue of 1000 data sets containing 1048 data points each, we~analogously calculated the relative precision of $D(z_\mathrm{l},z_\mathrm{s})$, leaving out the imprecision of $H_0$. 

As $D(z_\mathrm{l},z_\mathrm{s})$ is dependent on $z_\mathrm{l}$ and $z_\mathrm{s}$, the~best visualisation is obtained when fixing $z_\mathrm{l}$ and plotting the relative imprecision of $D(z_\mathrm{l},z_\mathrm{s})$ for all $z_\mathrm{s}$, as~shown in Figure~\ref{fig:D_zl}. 
In addition to the comparison at $z_\mathrm{l}=0.5$ established in~\cite{bib:Wagner5}, we also show the relative imprecisions for $D(z_\mathrm{l},z_\mathrm{s})$ at $z_\mathrm{l}=0.25$ and $z_\mathrm{l}=1.0$ as given by the confidence bounds of the simulated data~sets. 

By comparing the left- and the right-hand side of Figure~\ref{fig:D_zl}, we find that there are one to two orders of magnitude difference in the imprecision for the data-based (left) and Planck-model-based lensing distance ratio (right). 
The precision to which the data-based lensing distance ratio is determined is constrained by the measurement uncertainties of the SNe and the choice of the set of basis functions. 
Among the analytic sets of basis functions that we tested in~\cite{bib:Wagner5}, the~Einstein--de-Sitter basis with four basis functions was found to yield the tightest confidence bounds and the most accurate reconstruction of a flat $\Lambda$CDM model.
Its inaccuracies, determined in a simulation of a flat $\Lambda$CDM model, are contained within the confidence bounds of the imprecisions to avoid biases.
In addition, the~number of basis functions was determined in a $\chi^2$-parameter estimation with a reduced $\chi^2$-value of 0.9 to avoid overfitting.
Details about the quality test of all basis functions are given in~\cite{bib:Wagner5}. 
With further SNe data or a higher measurement precision, the~confidence bounds can be further reduced and the number of basis functions~increased.

\begin{figure}[H]
\centering
\includegraphics[width=0.43\textwidth]{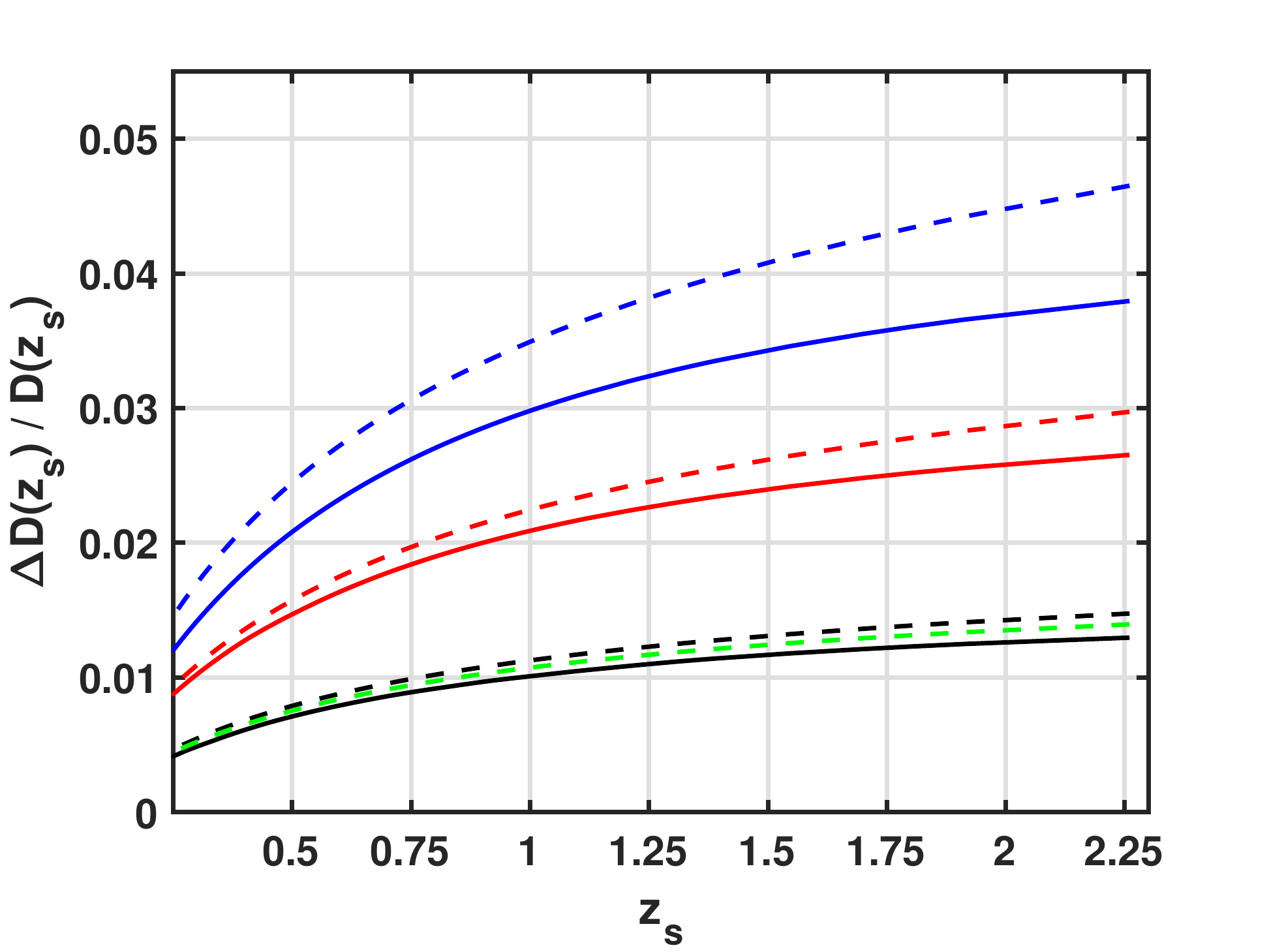} \hspace{-4ex}
\includegraphics[width=0.43\textwidth]{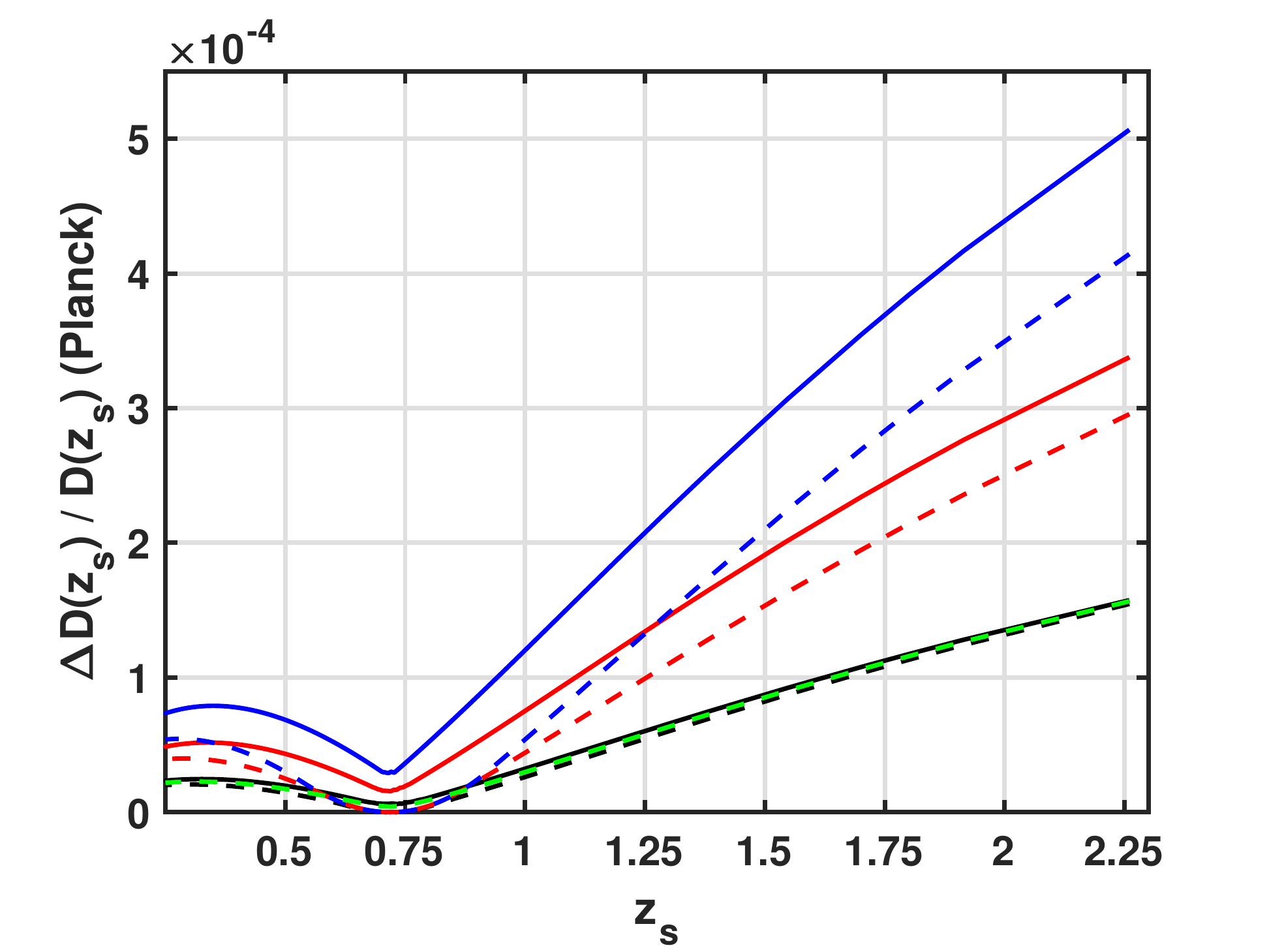} \hspace{-3ex}
\raisebox{10ex}{\includegraphics[width=0.19\textwidth]{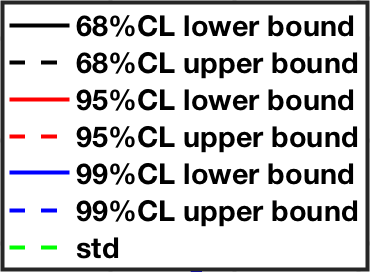}}
\caption{\textit{Cont}.}
\label{fig:D_zl}
\end{figure}   

\begin{figure}[H]\ContinuedFloat
\centering
\includegraphics[width=0.43\textwidth]{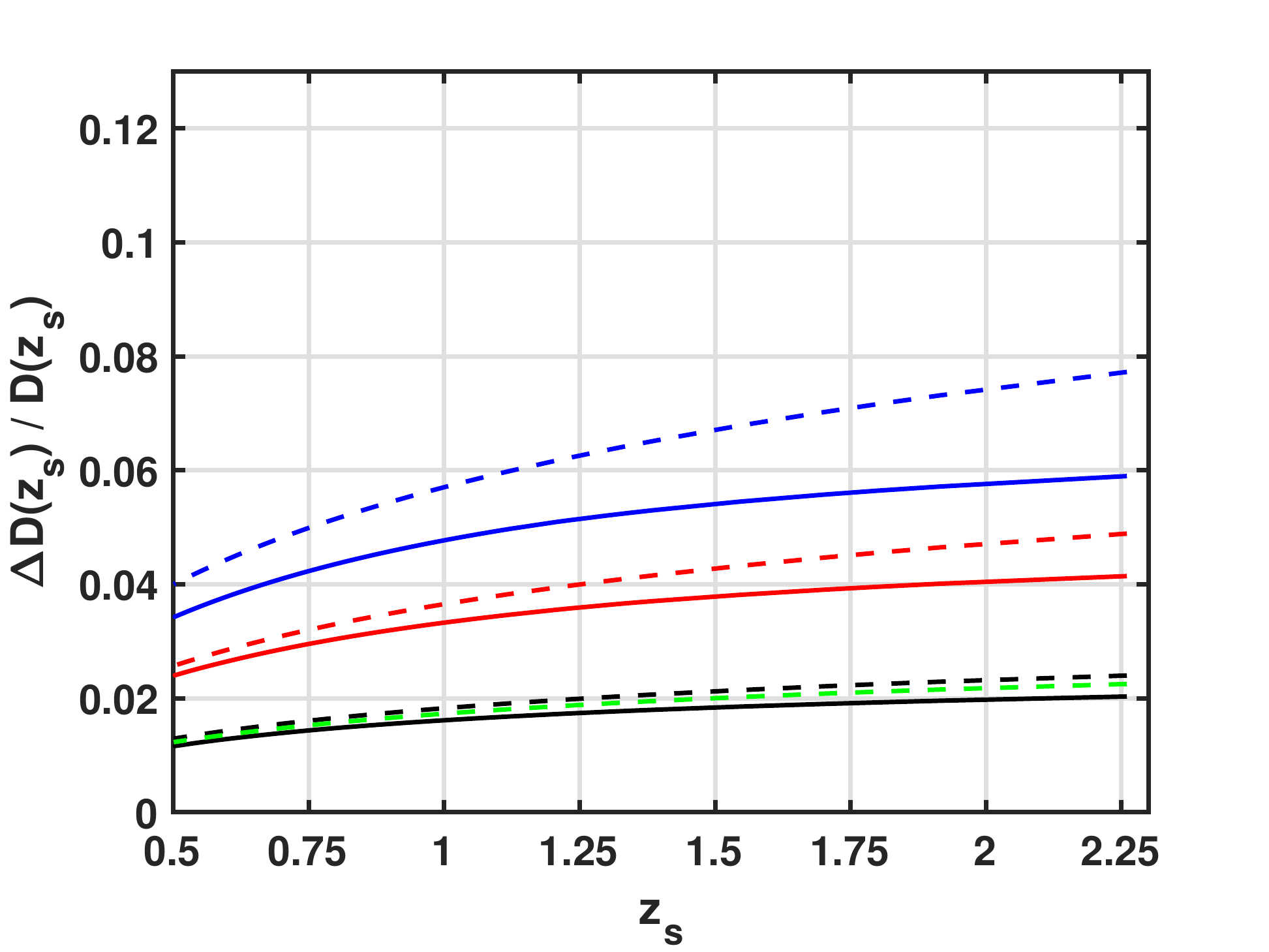} \hspace{-4ex}
\includegraphics[width=0.43\textwidth]{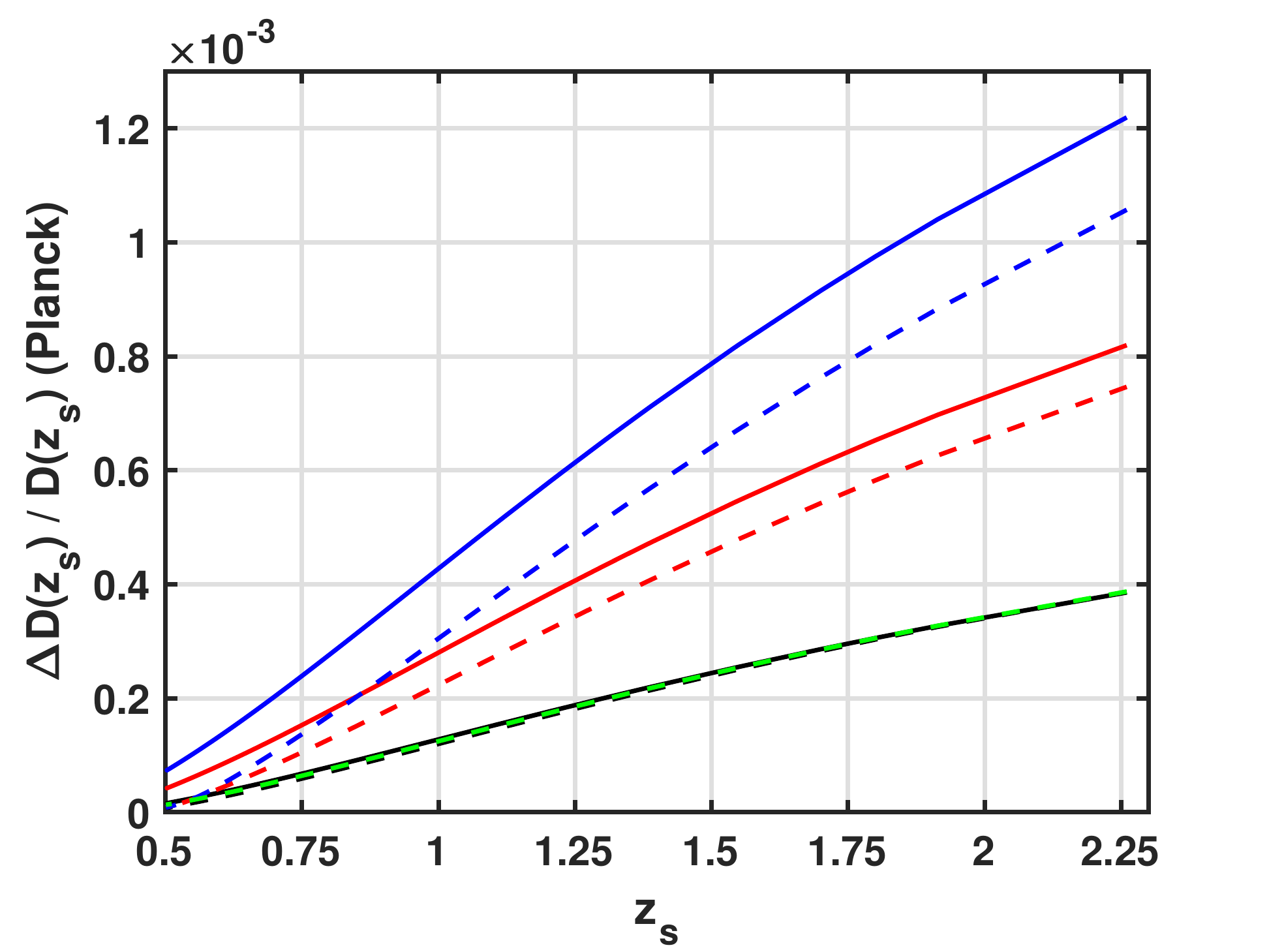} \hspace{-3ex}
\raisebox{10ex}{\includegraphics[width=0.19\textwidth]{distances_legend.png}}  \\
\includegraphics[width=0.43\textwidth]{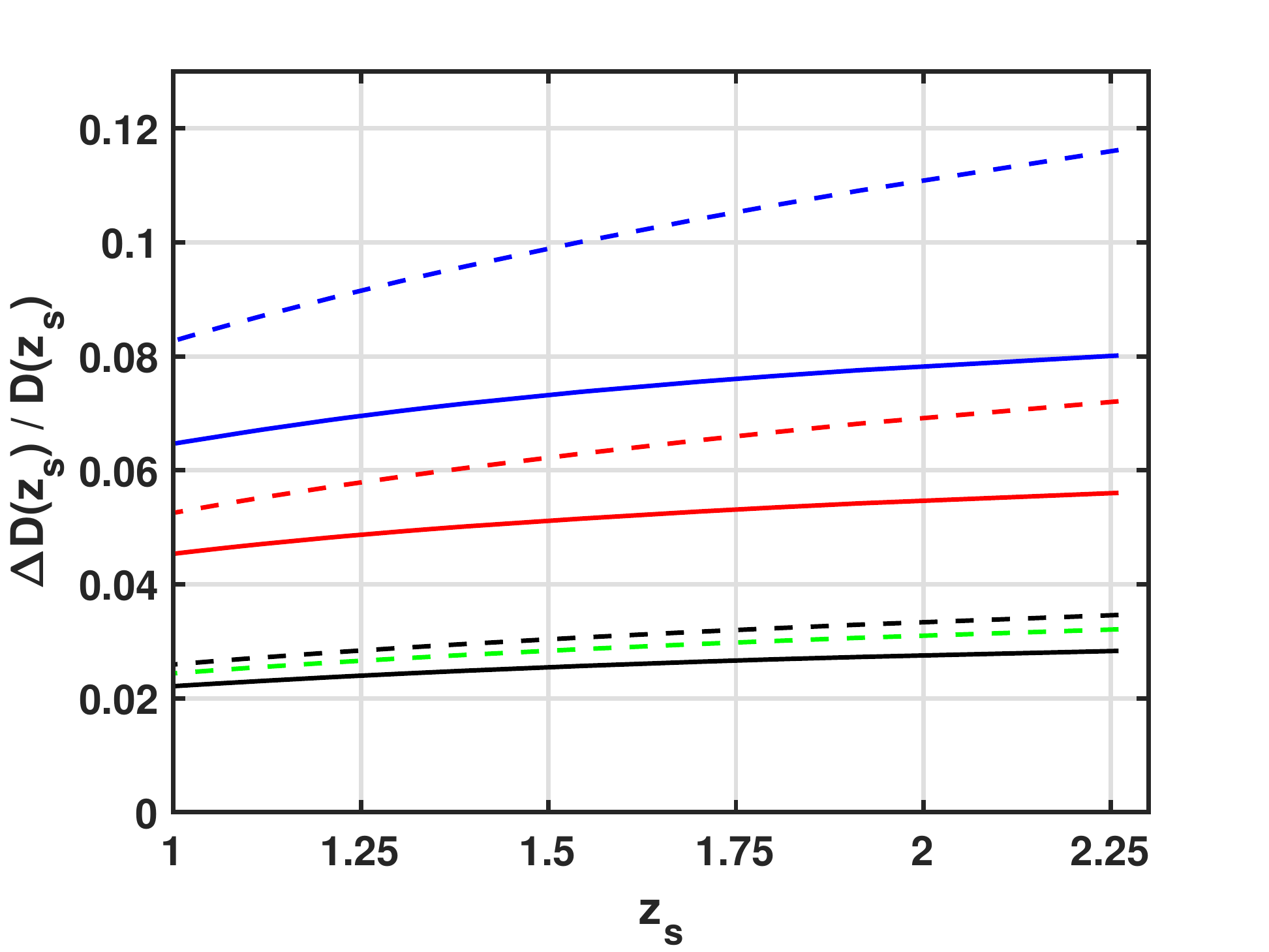} \hspace{-4ex}
\includegraphics[width=0.43\textwidth]{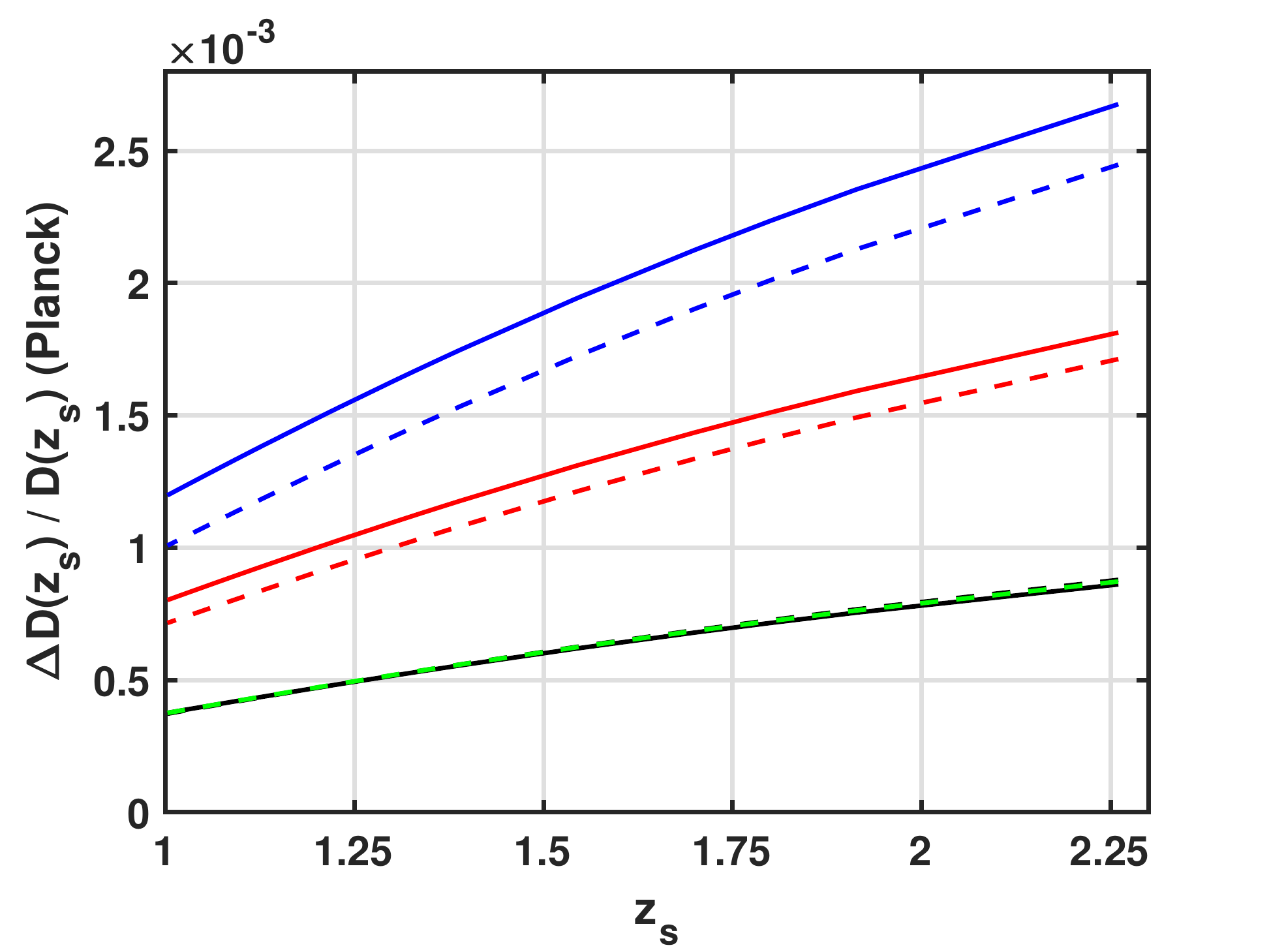} \hspace{-3ex}
\raisebox{10ex}{\includegraphics[width=0.19\textwidth]{distances_legend.png}} 
\caption{Comparison between the relative imprecision of the data-based lensing distance ratio (\textbf{left}) and the relative imprecision of the Planck-model-based distance ratio (\textbf{right}) for lens redshifts $z_\mathrm{l}=0.25$ (\textbf{top row}); $z_\mathrm{l}=0.5$ (\textbf{centre row}); $z_\mathrm{l}=1.0$ (\textbf{bottom row}). $D(z_\mathrm{s})$ is the lensing distance ratio defined in Equation~\eqref{eq:lensing_dr} for fixed  $z_\mathrm{l}$. Note the difference in the orders of magnitudes on the~ordinate.}
\label{fig:D_zl}

\end{figure}
%

The confidence bounds of the lensing distance ratio in a Planck-model-based cosmology are tighter and negligible compared to other sources of uncertainty in the time-delay equation (see Section~\ref{sec:app_H_0}). 
However,  the~tighter confidence bounds come at the cost that the model may be currently incomplete or it may become biased for an increasing amount of future data.
While the extensions of the $\Lambda$CDM model that can be included to amend the bias have a direct physical interpretation, e.g.,\@ an additional curvature term or any forms of dark energy, it is not clear which extension approach should be followed. 
Contrary to that, an~extension of the Einstein--de-Sitter basis with four basis functions is straightforward.
Assigning the basis functions a physical interpretation is the difficult part in this~approach.  

\subsection{Usage of the Time-Delay Equation to Infer $H_0$}
\label{sec:app_H_0}

As shown in~\cite{bib:Wagner6}, $H_0$ is uniquely determined by a time-delay-difference measurement in a FLRW background cosmology if the total integrated deflecting mass along the two light paths is known.
However,  observations to tightly constrain the total mass are rare, especially if we include the small-scale perturbing masses not associated with the most massive deflecting object along a light path.
In addition, parametrisations of the FLRW background cosmology are still subject to debates about potentially required extensions to the current concordance $\Lambda$CDM standard model (see Section~\ref{sec:app_distances}).
Therefore, current approaches to determine $H_0$ first solve the time-delay equation
\begin{equation}
\tau_{ij} = \dfrac{(1+z_\mathrm{l})}{c} D(z_\mathrm{l},z_\mathrm{s}) \left( \phi(\boldsymbol{x}_i, \psi) - \phi(\boldsymbol{x}_j,\psi) \right) \;,
\label{eq:time_delay}
\end{equation}
for the lensing distance ratio $D(z_\mathrm{l},z_\mathrm{s})$, Equation~\eqref{eq:lensing_dr}. For~$\tau_{ij}$, the~observed time-delay difference between light rays arriving from the multiple images $i$ and $j$ is inserted and $\phi_{ij} \equiv \phi(\boldsymbol{x}_i, \psi) - \phi(\boldsymbol{x}_j,\psi)$ is derived from a lens model. 
Since $D(z_\mathrm{l},z_\mathrm{s})$ contains the full information about the cosmological geometry of the multiple-image configuration, subsequently inferring $H_0$ from it conveniently allows to investigate different parametrisations of the FLRW background cosmology without repeating the lens~reconstruction.

In practice, reconstructions of $\phi_{ij}$ are subject to degeneracies. 
In our approach as outlined in Section~\ref{sec:materials}, a~local MSD at $\boldsymbol{x}_i$ and $\boldsymbol{x}_j$ remains to be broken, i.e.,~$\phi_{ij}$ between the two light rays can be scaled by a factor and $H_0$ can be rescaled accordingly to keep the observed time delay difference invariant. 
As, by~definition, $H_0$ is linked to the overall energy density of our universe today, this freedom of rescaling can be interpreted as redefining the split between the cosmological background energy density and the deflecting mass density of gravitational lensing on top of it. 
Other approaches usually split the total mass along the line of sight into a main deflecting mass and small-scale perturbers along the line of sight.
In this case, the~factor $\lambda$ appearing in an MSD consists of two parts: one part that can be associated with the small-scale perturbers and neighbouring masses of the main deflector, which is usually called external convergence $\kappa_\mathrm{ext}$, see e.g.,~\cite{bib:Suyu2}. 
The second part of the factor is the remaining MSD far away from the lens, in~the region where only the embedding cosmological background density remains. 
This~explains the argument stated in~\cite{bib:Schneider2} that the MSD cannot be solely caused by $\kappa_\mathrm{ext}$.

On the galaxy-cluster scale,~\cite{bib:Williams2} systematically investigated the relative precision to which $H_0$ can be determined from the well-resolved system of multiple images of the host galaxy of SN Refsdal and a plethora of multiple-image configurations from different background sources at different redshifts alone and in combination with different non-lensing observables and assumptions. 
They found that a combination of a flexible free-form with a consistent parametric lens reconstruction is capable of alleviating all occurring degeneracies to obtain $H_0$ to a relative imprecision of less than 10\%. 
The~precision of the inferred $H_0$-value greatly depends on the configuration of the multiple-image sets, the~measured time-delay differences but even more on the additional assumptions being made.

Galaxies as lenses are usually described by comparably simple models, yet the~high symmetry and the sparsity of multiple-image configurations also require additional non-lensing observables, like the velocity dispersions along the line of sight, to~be included in the model, see e.g.,\@ \cite{bib:Barnabe}.
On~both scales, potential biases due to the choice of a specific model (class) and degeneracies between the model parameters remain, see~\cite{bib:Wagner6} and references therein. 
To further decrease the imprecision in $H_0$, several multiple-image configurations with measured time delay differences can be jointly analysed, as~done e.g.,\@ in~\cite{bib:Saha3,bib:Oguri}, or~\cite{bib:Li}.


Having obtained a value for $D(z_\mathrm{l},z_\mathrm{s})$, the~definition of model-based angular diameter distances, Equation~\eqref{eq:D_A} usually using Equation~\eqref{eq:E} (or $E(z)$ for an extension of the $\Lambda$CDM model), is inserted to relate $D(z_\mathrm{l},z_\mathrm{s})$ to $H_0$. 
To determine $H_0$ from the time-delay equation in approaches based on Bayesian parameter inference, see e.g.,\@ \cite{bib:Grillo,bib:Bonvin}, parameters of the expansion function are either fixed, as~in Section~\ref{sec:app_distances}, or~considered as nuisance parameters and are marginalised over. 
Marginalising implies that the imprecision of $D(z_\mathrm{l},z_\mathrm{s})$ for Planck-model-based distances, is larger than the ones shown in Figure~\ref{fig:D_zl}. 
As found in~\cite{bib:Bonvin}, the~currently lowest imprecision for $H_0$ inferred from a single galaxy-scale lens is 3--4\% at 68\% confidence level. 
On galaxy-cluster level, the~lowest imprecision for $H_0$ is 4--5\% at 68\% confidence level obtained from the multiple images of SN Refsdal in MACS~1149,~\cite{bib:Grillo}.

To estimate the precision to which $H_0$ can be theoretically determined with a single time-delay-difference measurement when using the Pantheon-data-based distances, we rearrange Equation~\eqref{eq:time_delay}, such that
\begin{equation}
H_0 = \dfrac{\tilde{\Gamma}}{\tau_{ij}} \phi_{ij} 
\end{equation}
with $\tilde{\Gamma} = H_0 \Gamma$ (see Equation~\eqref{eq:abbreviations}). 
Then, we determine the relative imprecision of $H_0$ as
\begin{equation}
\dfrac{\Delta H_0}{H_0} = \sqrt{\left( \dfrac{\Delta D(z_\mathrm{l},z_\mathrm{s})}{D(z_\mathrm{l},z_\mathrm{s})} \right)^2 + \left( \dfrac{\Delta \tau_{ij}}{\tau_{ij}} \right)^2 + \left( \dfrac{\Delta \phi_{ij}}{\phi_{ij}} \right)^2}
\end{equation}
assuming that the relative imprecision in $\tilde{\Gamma}$ is dominated by the imprecision in $D(z_\mathrm{l},z_\mathrm{s})$ and the measurement uncertainty in $z_\mathrm{l}$ is negligible. 
From Figure~\ref{fig:D_zl}, we read off the relative imprecision of the data-based lensing distance ratio of about 2\% at 68\% confidence level. 
Constraints on the second term for different time-varying sources in galaxy-cluster-scale lensing can be found in~\cite{bib:Wagner_frb}. 
For a quasar, we~found a relative imprecision in the time delay difference of 1.5\%. 
In~\cite{bib:Liao} for quasars multiply-imaged by galaxy-scale lenses, it is stated that 3\% imprecision are obtained by the best algorithms.
For FRBs, the~second term is negligible with absolute measurement imprecisions on the order of milli-seconds compared to time delay differences of the order of days or~longer.

However,  by~far the largest source of uncertainties is the third term, see e.g.,\@  \cite{bib:Williams2,bib:Saha2,bib:Birrer2}. 
A systematic investigation for lens-modelling accuracy and precision on galaxy scale is currently being pursued by~\cite{bib:Ding}.
The work in Ref.~\cite{bib:Li} gives estimates for the accuracy and precision of models potentially achievable with HST in the 160W band of the Wide Field Camera 3 for exposure times of about 10,000~s and an astrometric precision of the order of sub-milliarcseconds in the angular positions, assuming the time-varying source is an FRB.
Including the line-of-sight external convergence to the lens modelling, their optimum estimate of relative imprecisions for the difference in the Fermat potential is about 2--3\%.  
Estimating an imprecision of 5\% for the difference of the Fermat potential\footnote{The reconstruction of $\phi_{ij}$ might include angular diameter distances between us, the~source, and~the lens, which correlates the imprecisions between $\phi_{ij}$ and $D(z_\mathrm{l},z_\mathrm{s})$. For~the following estimates, these correlations are ignored.} for currently available data, we obtain
\begin{align}
\left(\tfrac{\Delta H_0}{H_0}\right)_{\mathrm{cluster}} &= \sqrt{\left( 0.02 \right)^2 + \left( 0.015 \right)^2 + \left( 0.05 \right)^2} \approx 5.6\%\;, \\
\left(\tfrac{\Delta H_0}{H_0}\right)_{\mathrm{galaxy}} &= \sqrt{\left( 0.02 \right)^2 + \left( 0.030 \right)^2 + \left( 0.05 \right)^2} \approx 6.1\% \;.
\end{align}

At the current measurement and lens-reconstruction precision, the~imprecision of the Pantheon-data-based angular diameter distances is of the same order of magnitude as the measurement uncertainty of the time-delay-difference and smaller than the imprecision caused by the lens reconstruction.
Hence, the~cost for avoiding a parametrisation of $E(z)$ in terms of specific $\Omega_i$ is an increase in the imprecision in $H_0$ up to a factor of two for galaxy-scale lenses and by a factor of 1.4~for galaxy-cluster-scale lenses. 
With the currently increasing number of observations of multiple-image configurations with measured time-delay-differences, this disadvantage can be compensated with a joint analysis of several~configurations. 

Vice~versa, solving Equation~\eqref{eq:time_delay} for $\phi_{ij}$ for a given $H_0$, we found that the differences in the Fermat potential can be determined up to 2.7--3.7\%, if~$H_0$ can be measured with a relative imprecision of 1\%. 
We showed in~\cite{bib:Wagner_frb} by means of a simulation that even a single time-delay-difference observation can greatly increase the accuracy in the global lens reconstruction of a galaxy-cluster-scale lens, as~it fixes the global~MSD.

\section{Discussion and Conclusions}
\label{sec:discussion}

Thus far, constraints from observables of multiple images on the total deflecting mass-density distribution only sparsely cover the area of the latter. 
Attempting to reconstruct the total deflecting mass-density distribution in the entire lensing region on the sky is thus an under-constrained optimisation problem.
To solve it, additional assumptions are usually employed which capture the global shape and the overall smoothness of the deflecting mass-density distribution.
One way to capture the shape uses physically motivated parametric models of the mass-density profile, like a Navarro--Frenk--White profile, or~a superposition of parametric mass-density profiles.
Another common way decomposes the deflecting mass density into a set of orthonormal basis functions and determines the weights of the leading basis functions by means of the observables.
In~both methodological approaches, a~regularisation has to be introduced to avoid overfitting. 
The regularisation also sets the smoothness of the mass-density distribution. 
In Section~\ref{sec:introduction}, we gave examples of such approaches for galaxy-scale and galaxy-cluster-scale~lenses.

Contrary to these \emph{global} ansatzes that treat the reconstruction of the deflecting mass-density distribution as a global model fit and intertwine data-based evidence and model assumptions, we develop an approach to \emph{locally} characterise strong gravitational lenses solely by observables of the multiple images that they create. 
Our approach exploits the symmetries and correlations between the observables in the multiple images and thereby avoids introducing specific lens models. 
During its development, we were able to mathematically derive all degeneracies, like~the mass-sheet degeneracy, treating the single-lens-plane gravitational-lensing formalism as partial differential equations. Subsequently, we found a physical interpretation for the respective invariance transformations that explains why they~appear.

As our method does not reconstruct the global underlying mass density, it can be applied to any multiple-image configuration on any lens scale in the same way.
Thus far, we have employed relative distances between multiple images, features in the brightness profiles of resolved multiple images, multipole moments of the brightness profiles up to the quadrupole of unresolved multiple images, and~time delay differences, if~the source has a time-varying component. 
From these observables, we~calculate \emph{local} lens properties that are not subject to any degeneracies, if~time delay differences in a fixed background metric are used. 
For static sources, the~local equivalent of a mass-sheet degeneracy remains to be broken.
This is a consequence of setting up the standard gravitational lensing formalism as a theory of light deflection caused by inhomogeneities on top of a cosmic background metric and only observing angular quantities on the celestial~sphere.

To leading order, our approach determines ratios of deflecting mass densities, the~reduced shear, and~the linear and parabolic approximations of the critical curves in the vicinity of the multiple images. 
We showed the respective equations in Section~\ref{sec:materials}.
From simulations, we found inaccuracies of the order of and less than 10\% for extreme cases of large deflecting masses and highly elliptical density profiles. 
For realistic deflecting mass densities, we expect them to be lower and in the range of a few percent, which will be investigated in N-body simulated gravitational lenses. 
Imprecisions for some observational cases were found to be much larger, but~the imprecision greatly depends on the type of multiple image and the positions of the multiple images with respect to the scaled mass density isocontour $\kappa(\boldsymbol{x}) = 1$ or to the isocontour of $\phi_{11}(\boldsymbol{x})=0$, depending on the parametrisation that is~chosen. 

The set of local lens properties is the maximum leading-order local information in which all lens models coincide and that is directly constrained by the observables in the multiple images in the standard single-lens-plane gravitational-lensing formalism.
Formulated differently, our~model-independent approach partitions the lensing region on the sky into areas where local lens properties are directly inferred from observables, and~the remaining areas where global lens reconstructions are forced to make assumptions about the mass-density distribution because no constraints from multiple images are available.
With the continuously increasing number of observed multiple-image configurations in galaxy-cluster-scale lenses, areas where global assumptions are necessary are shrinking. 
As a consequence, the~confidence bounds on the local lens properties of our approach will tighten. 
Hence, our data-driven approach will become increasingly useful to determine the mass-density distribution of galaxy clusters.
As multiple images of different sources rarely overlap, the~local lens properties for newly discovered multiple-image sets can be easily added to the already known set of multiple images and their local lens characterisations. 
It is particularly advantageous for galaxy-cluster-scale lenses with their highly complex mass-density distribution to solicit properties of the deflecting mass distribution at certain positions that do not rely on global symmetry assumptions and that do not jointly utilise the information of several multiple-image~configurations.

On galaxy-scale, the~increasing amount of observed resolved multiple images of quasars in several radio bands reveals a more asymmetric mass-density distribution of their lenses than parametric, symmetric lens models can provide, so that the local lens properties determined by our approach are useful to sketch the asymmetries in the mass-density distribution.
In addition, with~increasing measurement precision to extract the shapes of the individual parts of the quasar, applying our approach on the scale of a quasar-image and on the scale of the individual parts of a quasar-image, small-scale lens properties may become~detectable.

Beyond constraining local lens properties, comparing the model-independent approach to global lens reconstructions corroborates or refutes the global assumptions that are employed in the lens reconstructions---for instance, the~hypothesis of whether luminous matter traces dark matter can be checked.
Alternatively, our local lens properties can be integrated as constraints in the global lens reconstructions, so that global lens reconstructions can be envisioned that extrapolate the local lens properties, i.e.,\@ the information that all models agree upon, into~areas where no multiple images are observed. 
Compared to existing approaches that directly intertwine observational evidence with model assumptions, the~extrapolating global lens reconstructions could separate the evidence from the assumptions. 
In this way, updating existing lens models with new observations becomes very efficient. 
In contrast, if~some of the multiple images are not spectroscopically confirmed, the~global non-extrapolating lens reconstruction has to be~redone.

In Section~\ref{sec:results}, we provided proof-of-principle example applications.
They showed that our method determines local lens properties for galaxy-scale and galaxy-cluster-scale lenses and in optical and radio bands alike.
For the galaxy-cluster-scale example, CL0024, we found that our approach obtained local lens properties in agreement with the local lens properties obtained by two different global lens reconstructions.
For the galaxy-scale example, B0128, this comparison is still outstanding at the time of writing, but~we already found that our approach and a third global lens reconstruction algorithm consistently hinted at an oversimplified parametric lens model that was assumed in previous~works.

On the galaxy-cluster scale, we found that the model-independent approach is also of great use for the reconstruction of the source from the observables of the respective multiple images. 
Our approach first determines the local lens properties in the vicinity of the multiple images and then infers the source by a local back-projection of the multiple images. 
Hence, the~reconstructed source does not depend on a lens model. 
This enables us to study the morphologies of faint and distant galaxies to high redshifts free of potential biases due to a lens model.

For galaxy-scale lenses that show giant arcs, we could establish a simple set of equations that clearly show the degeneracies between an axisymmetric main lens and a smaller perturbing mass density. 
These equations explain the degeneracies between the main lens and the perturber which were previously discovered in simulations. 
Furthermore, they show that most of the observed examples of perturbed, symmetric giant-arc image configurations do not allow us to simultaneously constrain the mass-density distribution in the main lens and properties of the perturber, like its position along the line of sight or its total mass. 
With Bayesian lens-modelling algorithms, the~most likely composition of the deflecting mass-density distribution can be determined. 
Prior to an elaborate modelling and Monte Carlo simulations, our set of model-independent equations is an efficient means to exclude classes of models and parameter sets that are incompatible with the~observations. 

Strong gravitational lenses are also used to infer the Hubble-Lemaître constant, $H_0$. 
As the ratio of angular diameter distances in the time-delay equation is dependent on the cosmological background model, calculating $H_0$ from this equation implies to either fix the other parameters of the cosmological model by complementary observations, or~marginalise over them. 
Both ways are difficult to realise in particular due to the unknown form of the parameter attributed to a cosmological constant or dark energy. 
To avoid this problem, we set up a ratio of angular diameter distances that is based on the standardisable supernovae type Ia from the recent Pantheon sample. 
The ratio of angular diameter distances is thus determined up to an overall scale factor, $H_0$, and~all distances (up to this overall scale) are expressed in terms of a basis-function decomposition. 
Purely data-based distance measures cannot be established due to the sparsity of the supernovae.
However,  the~flexible set of orthonormal, analytic basis functions captures  all characteristics of the data set. 
In addition, it avoids overfitting as the number of basis functions is limited by the measurement uncertainties of the supernovae.
Using these data-based angular diameter distances in the lensing time-delay equation, the~imprecision on $H_0$ only grows about a factor of two for the current measurement uncertainty and imprecision in the lens reconstruction compared to using angular diameter distances based on the currently best-fit cosmological standard model.
This imprecision can be compensated for by increasing the number of jointly evaluated lensing configurations.
If multiply-imaged fast radio bursts are observed in the future, they can replace the commonly used quasar sources, which will also yield an increase in the precision of $H_0$ inferred from time-delay difference~measurements.

On the whole, we have introduced a new approach to strong gravitational lensing with the aim to separate model-based assumptions from data-based evidence which is a complementary and useful tool to analyse large data sets in the most efficient and objective way. 
A detailed discussion of the results achievable with current observations can be found in Section~\ref{sec:results}.
We plan to further extend the approach including higher-order features of the multiple images and additional non-lensing~observables. 

\begin{figure}[H]
\centering
\includegraphics[width=\textwidth]{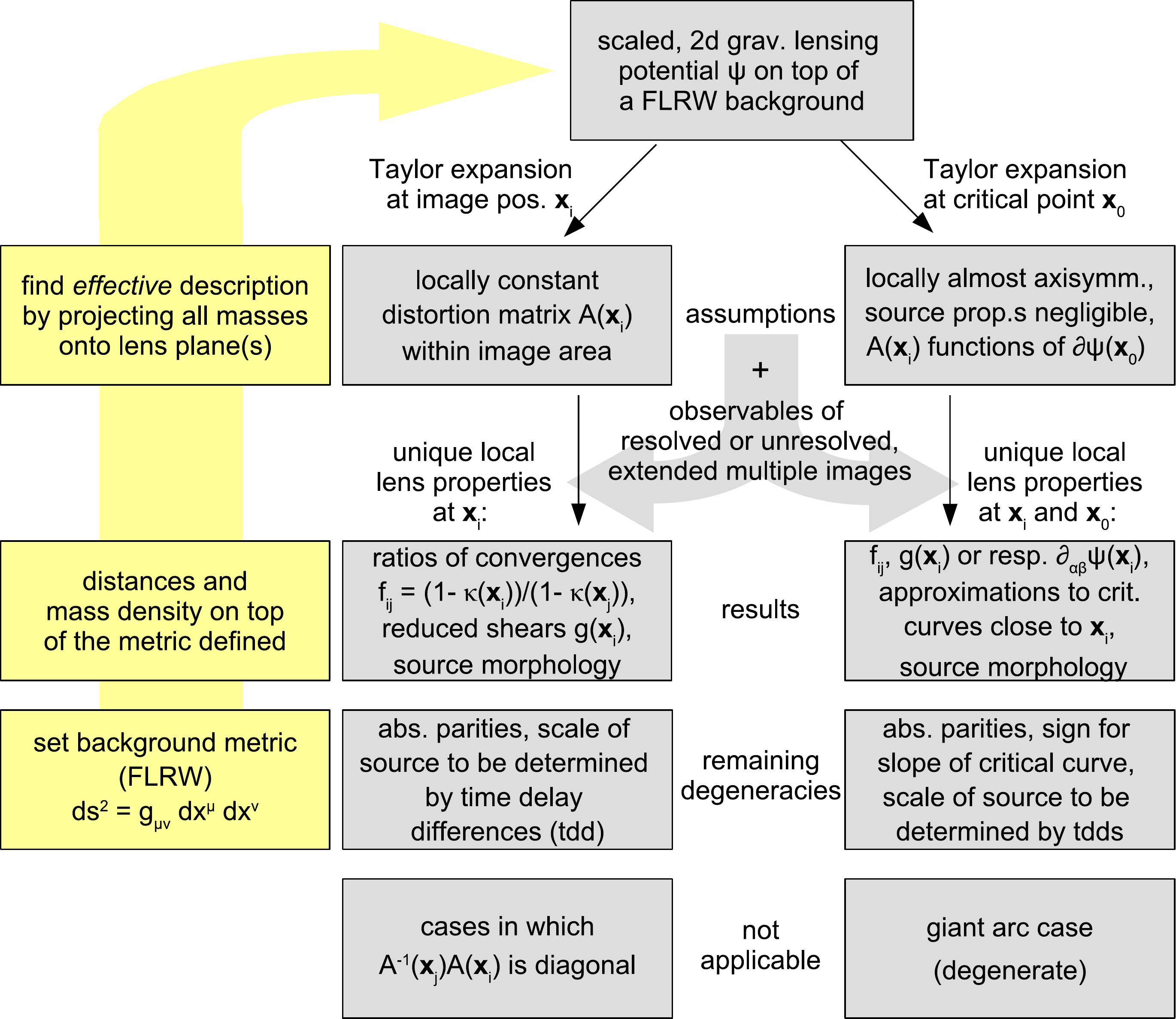}
\caption{Summary of our approach as outlined in Section~\ref{sec:materials}, its prerequisites (yellow boxes), working~principles and resulting lens properties (grey boxes). Each result also has its remaining degeneracies and there are cases to which the approach cannot be~applied.}
\label{fig:summary}
\end{figure}

Figure~\ref{fig:summary} summarises our approach (grey boxes) with its prerequisites (yellow boxes) as outlined in Section~\ref{sec:materials}. 
It shows the underlying working principle and the resulting local lens properties that can be obtained from the single-lens-plane standard gravitational-lensing formalism without using a specific lens model. 
The remaining degeneracies and the way to break them are also listed in addition to those cases to which the algorithm cannot be~applied.

\vspace{6pt} 




\funding{This research has been funded by the Deutsche Forschungsgemeinschaft (DFG) Grant Nos. WA3547/1-1 and WA3547/1-3.}

\acknowledgments{I would like to thank everybody who contributed to this research project, especially my collaborators, colleagues, friends, and~anonymous referees for stimulating discussions, their continuing support, and~contributions to develop this new approach to gravitational lensing. I highly appreciate being a part of the very friendly and supportive gravitational lensing community and the continuous funding by the~DFG.}

\conflictsofinterest{The author declares no conflict of~interest.}

%


\abbreviations{The following abbreviations are used in this manuscript:

\noindent 
\begin{tabular}{@{}ll}
$A(\boldsymbol{x})$ & distortion matrix around point $\boldsymbol{x}$ in the lens plane \\
$\boldsymbol{\alpha}(\boldsymbol{x})$ & deflection angle in the lens plane \\
$c$ & speed of light \\
$D(z_\mathrm{l}, z_\mathrm{s})$ & lensing distance ratio, s.\@ Equation~\eqref{eq:lensing_dr}\\
$D_\mathrm{A}(z_1, z_2)$ & angular diameter distance between two redshifts, s.\@ Equation~\eqref{eq:D_A} \\
$D_\mathrm{L}(z_1, z_2)$ & luminosity distance between two redshifts \\
$E(z)$ & cosmic expansion function dependent on redshift $z$ \\
$f_{ij}^{(\kappa)}$ & ratio of $1-\kappa(\boldsymbol{x})$ between multiple image $i$ and $j$, s.\@ Equation~\eqref{eq:kappa_fgs} \\
$f_{ij}^{(\phi)}$ & ratio of $\phi_{11}(\boldsymbol{x})$ between multiple image $i$ and $j$, s.\@ Equation~\eqref{eq:phi_fgs} \\
FLRW metric & Friedmann--Lemaître--Robertson--Walker metric \\
FRB & fast radio burst \\
$\boldsymbol{g}(\boldsymbol{x})$ & reduced shear at position $\boldsymbol{x}$, s.\@ Equation~\eqref{eq:kappa_fgs} \\
$\boldsymbol{\gamma}(\boldsymbol{x})$ & shear at position $\boldsymbol{x}$ \\
$\Gamma$ & ratio describing the lensing geometry along the line of sight, s.\@ Equation~\eqref{eq:abbreviations} \\
$H_0$ & Hubble--Lemaître constant \\
HST & Hubble Space Telescope \\
$\kappa(\boldsymbol{x})$ & convergence at position $\boldsymbol{x}$ \\
$\Lambda$CDM & cold-dark-matter model with cosmological constant $\Lambda$ \\
LTM & light traces mass (assumption) \\
MERLIN &  Multi-Element Radio Linked Interferometer Network \\
MIS & multiple-image set \\
MLR & mass-to-light ratio \\
MSD & mass-sheet degeneracy \\
NFW & Navarro--Frenk--White (lens model), s.\@ Appendix~\ref{appendix:NFW} \\
$\Omega_K$ & curvature density parameter of the $\Lambda$CDM at $z=0$ \\
$\Omega_\Lambda$ & cosmological constant parameter of the $\Lambda$CDM at $z=0$ \\
$\Omega_m$ & matter density parameter of the $\Lambda$CDM at $z=0$ \\
$\Omega_r$ & radiation density parameter of the $\Lambda$CDM at $z=0$ \\
$\phi(\boldsymbol{x},\psi)$ & Fermat potential, s.\@ Equation~\eqref{eq:abbreviations} \\
PSF & point spread function \\
$\psi(\boldsymbol{x})$ & projected Newtonian deflection potential in the lens plane\\
QSO & quasi-stellar object, quasar \\
SIE & singular isothermal ellipse (lens model), s.\@ Appendix~\ref{appendix:SIE}\\
SIS & singular isothermal sphere (lens model) \\
SN  & supernova \\
SNR & signal-to-noise ratio \\
VLA & Very Large Array \\
VLBI & very-long-baseline interferometry \\
$\boldsymbol{x}$ & positions in the lens plane \\
$\boldsymbol{x}_0$ & critical point (either a cusp or a fold singularity) \\
$\boldsymbol{x}_\mathrm{c}$ & cusp critical point \\
$\boldsymbol{x}_\mathrm{f}$ & fold critical point \\
\end{tabular}

\noindent 
\begin{tabular}{@{}ll}
$\boldsymbol{x}_{i\alpha}$ & reference point $\alpha$ in image $i$ \\
$\boldsymbol{x}_\mathrm{t}$ & expansion point for Taylor series of the Fermat potential \\
$\boldsymbol{y}$ & positions in the source plane \\
$\boldsymbol{y}_0$ & caustic point (either a cusp or a fold singularity) \\
$z_\mathrm{l}$ & redshift of the lens \\
$z_\mathrm{s}$ & redshift of the source object
\end{tabular}}

\appendixtitles{yes} 
\appendix
\section{The Fermat Potential at Critical Curves of Axisymmetric and Elliptical~Lenses}
\label{appendix:lens_model_examples} 

In this appendix, we calculate the derivatives of the Fermat potential at the critical points for Equation~\eqref{eq:taylor_series} for two example lenses.
Appendix~\ref{appendix:NFW} determines the derivatives of the Fermat potential at the fold critical radius of a Navarro--Frenk--White (NFW) lens model.
Subsequently, Appendix~\ref{appendix:SIE} shows the derivatives of the Fermat potential at the cusp critical point on the semi-minor axis of a singular isothermal elliptical (SIE) lens~model. 

\subsection{Axisymmetric~Lens}
\label{appendix:NFW}

The deflection potential for an NFW lens model in polar coordinates $(r,\varphi)$ is given by
\begin{equation}
\psi(r) = 4 \kappa_\mathrm{s} \left[ \tfrac12 \left( \log(\tfrac{r}{2}) \right)^2  - 2 \left( \text{artanh}\left( \sqrt{\tfrac{1-r}{1+r}}\right) \right)^2\right] 
\label{eq:NFW}
\end{equation}
as derived in~\cite{bib:Meneghetti2}. 
$\kappa_\mathrm{s}$ is a scaling convergence. $r$ is the scaled radius in the lens plane, given by the radial distance from the centre of the lens $\xi$, scaled by the scale radius $r_\mathrm{s}$.  
Thus, measuring the angular radius $x = \xi/D(0,z_\mathrm{l})$ on the sky, we obtain $r = x D(0,z_\mathrm{l})/r_\mathrm{s}$.

As a quantitative example, we insert $\kappa_\mathrm{s}=0.4$, such that the fold critical radius $r_\mathrm{f}=0.14$.
The~vanishing derivatives of the Fermat potential for this example are also zero for other values because $r_\mathrm{f}$ and thus the lens monotonically increase for increasing $\kappa_\mathrm{s}$ without changing the topology of the critical curves. 
We obtain
\begin{align}
\phi_{11}(\boldsymbol{x}_\mathrm{f}) &= -0.76 \;, \\
\phi_{112}(\boldsymbol{x}_\mathrm{f}) &= \phantom{-}5.52 \;, \quad  \phantom{1}\phi_{222}(\boldsymbol{x}_\mathrm{f}) = \phantom{-1}5.12 \;, \\
\phi_{1111}(\boldsymbol{x}_\mathrm{f}) &= 119.78 \;, \quad  \phi_{1122}(\boldsymbol{x}_\mathrm{f}) = -42.79 \;, \quad  \phi_{2222}(\boldsymbol{x}_\mathrm{f}) = -43.61 \;.
\end{align}

All other derivatives vanish. 
We insert these derivatives into Equation~\eqref{eq:taylor_series} and take into account that the image positions $r$ close to $r_\mathrm{f}=0.14$ are smaller than one and that the coordinate system can be chosen such that $x_2 \gg x_1$.  
As $r^n > r^{n+1}$ for $r < 1$, the~approximation shown in Table~\ref{tab:taylor_configs} can be used for fold configurations of multiple images created by an NFW mass-density~distribution.

\subsection{Elliptical~Lens}
\label{appendix:SIE}

According to~\cite{bib:Kormann}, an~SIE model for a gravitational lens is given by the deflection potential in polar coordinates $(r,\varphi)$
\begin{equation}
\psi(r,\varphi) = r \sqrt{\tfrac{f}{1-f^2}}  \left[ \left| \sin(\varphi) \right| \text{arccos}(\Delta(\varphi)) + \left| \cos(\varphi) \text{arcosh} \left( \tfrac{\Delta(\varphi)}{f}\right) \right| \right]
\label{eq:SIE}
\end{equation}
with $\Delta(\varphi) = \sqrt{\cos(\varphi)^2 + f^2 \sin(\varphi)^2}$ and the axis ratio $f$ between the semi-minor and the semi-major axis of the ellipse. 
As in Appendix~\ref{appendix:NFW}, the~radius is given by the radial distance to the lens centre scaled by a scale radius $r_\mathrm{s}$. 
We determine the derivatives of the Fermat potential for the cusp critical point on the $x_1$-axis.
This is the end point of the semi-minor axis of the critical curve.
With the same argumentation as in Appendix~\ref{appendix:NFW}, the~topology does not change for increasing $f$. 
The calculation for $f=0.65$ yields
\begin{align}
\phi_{11}(\boldsymbol{x}_\mathrm{c}) &=  \phantom{-}1.00\;, \\
\phi_{122}(\boldsymbol{x}_\mathrm{c}) &= \phantom{-}1.24 \;, \\
\phi_{1122}(\boldsymbol{x}_\mathrm{c}) &= -3.08 \;, \quad  \phi_{2222}(\boldsymbol{x}_\mathrm{c}) = 3.73 \;.
\end{align}

All other derivatives vanish. 
We insert these derivatives into Equation~\eqref{eq:taylor_series} and take into account that the image positions $r$ close to $r_\mathrm{c}=0.81$ are smaller than one and that the coordinate system can be chosen such that $x_2 \gg x_1$.  
As $r^n > r^{n+1}$ for $r < 1$, the~approximation shown in Table~\ref{tab:taylor_configs} can be used for cusp configurations of multiple images created by an SIE mass-density~distribution.

\section{Transformation of a Cusp Configuration into Its Special Coordinate~System}
\label{sec:appendix_cusp_rotation}

Let $\vartheta$ be the angle that the distance $\delta_{AB}$ has with the positive $x_1$-axis of the special coordinate system.
As derived in detail in~\cite{bib:Wagner1}, we can employ the lensing equations and approximations of the $A(\boldsymbol{x}_i)$, $i=A, B, C$, to~determine $\vartheta$ from the observables defined in Section~\ref{sec:observables}.
$\vartheta$ is obtained by (numerically) solving
\begin{equation}
\dfrac{-2}{\tan(\vartheta)} = \dfrac{(r_A-1) \tan \left( \vartheta + \tilde{\varphi}_A \right)}{r_A + \tan^2 \left( \vartheta + \tilde{\varphi}_A \right)} + \dfrac{(r_B-1) \tan \left( \vartheta + \tilde{\varphi}_B \right)}{r_B + \tan^2 \left( \vartheta + \tilde{\varphi}_B \right)} \;,
\end{equation}
where the orientation angles between the semi-major axis of images $A$ and $B$ are called $\tilde{\varphi}_A$ and $\tilde{\varphi}_B$, respectively (see Table~\ref{tab:image_configurations}). 
They are measured with respect to $\boldsymbol{\delta}_{AB}$. 

From Equation~\eqref{eq:cusp_A}, we can determine the coordinates of $\boldsymbol{x}_A$ with respect to $\boldsymbol{x}_\mathrm{c}=(0,0)$ in the coordinate system shown in Table~\ref{tab:taylor_configs} as
\begin{myequation}
\begin{array}{ll}
x_{A1} &= \dfrac{1}{\tilde{\phi}_{122}(\boldsymbol{x}_\mathrm{c})} \cdot \left( \tilde{\phi}_{22}(\boldsymbol{x}_A) - \dfrac12 \tilde{\phi}_{2222}(\boldsymbol{x}_\mathrm{c})x_{A2}^2\right) = \dfrac{1}{\tilde{\phi}_{122}(\boldsymbol{x}_\mathrm{c})} \cdot \left( \mp \dfrac{r_A + \cot^2(\varphi_A) }{r_A \cot^2(\varphi_A) + 1} - \dfrac12 \tilde{\phi}_{2222}(\boldsymbol{x}_\mathrm{c})\right) \;, \label{eq:xA1} 
\end{array}
\end{myequation}
\begin{equation}
\begin{array}{ll}
x_{A2} &= \dfrac{1}{\tilde{\phi}_{122}(\boldsymbol{x}_\mathrm{c})} \cdot \tilde{\phi}_{12}(\boldsymbol{x}_A) = \dfrac{1}{\tilde{\phi}_{122}(\boldsymbol{x}_\mathrm{c})} \cdot \dfrac{\cot(\varphi_A) (r_A - 1) }{r_A \cot^2(\varphi_A) + 1} \;. \label{eq:xA2}
\end{array}
\end{equation}

The sign in Equation~\eqref{eq:xA1} is chosen according to the parity of image $A$. 
The negative sign belongs to the positive cusp, i.e.,~image $A$ is a saddle point with negative parity. 
The positive sign belongs to the negative cusp with image $A$ of positive~parity.

\section{Brief Characterisation of \textsc{Lenstool} and \textsc{Grale}}
\label{appendix:lens_reconstructions}
\vspace{-6pt}

\subsection{\textsc{Lenstool}}
The version of \textsc{Lenstool} that we used reconstructs the mass-density distribution of the gravitational lens as a superposition of smooth, analytical, large-scale dark-matter halo profiles of a specific type previously set by the user. 
Apart from the dark halos, it takes into account a catalogue of luminous cluster member galaxies provided by the user. 
The member galaxies are all modelled with the same mass profile and their individual parameters are determined using the LTM assumption and the Tully--Fisher and Faber--Jackson scaling relations. 
The MLR of these member galaxies is a free parameter of the model to be set by the~user.

To constrain local lens properties by \textsc{Lenstool} with tight confidence bounds, using a lot of MISs from different sources scattered over the entire cluster area leads optimum results. 
Since the number of parameters to be determined from the observables is rather small (compared to the free-form reconstructions, as~e.g.,\@ \textsc{Grale}), the~quality of fit of such models to the observations shows how well the complexity of the data are captured employing the optimum model parameter~set.

\subsection{\textsc{Grale}}
\textsc{Grale} is a free-form lens reconstruction method, which makes no assumptions whether light traces mass or not. 
It describes the deflecting mass-density distribution (or the deflection potential) as a superposition of a number of basis functions. 
Their number, location, and~parameters are determined on a grid that is iteratively refined in regions of high deflecting mass density according to the following requirements.
As one optimisation criterion, called ``fitness measure'', the~overlap of all back-projected multiple images of a common source is maximised in the source plane.
Additionally, it employs a nullspace fitness measure, i.e.,\@ the reconstruction algorithm should not generate images in regions where no images have been observed. 
The third fitness measure that is implemented in \textsc{Grale}, but~not used in the analysis of CL0024, requires that no critical line intersects non-merging multiple images. 
A fourth one, also not used for CL0024, is the time-delay fitness measure that constrains the model-predicted time-delay differences by observed ones.
In order to determine the optimum number of basis functions and their weights, a~genetic algorithm samples the space of all potential solutions to this multi-objective optimisation problem on the current grid configuration. 
The refinement of the grid is subsequently performed until an acceptable reconstruction is~obtained. 

The number of parameters to be determined by the data are usually larger than for parametric approaches, as~e.g.,\@ for \textsc{Lenstool}.
Therefore, allowing for local fine-tuning, e.g.,~by adding masses in regions without multiple images, free-form methods are more prone to overfitting than to introducing biases. 
A good estimate of the quality of fit of such parameter-free models to the sparse amount of observations is the size of the set of all possible solutions allowed by the observables.
The smaller this set and the tighter the fit, the~more constraining the data. 
\textsc{Grale} averages over several acceptable solutions generated by the genetic algorithm to arrive at the final mass-density reconstruction, such that the standard deviation from this average is a measure for the constraining power of the observables.
Hence, to~optimally constrain local lens properties with \textsc{Grale}, the~tightest confidence bounds are achieved when using constraints from a lot of reference points in the multiple images of the same MIS close to the positions of interest\footnote{Similarly, extended unresolved multiple images can be used to arrive at the same result.}.
This is the opposite way to obtain the tightest confidence bounds compared to \textsc{Lenstool}.



\reftitle{References}

%
%



\end{document}